\documentclass{article}
\usepackage{arxiv}
\usepackage[utf8]{inputenc} 
\usepackage[T1]{fontenc}    
\usepackage{hyperref}       
\usepackage{url}            
\usepackage{booktabs}       
\usepackage{amsmath,amsfonts,amsthm}
\usepackage{nicefrac}       
\usepackage{microtype}      
\usepackage{cleveref}       
\usepackage{lipsum}         
\usepackage{graphicx}
\usepackage{natbib}
\usepackage{doi}

\usepackage{float}
\usepackage{graphicx}
\usepackage[english]{babel}
\usepackage{caption}
\usepackage{subcaption}
\usepackage{rotating}
\usepackage{pdflscape}
\usepackage{afterpage}
\usepackage{capt-of}
\usepackage{booktabs}
\usepackage{threeparttable}
\floatstyle{plaintop}
\RequirePackage{natbib}
\usepackage{threeparttable}
\usepackage{rotating}
\usepackage{hyperref}
\hypersetup{
  colorlinks=true,
  linkcolor=blue,
  citecolor=blue,
  urlcolor=magenta,
  hyperfootnotes=true,
}

\usepackage{enotez}
\let\footnote=\endnote
\usepackage{graphicx}
\usepackage[title]{appendix}
\usepackage{color}
\usepackage{graphicx,multicol}

\title{Academic resilience in the Latin America region post COVID-19 pandemic---an explainable machine learning  analysis of its determinants and heterogeneity using alternative definitions}

\date{}

\usepackage{authblk}

\newbox{\orcid}\sbox{\orcid}{\includegraphics[scale=0.06]{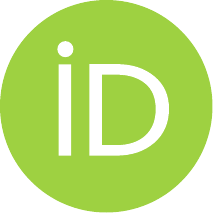}}
\author[1,2]{%
	\href{https://orcid.org/0000-0001-9333-8331}{\usebox{\orcid}\hspace{1mm}Marcos Delprato\thanks{Corresponding author. Email: \texttt{md2645@bath.ac.uk} and \texttt{marcos.a.delprato@gmail.com}.}}%
}
\author[1]{%
	\href{https://orcid.org/0000-0002-4106-062X}{\usebox{\orcid}\hspace{1mm}Andr{\'e}s Sandoval-Hern{\'a}ndez\thanks{Email: \texttt{ash22@bath.ac.uk}.}}%
}
\affil[1]{Department of Education,  University of Bath, UK}
\affil[2]{Instituto de Investigaciones Educativas, Universidad Nacional de Chilecito, Argentina}

\renewcommand{\shorttitle}{Students academic resilience in Latin America-post pandemic}

\hypersetup{
pdftitle={Academic resilience in the Latin America region post COVID-19 pandemic---an explainable machine learning  analysis of its determinants and heterogeneity using alternative definitions},
pdfsubject={econ.GN},
pdfauthor={Marcos Delprato},
pdfkeywords={academic resilience, explainable machine learning, SHAP values, Latin America, PISA 2022},
}

\begin{document}
\maketitle

\begin{abstract}
The learning crisis in the Latin American region (i.e., higher rates of students not reaching basic competencies at secondary level) is worrying, particularly post-pandemic given the stronger role of inequality behind achievement. Within this scenario, the concept of student academic resilience (SAR), students who despite coming from disadvantaged backgrounds reach good performance levels, and an analysis of its determinants, are policy relevant. In this paper, using advancements on explainable machine learning methods (the SHAP method) and relying on PISA 2022 data for 9 countries from the region, we identify leading factors behind SAR using diverse indicators. We find that household inputs (books and digital devices), gender, homework, repetition and work intensity are leading factors for one indicator of academic resilience, whereas for other indicator leading drives fall into the school domain: school size, the ratio of PC connected to the internet, STR and teaching quality proxied by certified teachers and professional development rates and school type. Also, we find negative associations of SAR with the length of school closures and barriers for remote learning during the pandemic. The paper's findings adds to the scare regional literature contributing to future policy designs where key features behind SAR can be used to lift disadvantaged students from lower achievement groups towards being academic resilient.
\end{abstract}



\newpage
\section{Introduction}
\label{section1}

The unfolding of the COVID-19 pandemic intensified the Latin America and the Caribbean (henceforth: LAC) existing learning crisis (\citealp{azevedo21,betthauser23,neidhofer21}) since, on top of the poor connectivity and wealth-driven educational attainment (\citealp{attanasio25,cepal22,munguia23}), schools remained closed much longer than in other regions. In the LAC region school shutdowns extended for around 270 days on average between 2020-2012, representing a disruption of 1.42 years of schooling (\citealp{bracco25}). Thus, the pandemic shifted large cohorts of LAC disadvantaged students towards the end of the learning distribution (\citealp{betthauser23}), with estimates indicating that the region lost around 0.9-1.1 years of schooling (\citealp{azevedo21}). Regrettably, vulnerable students experienced the greatest losses, with studies suggesting a mean loss years of education for a student in the bottom decile of the income distribution of 81\%, nearly four as large than a student in the top decile, losing 22\% (\citealp{bracco25}). Within this bleak education scenario post-pandemic, it is vital to explore ways to counterbalance this escalation of educational inequality in the LAC region.

Resilience refers to successful adaption to situations despite risks that put someone at a disadvantage or adversity (\citealp{windle11}). When applied to the educational domain, the concept of academic resilience denotes students who, despite coming from relatively disadvantaged socioeconomic backgrounds, reach higher education achievement (\citealp{agasisti18,rudd21,tudor17,vicente21}). In other words, academic resilience denotes the capacity of students to perform well in school despite a disadvantaged background which place them into a heightened risk of school failure (\citealp{oecd11}). Assessing the drivers of students academic resilience (henceforth: SAR) can be understood as an initial avenue to lessen post-pandemic educational disparities in the LAC region because resilient students have more opportunities to develop their potential, greater likelihood of social growth, and lower risks of poverty (\citealp{oecd16}). Mapping out which regional factors are either enablers or barriers for students' academic resilience is key for educational policies directed at lifting disadvantaged students from lower achievement groups.

Using as a benchmark the bottom two quintiles of the socio-economic in the distribution of the economic, social and cultural status (ESCS) index of households in the region to capture disadvantaged students and level 2 as the cutoff used for basic performance, this paper contributes to the body of literature on academic resilience by presenting new empirical evidence for the LAC region because studies using a large sample of countries do not focus on the LAC region (e.g., \citealp{agasisti18,cheung24,clavel22,erberber15,garcia22}). Thus, as far as we are aware, this is the first study investigating the determinants of students academic resilience for the LAC region.

The paper also innovates in other directions: it relies on four alternative definitions of students academic resilience given the lack of agreement on how to measure it (\citealp{rudd21,xenofontos23}); it focuses on drivers by education sub-systems; it looks into the linkage of academic resilience with relevant COVID-19 background variables; and, additionally, the paper relies on novel explainable artificial intelligence (XAI) methods. Specifically, based on the latest wave of PISA data (year: 2022), and relying on a sample of nine countries from the LAC region and recent developments in the field of explainable machine learning --i.e., the SHAP method (\citealp{chen23,lundberg17}), the paper's aim is to answer the following research questions:

\begin{enumerate}
  \item Globally, which are the principal determinants of SAR for the whole LAC region, and are these determinants different per SAR indicator?
  \item How do SAR's determinants vary across four education sub-systems (namely: private against public schools, and urban against rural schools) of the LAC region?
  \item Locally, which are the academic resilience profiles of students with the highest and lowest chances of being academic resilient in the LAC region?
\end{enumerate}

The paper is organised as follows. Section \ref{section2} contains a brief review of students academic resilience literature and definitions of the array of indicators of SAR indicators employed in the analysis. We present the data employed for the analysis and leading covariates used in Section \ref{section3}. Section \ref{section4} contains the methodological approach followed --i.e., the ML model used for estimations, gradient boosted trees-- as well as the explainable ML method followed (the SHAP method). Section \ref{section5} shows the paper's results. A discussion of the results is included in Section \ref{section6} and concluding remarks are shown in Section \ref{section7}.


\section{Background}
\label{section2}

\subsection{Students' academic resilience review}
\label{section21}

The concept of resilience has its origins in psychology (\citealp{tudor17,xenofontos23}). It has been categorised as a feature of certain individuals who have good psychological outcomes, despite exposure to acute or chronic stressors that are associated with negative outcomes (\citealp{luthar00,rutter06}), or as the variation in individual responses to personal stress, seeking to pin down determinants of positive adaptation in the face of adversities (\citealp{rutter12}). There are various definitions of resilience, although there are two common elements which all definitions are based on:  adversity and positive adaptation (\citealp{windle11}). More recently, definitions have shifted to understanding it as a process linked to positive adaptation, where individuals interact with their environment (\citealp{luthar00,rutter12}). The psychological literature shows that resilient students share certain characteristics, such as high levels of self-esteem, self-efficacy and motivation (\citealp{wang12}).

When resilience is applied to the context of education, it has been defined as a process, capacity or outcome of successful adaptation during exposure to adversity or risk (\citealp{allan14}). \cite{wang12} define academic resilience as the heightened likelihood of success in school despite environment adversities brought about by early traits, conditions, and experiences. Students who demonstrate academic resilience are those who have been exposed to adverse circumstances, such as low socioeconomic status (SES), that put them at a heightened risk of school failure or performing poorly, yet they demonstrate continued high levels of academic performance (\citealp{rudd21}). When the concept is empirically applied within international learning surveys (ILAs), \cite{oecd11}'s study defines students' resilience as the odds that a student does well academically despite their disadvantaged background (by using the PISA index of economic, social and cultural status, ESCS) to identify the `adverse circumstances', and what is a good student's performance result is given by learning scores thresholds (\citealp{agasisti21,ye21}).

Although there is a consensus on how SAR should be conceptualised, there is a lack of agreement about how to measure it (\citealp{rudd21,xenofontos23}). Among SAR groups, protective factors include elements such as high self-esteem, self-efficacy and autonomy (\citealp{wang12}), engagement in school (\citealp{martin22}) and environmental factors such as parent involvement (\citealp{garcia21}) and classroom environment (\citealp{agasisti18,escalante20,frisby20}). Individual factors of academic resilience are to some extent related to what is known as the 5Cs framework for academic buoyancy: Confidence (self-efficacy), Co-ordination (planning), Control (low uncertain control), Composure (low anxiety) and Commitment (persistence) (\citealp{martin06}).

A large body of SAR quantitative studies follows a person-centred approach by exploring differences between those students identified as resilient and those students as non-resilient, and identify protective factors (\citealp{rudd21}). This is the approach followed in this paper. A detailed account of quantitative analyses of SAR can be found in the reviews of \cite{rudd21}, \cite{tudor17}, \citealp{xenofontos23} and \cite{zheng24}. Among the leading factors at the student$/$family level for SAR found in quantitative studies are gender, stable emotional state and low anxiety, high levels of self-efficacy, aspirations and academic expectations, low chances of truancy, enjoyment of learning a subject, a positive family environment, and students immigrant background (\citealp{cheung24,choi24,clavel22,erberber15,gabrielli22,garcia22,wang22,yavuz16,zhang24}). Moreover, school main factors related to academic resilience are school environment (classroom climate, a high number of extracurricular activities,  additional time for instruction and teachers' quality; \citealp{agasisti18}, \citealp{agasisti21}) and the quality and quantity of school resources (\citealp{vicente21}).

\subsection{Student academic resilience--working definitions}
\label{section22}

Since the paper's purpose is to establish the leading drivers of SAR, methodologically two elements need to be defined: (i) the cutoff used for good performance (an indicator of positive adaption; \citealp{tudor17}), and (ii) how to capture students' disadvantage. For the former element, and following the literature linked to the highly segregated nature and learning crisis of the LAC region's education systems (\citealp{arias23,arias24,brunori24,delprato15,fernandez24}), we employ level 2 as the cut-off point as it is a more appropriate definition for the region's competency$/$baseline achievement level, with students achieving below level 2 referred to as 'low performers' (\citealp[p.~89]{oecd23a}). What is different here to other studies is that we combine proficiency in the three subjects assessed (math, reading and science) so that student academic resilience measurement is not subject-specific, which is the norm in SAR empirical studies (e.g., \citealp{agasisti18,cheung24,garcia22,zhang24}). For the latter element, we restrict the analysis to the two bottom quintiles in the distribution of the economic, social and cultural status (ESCS) index of households in the region, thereby academic resilience is measured among the 40\% poorer students of the LAC region.

The paper relies on four definitions of SAR which fall into the definition-driven approach. That is, all SAR indicators are data-driven as they explicitly identify the resilient sub-sample before conducting statistical analyses. They all fall within what \cite{ye21} call fixed background and either fixed or relative outcome thresholds. The data-driven approach contains two measurement sub-approaches: (i) a high-risk and high achieving group membership and (ii) resilience residuals (\citealp{rudd21}). The four SAR binary indicators employed can be classified into these two sub-approaches. In particular, we focus on two main SAR indicators (SAR1 and SAR2), whereas their extended versions (SAR3 and SAR4) apply stricter conditions for a student being deemed as academic resilient.

Let $i$ denote a student, $j$ a school and $k$ a country. SAR1 is defined as students reaching level 2 or above (in the three subjects) among the LAC's region bottom two SES quintiles. Formally:

\begin{equation}\label{sar1}
\text{SAR1}_{ijk} =  \mathbf{1} \Big[ \big( y_{ijk,\text{math}} \land y_{ijk,\text{reading}} \land y_{ijk,\text{science}} \big) \ge \text{level2} \Big]
\end{equation}

for student $i$'s $\text{SES} \in \big( \text{SES}_{\text{Q1}} \text {  or  }  \text{SES}_{\text{Q2}} \big)$, and 0 otherwise for students who are not academic resilient (or NSAR), and $y_{i}$ denotes students' learning scores. This formulation, with minor variations, has been used in earlier research (\citealp{agasisti18,agasisti17,erberber15}). Its supplementary version (i.e., SAR3) implies a further restriction which, adds as layer of classification for academic resilience, school-driven inequality measured by schools' correlation coefficient of average performance with family SES. Only those students achieving at least level 2 within the top half most unequal schools are considered academic resilient. That is:

\begin{equation}\label{sar3}
\text{SAR3}_{ijk} \equiv \text{SAR1}_{ijk} \land \big[ \rho_{j,\text{yscore,family-SES}} \ge \overline{\rho}_{\text{yscore,family-SES}} \big]
\end{equation}

When moving to SAR indicators of the second sub-approach (i.e., SAR2 and SAR4), they entail some prior estimations on the probability of students achieving at least level 2 and then placing a threshold on estimated probabilities to define whether or not a students is academic resilient. This approach is in line with earlier studies (e.g., \citealp{cheung17,clavel22,garcia19,wills19}), with the main difference being that we account for contextual disadvantage, i.e., average school SES, as an additional control. SAR2 is obtained by running a 3-level logit multilevel model for $y_{ijk}$ (the composite binary indicator equals to 1 if a student achieves level 2 or above, and 0 otherwise), where students are nested in schools which are nested in countries:

\begin{equation}\label{sar2}
\begin{split}
\text{logit}\Big(\text{Pr}\big(y_{ijk}=1 \bigl\lvert \textbf{x}_{ijk},\mu^{(2)}_{jk},\mu^{(3)}_{k}\big)\Big) & = \text{SES}_{ijk}\beta_{1} + \overline{\text{SES}}_{jk}\beta_{2}  \\
& +\mu^{(2)}_{jk} + \mu^{(3)}_{k} + \epsilon_{ijk} \\
 \text{SAR2}_{ijk} & = 1 \text{    if    } \widehat{y}_{ijk} \in \big( \widehat{y}_{\text{Q4}} \text{    or     } \widehat{y}_{\text{Q5}} \big)
\end{split}
\end{equation}

where $\mu^{(2)}_{jk}$ and $\mu^{(3)}_{k}$ are schools's and country's random intercepts, and $\epsilon_{ijk}$ is the level 1 error term following a logistic distribution with variance $\pi^{2}/3$ ($\approx$ 3.29). So, SAR2 measures as academic resilient a student whose estimated probability (after both SES controls at the family and school levels, as well as net of schools' and countries' unobserved random effects) falls into the top 40\% of the estimated probability of the level 2 achievement distribution. SAR4 is a further tuning of SAR2 excluding students from the most efficient schools (those with large random intercepts). That is, among the group of top 40\% students estimated probabilities, we exclude students coming from the schools whose random intercepts are in top 20\% (or quintile 5) of the $\mu^{(2)}_{jk}$ distribution. SAR4 is then defined as:

\begin{equation}\label{sar4}
\begin{split}
\text{SAR4}_{ijk} \equiv \text{SAR2}_{ijk} \land \big[ \widehat{\mu}^{(2)}_{jk} \notin  \widehat{\mu}^{(2)}_{jk,\text{Q5}} \big]
\end{split}
\end{equation}


\section{Data}
\label{section3}

The analysis is based on the latest wave of PISA for the year 2022. The paper is focused on a sample of nine countries from the LAC region (i.e., Argentina, Brazil, Chile, Colombia, Dominican Rep., Mexico, Panama, Peru and Uruguay).\footnote{We exclude two additional LAC countries from the analysis, i.e., Costa Rica (because its dataset does not contain the key covariate household SES) and Paraguay (soft-skills covariates missing).}  Table \ref{table1} includes the four SAR rates for the whole sample and by sub-samples.  Among the bottom 40\% poorer students for the whole LAC region, 21.2\% are academic resilient (SAR1) and 11.2\% according to SAR3 definition (column 1); also, the table SAR's rates shows how SAR3 and SAR4 are subsets of SAR1 and SAR2, respectively, with values of 11.7\% (SAR2) and 5.2\% (SAR4). What is notable is the substantial heterogeneity of SAR rates across education sub-systems (private$/$public schools, and urban$/$rural schools; columns 2 to 5 of Table \ref{table1}). Regardless of the SAR formulation used, students attending either private or urban schools are more likely to be academic resilient; private-public schools gaps on SAR rates vary between 11\%-25\% (and by 5\%-12\% between urban and rural schools).

\medskip
\begin{center}
  [Table \ref{table1} here]
\end{center}
\medskip

Students' chances of being academic resilient depend not only on closest individual features but also their proximal context. The empirical relevance of students and family factors as pathways for academic resilience is shown by covariates' summary statistics for SAR and non-SAR groups in Table \ref{table2}, focusing on SAR1 and SAR2 indicators.\footnote{Similar results for the other two SAR indicators, SAR3 and SAR4, can be obtained from the authors upon request.} Nearly all students' background characteristics are statistically different for these two groups. Using SAR1 as benchmark (Table \ref{table2}, columns 1 to 3), compared to students who are not academic resilient, resilient students are more likely to be male and carry out more homework, have a higher number of digital devices (= +1.36) and books (= +0.39), being considerably less likely to be a immigrant or speak other language at home, and come from relatively better-off households of smaller size with higher parental education levels. Likewise, in comparison to the non-SAR group, students from the SAR group are less prone to engage in both paid and unpaid work. Educational risk factors in the form of earlier weaker achievement are significantly lower for the SAR group. Repetition for resilient students --either at primary or lower secondary-- is much lower (3.6\% against 22.6\% and 4.2\% against 17.2\%) as well as the chances of missing or skipping school, with school climate and contextual safety being slightly better. Furthermore, the range of specific personality traits (i.e., assertiveness, curiosity, emotional control and empathy, and stress management) are all superior, with larger indices, for the SAR group. Gaps on students and family characteristics based on SAR2 are similar (column 6 of Table \ref{table2}).

\medskip
\begin{center}
  [Table \ref{table2} here]
\end{center}
\medskip

In Panel A of Table \ref{table3} (based on SAR1, columns 1 to 3) it can be seen, and as previously shown in Table \ref{table1}, the higher chances of being academic resilient in the most advantageous private (+0.10) and urban (+0.17) schools, as well as in larger schools and schools facing more competition. The rates of students with a second language and disadvantaged students are lower by 4\% and 7\% in the SAR1 group (column 3), which also has a somewhat higher PC connected web-teacher ratio. Vitally, in schools where parental involvement is actually stimulated by schools, students have a larger chances of being academic resilient. Additionally, summary statistics for COVID-related school variables (Panel B, Table \ref{table3}) show there is an increasing probability of a student being classified as academic resilient if he$/$she attends a school with shorter shutdown period during the pandemic (by around 10 days in average), and a school with lower overall barriers for remote instruction, and with teachers having better opportunities though the use of learning support platforms and better skills for remote learning.

\medskip
\begin{center}
  [Table \ref{table3} here]
\end{center}
\medskip


\section{Methods}
\label{section4}


\subsection{Models' comparison}
\label{section41}

Given that SAR dependent variables are dichotomous (taking the value of 1 for students classified as academic resilient, and 0 otherwise), the analysis is based on binary classification ML models. In particular, we rely on three ML models: logistic regression (logit), neural networks (NN) and gradient boosted trees (GBT) (for details, see: \citealp{hastie09,kumari17,natekin13,raschka19}). Training of each model follows a stratified 5-fold cross validation with a 80\%-20\% random division of the dataset into training and testing sets, correspondingly, assuring a balanced data within each fold. The optimisation of hyperparameters uses a grid search to arrive to the best performing version of each model. Following that, we select the final ML model relying on two performing metrics: the area under the receiver operator characteristic curve (AUROC), and the area under the precision recall curve (AUPRC) (\citealp{naidu23,rainio24}).

For the logit model we consider six specifications based on two penalties terms (L1-Lasso regularisation based on the absolute value of coefficients, and L2-ridge: penalty based on the square of the coefficients) and three inverse of regularisation strengths (C = 0.1, 1, 10). For the NN model, we use nine specifications, three hidden layer size sets: \{200\}, \{100, 100\}, \{200, 100, 50\}, and three activation rules (relu, tanh, logistic). For the GBT model, we evaluate several specifications based on combinations of number of estimators (100, 500, 1000, 5000), sub-sample ratios (0.5, 0.7, 0.9), trees of maximum depth (3, 5, 7, 9) and three learning rates (0.001, 0.01, 0.1).

A comparison for the three ML models and four SAR indicators is shown in Table \ref{table4} for the whole sample.\footnote{The same exercise is carried out for the four sub-samples (private, public, urban and rural schools). Results are included in the Appendix (Table \ref{tableC1}). For the final 16 models, a comparison of evaluation metrics leads to the selection of the GBT model as the more efficient$/$chosen estimation method.} Across the 12 ML models for the four SAR outcomes, evaluation metrics show a clear performance dominance of GBT. In the case of SAR1, for instance, AUROC and AUPRC scores for the GBT model are 0.911 and 0.742, respectively, a gap on performance of nearly 11\%-23\% in comparison to the logit model and of around 5\%-10\% with respect to the NN model (columns 2 and 3 of Table \ref{table4}). Equivalently, for SAR2, based on metrics, the better preforming model is the GBT with differential performance between 6\% to 17\% (AUROC score) and 25\%-58\% (AUPRC score). This is not unexpected because tree-based model such as the GBT can achieve higher predictive power, capturing non-linear and interaction effects between predictors (\citealp{ke17,qiu22}). Therefore, we present below the GBT model which will be used for initial estimations and, in turn, their estimated values to obtain the ML explainable results based on the Shapley Additive Explanations (SHAP) method (\citealp{lundberg20,lundberg17}),

\medskip
\begin{center}
  [Table \ref{table4} here]
\end{center}
\medskip


\subsection{Estimation--gradient boosted trees}
\label{section42}

Gradient boosted trees (GBT) is a nonparametric ML method composed of iteratively trained decision trees (\citealp{krauss17}). GBT is highly efficient and flexible with a high performance because the final ensemble of trees is able to capture non-linear and interaction effects between predictors; hence, it is not surprising the higher performance found when comparing it to the other two ML models. For the analysis, we use the XGBoost implementation of the GBT model (\citealp{chen16}).\footnote{See: \href{https://xgboost.readthedocs.io/en/latest/python/index.html}{link-XGBoost}.} XGBoost's objective function captures deviations from the model and introduces regularization to prevent overfitting (\citealp{ntamwiza25}). A tree ensemble model uses $K$ additive functions to predict the output (\citealp{chen16}):

\begin{equation}\label{xgboost1}
  \hat{y}_{i} = \phi(\mathbf{x}_{i}) = \sum_{k=1}^{K} f_{k}(\mathbf{x}_{i})
\end{equation}

Here, $\hat{y}_{i}$ denotes the prediction label for a given sample $f(\mathbf{x}_{i})$, and $f_{k}(\mathbf{x}_{i})$ represents the predicted score for the sample, an independent tree structure $q$ and leaf weights $\omega$, and $f(\mathbf{x})$ denotes the value of a leaf (using $\omega_{i}$ to represent the score of the $i$-th leaf), and where $f_{k} \in \mathcal{F}$ is the space of regression trees: $\mathcal{F} = \{f_{k}(\mathbf{x}) = \omega_{q(x)}\} (q: \mathbb{R}^{m} \mapsto T,  \omega \in \mathbb{R}^{T})$, where $T$ is the number of leaves in the tree.

To learn the set of functions used in the model, the following regularized objective is minimised:

\begin{equation}\label{xgboost2}
\begin{aligned}
  \mathcal{L}(\phi) &= \sum_{i} l( \hat{y}_{i}, {y}_{i}) + \sum_{k} \Omega (f_{k})    \\
  \text{and  } \Omega(f) &= \gamma T + \frac{1}{2} \lambda \|\omega\|^{2}
\end{aligned}
\end{equation}

where $l$ measures the distance between the prediction and the target dependent variable, and $\Omega(.)$ is the penalisation term minimising the risk of overfitting. An additive form for optimisation is used when training the model:

\begin{equation}\label{xgboost3}
  \mathcal{L}^{(t)} = \sum_{i=1}^{N} l \big(y_{i}, \hat{y}^{t-1}_{i}, {y}_{i}) + f_{t}(\mathbf{x}_{i})\big) + \Omega (f_{t})
\end{equation}

where $\hat{y}^{t-1}_{i}$ is the prediction of the $i$-th instance at the $t$-th iteration and $f_{t}$ is added to improve the model according to Eq. (\ref{xgboost2}).


\subsection{Explainable ML estimates--the SHAP method and partial dependence plots}
\label{section43}

Due to the latest reintroduction of Shapley values as a ML explanatory framework (\citealp{lundberg17,vstrumbelj14}), this method has become one of the most common approach within the explainable artificial intelligence (XAI) field. And, particularly for educational studies, there has been a recent growth in the application of Shapley values to identify main drivers of educational outcomes (for a detailed review, see: \citealp{ersozlu24,hilbert21}). Shapley values have their origin in game theory (\citealp{shapley53}), measuring the average marginal contribution of a player in a cooperative game. In the context of ML, the player is interpreted as a characteristic or covariate (an attribute), and the cooperative game becomes the prediction task performed by the model.

We follow the SHAP method (\citealp{chen23,lundberg17}) as an explanatory method of estimates of the GBT model. Computationally, we use the TreeExplainer (\citealp{lundberg20}) package which offers an exact calculation of SHAP values for tree-based models. SHAP values offer a local explanation of a model output by calculating the degree to which each feature (or variable) contributes to a given prediction value when conditioning on that feature. Formally, the change on prediction or SHAP value $\phi_{i}$(.) for feature (or covariate) $i$ in model $f$ for data point $x$ is given by:

\begin{equation}\label{shap1}
  \phi_{i}(f, x) = \sum_{R \in \mathcal{R}}\frac{1}{M!}\big(f_{x}(P^{R}_{i} \cup i) - f_{x}(P^{R}_{i})\big)
\end{equation}

where $\mathcal{R}$ is the set of all feature permutations, $P^{R}_{i}$ is the set of all features before $i$ in the ordering $R$, $M$ is the number of input
features, and $f_{x}$ is an estimate of the conditional expectation of the model’s prediction. The additive property of SHAP values means that SHAP values add to the output of the model:

\begin{equation}\label{shap2}
  f(x) = \phi_{0}(f) + \sum_{i=1}^{M}\phi_{i}(f, x)
\end{equation}

Attributions based on the Shapley interaction index result in a matrix of feature attributions. SHAP values main effects (on the matrix's diagonal) are obtained as the differences of SHAP values and the off-diagonal elements (i.e., SHAP interactions) for a given feature, and are defined as:

\begin{equation}\label{shap3}
  \Phi_{i,i} = \phi_{i}(f,x)-\underbrace{\sum_{j \neq i}^{}\Phi_{i,j}(f,x) }_{\text{SHAP interactions}}
\end{equation}

Moreover --as well as estimating SHAP values' main effects for the whole range of covariates of Tables \ref{table2} and \ref{table3}-- we rely on dependence plots when explaining specific findings linked to the impact of COVID-19 background variables (CBV) and students' personality traits or soft skills (SK) on the chances of a student's being academic resilient. Partial dependence plots show the marginal effect of a given covariate (feature) on the prediction of the ML model by calculating the relative probability (RP) of SAR outcomes taking the value 1 (a student falling into the SAR group). The RP is obtained as:

\begin{equation}\label{RP}
  \text{RP} = \frac{f_{\mathcal{S}}(x_{\mathcal{S}})}{(1/n) \sum_{i=1}^{n}f(x^{(i)})}
\end{equation}

which is the mean value of the GBT model predicted probability when fixing the specific feature (CBV or SK) divided by the mean value of the model's predicted probability, and $f_{\mathcal{S}}(x_{\mathcal{S}})$ is the partial function of the specific variable.


\section{Results}
\label{section5}


\subsection{Whole sample estimations}
\label{section51}

\subsubsection{Overall determinants}
\label{section511}

\begin{figure}[ht!]
\centering
\subfloat[\footnotesize{SAR1}]{\includegraphics[width=0.49\textwidth]{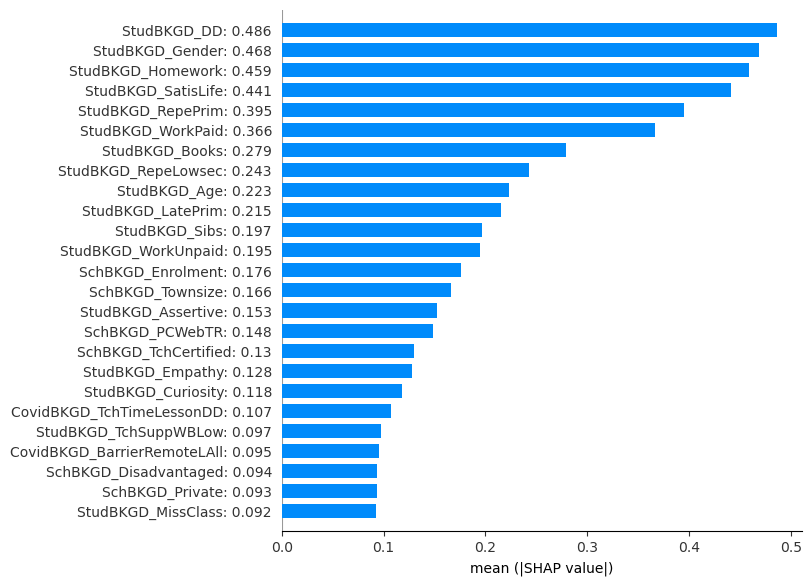}\label{figure1a}}
\subfloat[\footnotesize{SAR2}]{\includegraphics[width=0.49\textwidth]{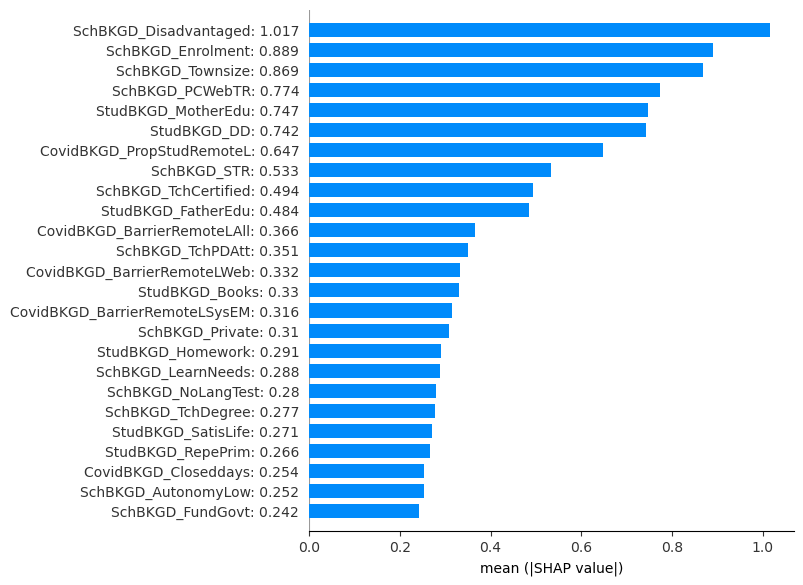}\label{figure1b}}
\\
\subfloat[\footnotesize{SAR3}]{\includegraphics[width=0.49\textwidth]{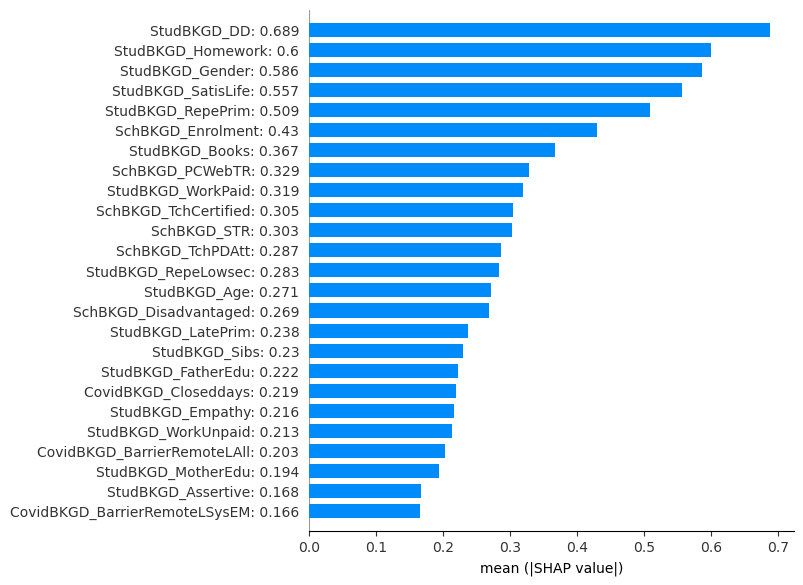}\label{figure1c}}
\subfloat[\footnotesize{SAR4}]{\includegraphics[width=0.49\textwidth]{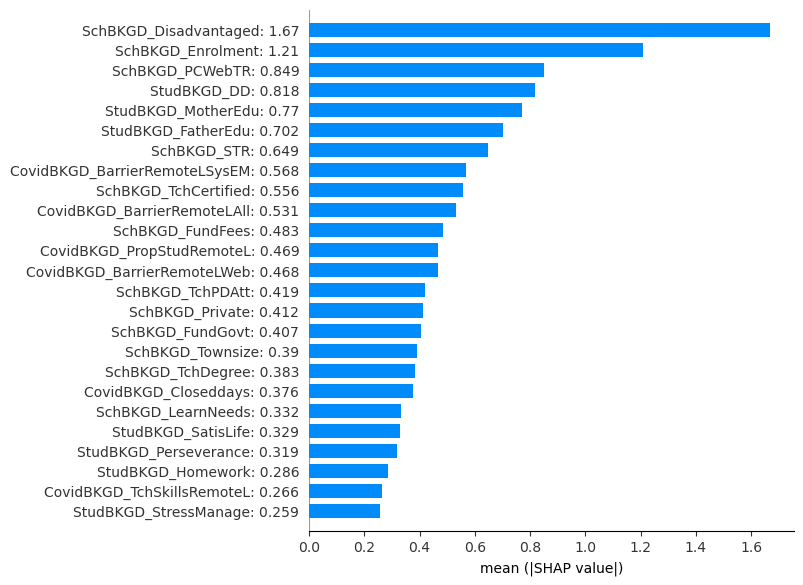}\label{figure1d}}
\\
\subfloat[\footnotesize{SAR1 versus SAR3}]{\includegraphics[width=0.49\textwidth]{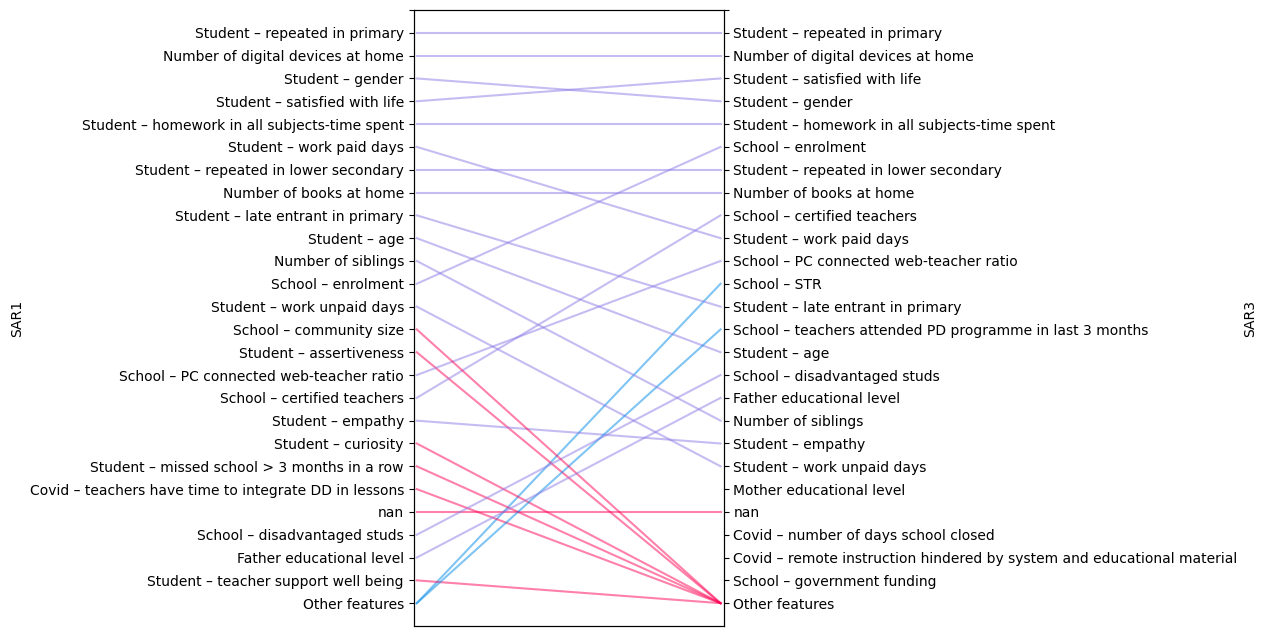}\label{figure1e}}
\subfloat[\footnotesize{SAR2 versus SAR4}]{\includegraphics[width=0.49\textwidth]{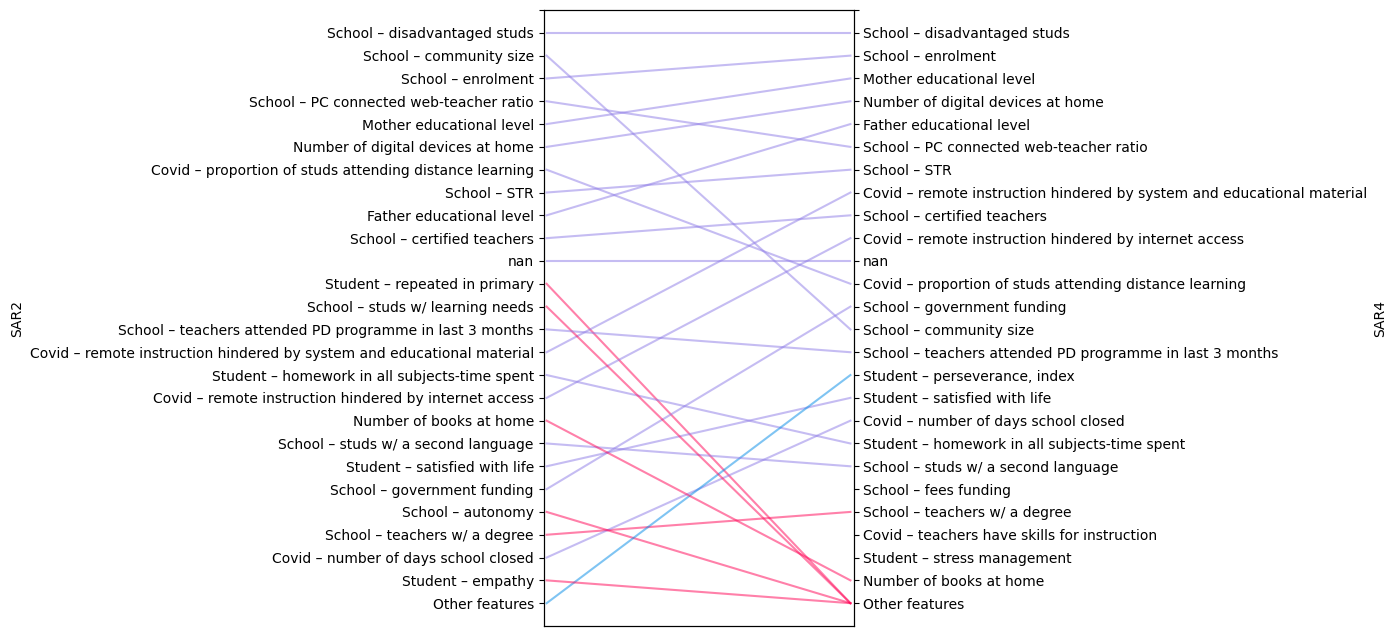}\label{figure1f}}
\caption{SHAP values of leading determinants}
\label{figure1}
\medskip
\begin{minipage}{0.99\textwidth}
\footnotesize{
Notes: (1). Importance plots show ranked average absolute SHAP values per covariate (top 25 covariates). (2) Details about covariates' definition can be found in Table \ref{tableA1}.
}
\end{minipage}
\end{figure}

Out of the whole range of students$/$family and school characteristics (Tables \ref{table2} and \ref{table3}, 65 covariates in total), Figure \ref{figure1} displays the top 25 leading covariates of GBT estimates in predicting each SAR outcome by ranking covariates importance based on their average absolute SHAP values across observations. The plot includes variables' acronyms, with full details on covariates' definitions included in Table \ref{tableA1}. First, among the top 25 powerful predictors shown in Figure \ref{figure1a} for the SAR1 dependent variable, most features are from to the student's group and with higher ranked larger SHAP values. Leading predictive drivers with an average importance over 0.24 are the number of digital devices at home, gender, homework, satisfaction with life, primary repetition, engagement with paid work, books at home, and lower secondary repetition. Students' personal traits (e.g., assertiveness, curiosity and empathy), though lower ranked with smaller mean SHAP values, still carry out some predictive importance for the probability of students' academic resilience, as well as some school-pandemic relate covariates (i.e., barriers to for remote learning and teachers having time to integrate digital devices in lessons) and school type.

\medskip
\begin{center}
  [Figure \ref{figure1} here]
\end{center}
\medskip

Second, for the alternative SAR2 formulation (Figure \ref{figure1b}), compared to SAR1, there is a shift in the most influential determinants towards the school domain. This is expected as SAR2 is the predicted probability of academic resilience net of family and school SES effects. Specifically, SAR2's prominent covariates are the rate of disadvantaged students at school, school and town size, the ratios of PC connected to the internet and students$/$teacher ratio, parental education, digital devices at home and the prevalence of qualified teachers (all with mean $|\text{SHAP}|$ values $\ge$ 0.48). Few school-pandemic variables (different kind of barriers for online$/$remote learning) are located in the next range (above 0.30). Altogether, this indicates the empirical relevance of how a SAR outcome is constructed given the dissimilar rankings of drivers found in Figures \ref{figure1a}-\ref{figure1b}.

Estimates for the other versions of the SAR1 and SAR2 indicators are shown in Figure \ref{figure1c} (SAR3) and in Figure \ref{figure1d} (SAR4). Recall that these two latter indicators have smaller rates as they are subsets of SAR1 and SAR2 (that is: SAR3 = 11.2\% whereas SAR1 = 21.2\%, and SAR4 = 5.2\% versus SAR2 = 11.7\%; Table \ref{table1}). On the one hand, in the case of SAR3 and SAR1, when further restricting the population of academic resilient students to those attending the most unequal schools, determinants' comparison points towards a superior relevance of school's characteristics for SAR3 than for SAR1 (Figure \ref{figure1e}), with school enrolment, level of certified teachers, connected PC, student-teacher ratio (STR) going up in the rank. On the other hand, when comparing SAR4 (which excludes the most efficient schools) to SAR2, the chances of a student being academic resilient are more powerfully influenced by family inputs (mother's and father's education), students' perseverance and life satisfaction as well as some COVID-19 variables (barriers for remote instruction and length of school closure during the pandemic), once more suggesting why the definition of the SAR indicator empirically matters.

\begin{figure}[ht!]
\centering
\subfloat[\footnotesize{SAR1}]{\includegraphics[width=0.49\textwidth]{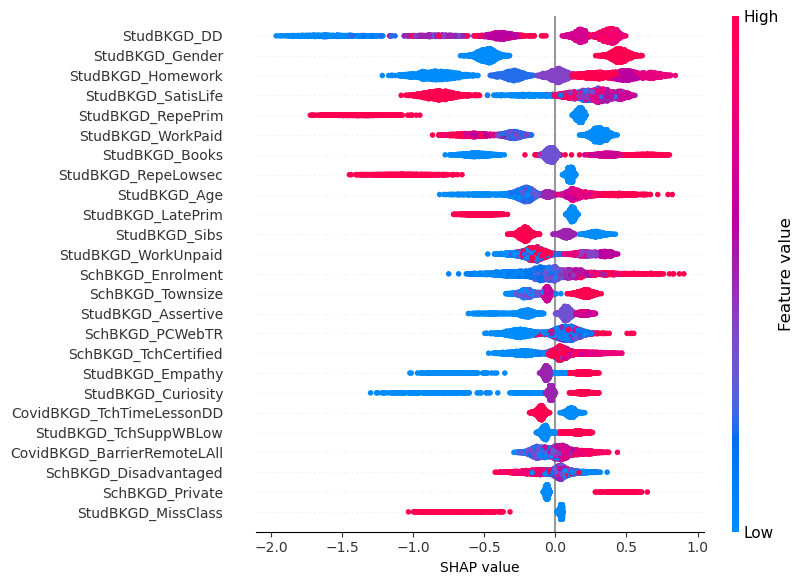}\label{figure2a}}
\subfloat[\footnotesize{SAR2}]{\includegraphics[width=0.49\textwidth]{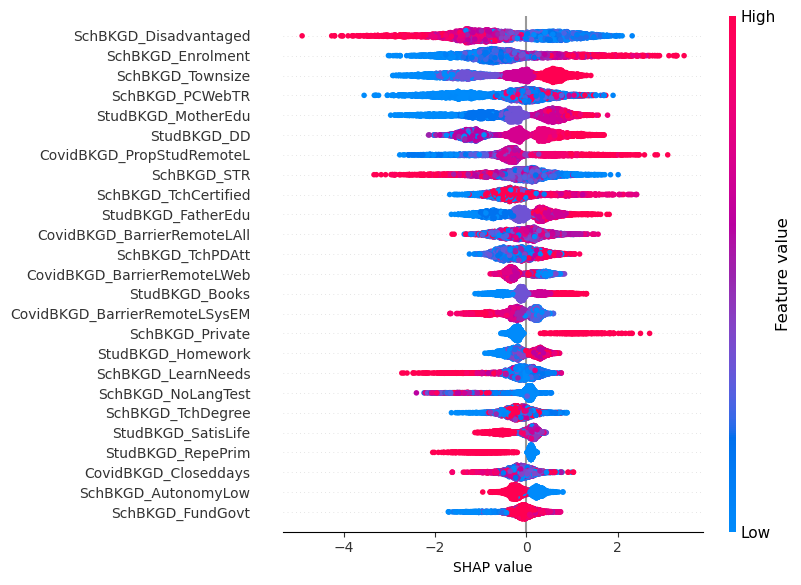}\label{figure2b}}
\\
\subfloat[\footnotesize{SAR3}]{\includegraphics[width=0.49\textwidth]{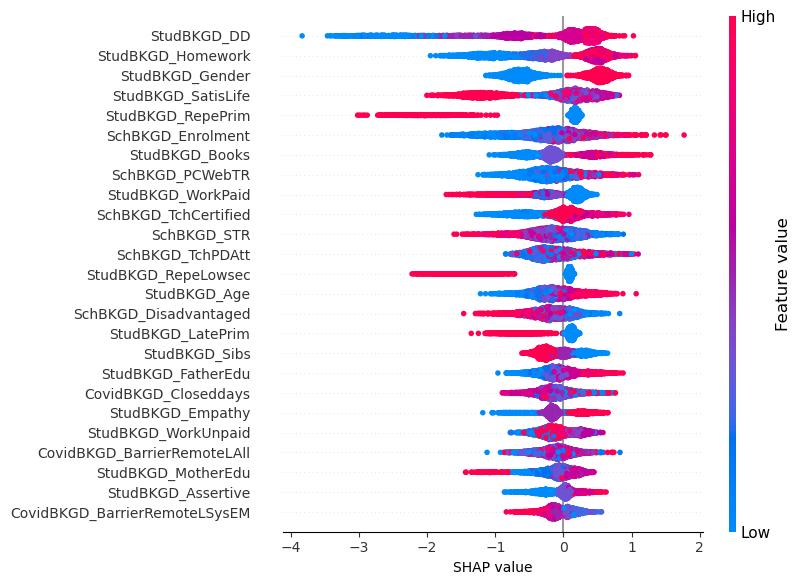}\label{figure2c}}
\subfloat[\footnotesize{SAR4}]{\includegraphics[width=0.49\textwidth]{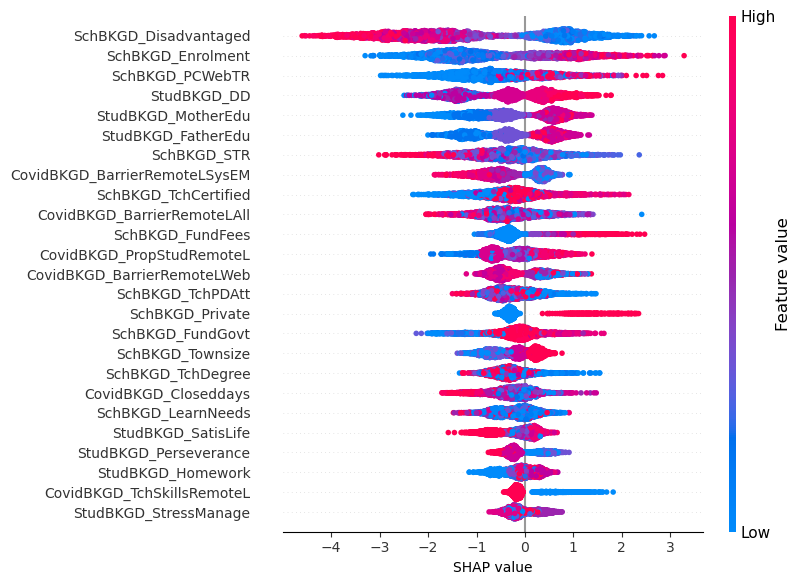}\label{figure2d}}
\caption{Beeswarm plots}
\label{figure2}
\medskip
\begin{minipage}{0.99\textwidth}
\footnotesize{
Notes: (1) Beeswarm plots show point-wise SHAP values. Red dots represent covariates taking values equal to = 1 and, blue dots (= 0) otherwise.
}
\end{minipage}
\end{figure}

To further decompose the contribution of each covariate to SAR, Figure \ref{figure2} reveals the magnitude and direction of SHAP values through beeswarm plots. These plots display the distribution of SHAP values for each student and covariate, offering a more granular approach than the aggregated SHAP values.\footnote{Beeswarm plots are essentially dot plots where each student's specific SHAP value has one dot on the line of each covariate, and dots are stacked horizontally showing the SHAP value density for each $X$ (explanatory variable) arranged from low to high values.} Here, we discuss central results and for the two main outcomes (i.e., SAR1 and SAR2). For SAR1, Figure \ref{figure2a} shows how high values of digital devices (DD) have a positive contribution on the probability of SAR (SHAP values $\ge$ 0.3), whereas the group of students who are not academic resilient (NSAR) have low numbers of DD have with SHAP values below 1. Gender-wise, the plot shows that male students tend to be SAR (in red with SHAP values $\approx$ 0.5), but female students tend to fall into the NSAR group (in blue, SHAP value $\le$ -0.5). Other covariates with important gaps on their SHAP values are primary and lower secondary repetition rates (with repeaters having significant lower chances of being academic resilient; SHAP values gaps of +1.50) and paid work intensity (StudBKGD\_WorkPaid), with students working several day per week having negative SHAP values of -0.7, and positive (of +0.4) for those with minimal engagement in paid work. Increasing students' curiosity and empathy are also linked to higher SAR probabilities (red dots with SHAP values $\approx$ 0.3), whilst students from public schools (blue dots for feature SchBKGD\_Private) have SHAP values negative close to zero, but private school students having larger SHAP values (around +0.5), highlighting the decisive role of school type behind whether or not a student is academic resilient.

\medskip
\begin{center}
  [Figure \ref{figure2} here]
\end{center}
\medskip

Additionally, SAR2 estimated SHAP values (Figure \ref{figure2b}) are more dispersed than for SAR1 (but continuously distributed without gaps across dots), with a range between -4 and +2. High availability of digital resources for educational purposes contributes to augmenting SAR chances (SHAP value $\le$ 2) and, on the contrary, NARS tend to have lower DD (SHAP values close to -2). Increasing prevalence of students engaged in remote learning during the pandemic and smaller STR greatly contribute to the chances of students falling into the SAR2 group (SHAP values in the interval +2, +2.5), with SAR-NSAR differences of +5 for SHAP values for these covariates between high (red dots) and low values (blue dots). The proportion of certified teachers and parental education are also important positive influences on academic resilience (SHAP values $\ge$ 2).

\subsubsection{COVID-19 background variables}
\label{section512}

\begin{figure}[ht!]
\centering
\subfloat[\scriptsize{SAR1-closure days}]{\includegraphics[width=0.24\textwidth]{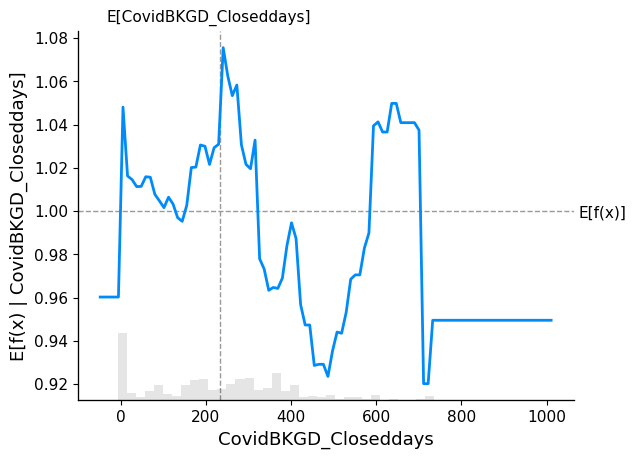}\label{figure3a}}
\subfloat[\scriptsize{SAR1-study remotely}]{\includegraphics[width=0.24\textwidth]{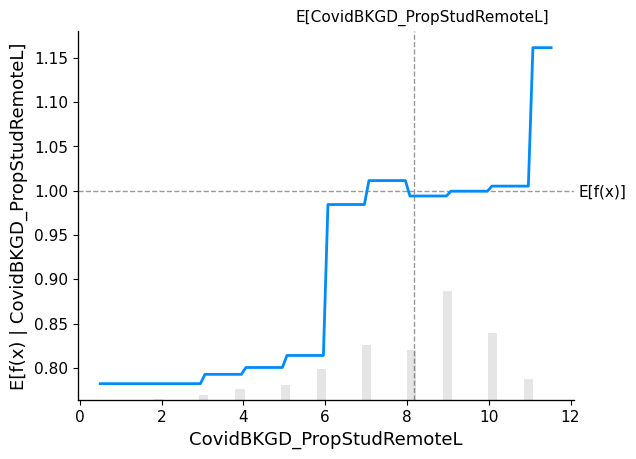}\label{figure3b}}
\subfloat[\scriptsize{SAR1-barrier: internet}]{\includegraphics[width=0.24\textwidth]{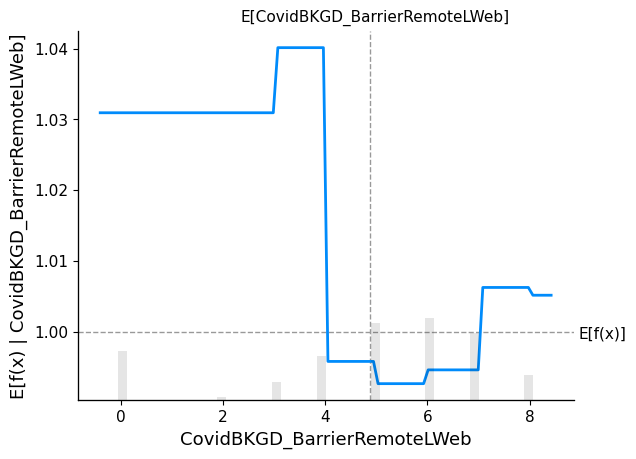}\label{figure3b}}
\subfloat[\scriptsize{SAR1-barrier: system$/$material}]{\includegraphics[width=0.24\textwidth]{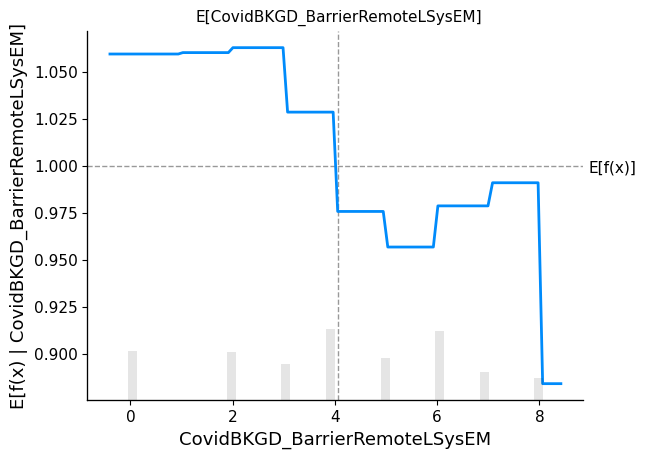}\label{figure3d}}
\\
\subfloat[\scriptsize{SAR2-closure days}]{\includegraphics[width=0.24\textwidth]{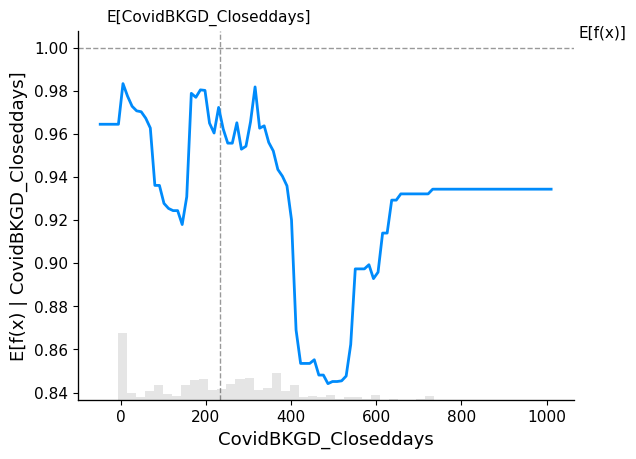}\label{figure3e}}
\subfloat[\scriptsize{SAR2-study remotely}]{\includegraphics[width=0.24\textwidth]{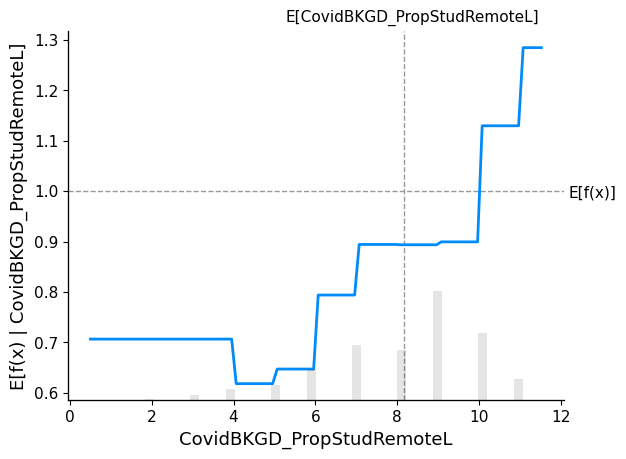}\label{figure3f}}
\subfloat[\scriptsize{SAR2-barrier: internet}]{\includegraphics[width=0.24\textwidth]{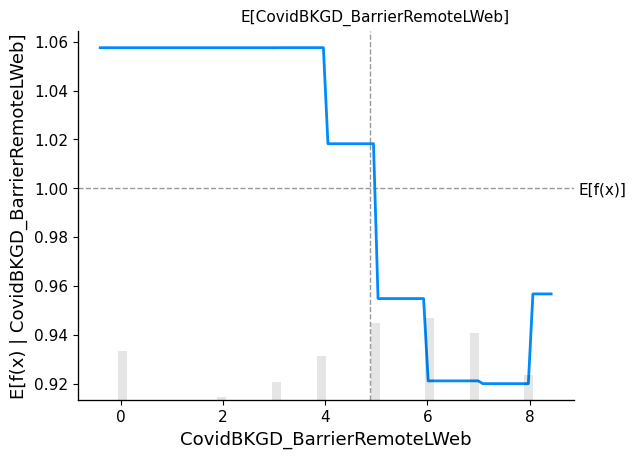}\label{figure3g}}
\subfloat[\scriptsize{SAR2-barrier: system$/$material}]{\includegraphics[width=0.24\textwidth]{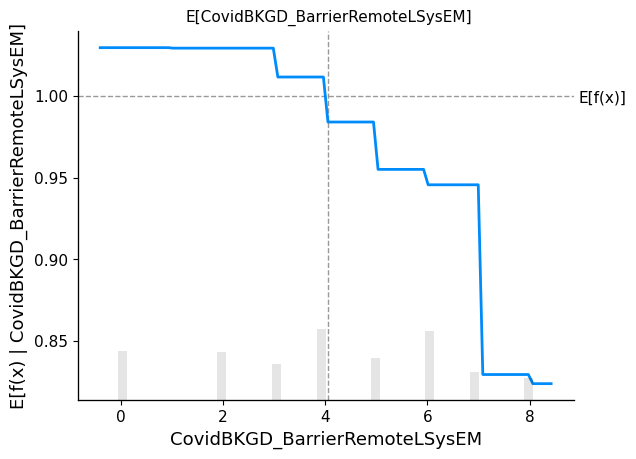}\label{figure3h}}
\\
\subfloat[\scriptsize{SAR3-closure days}]{\includegraphics[width=0.24\textwidth]{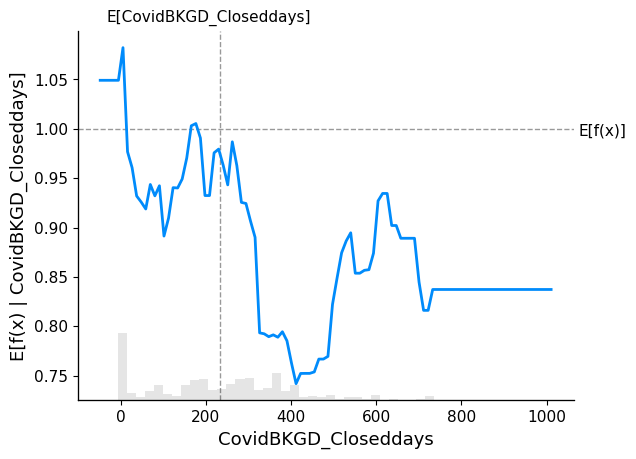}\label{figure3i}}
\subfloat[\scriptsize{SAR3-study remotely}]{\includegraphics[width=0.24\textwidth]{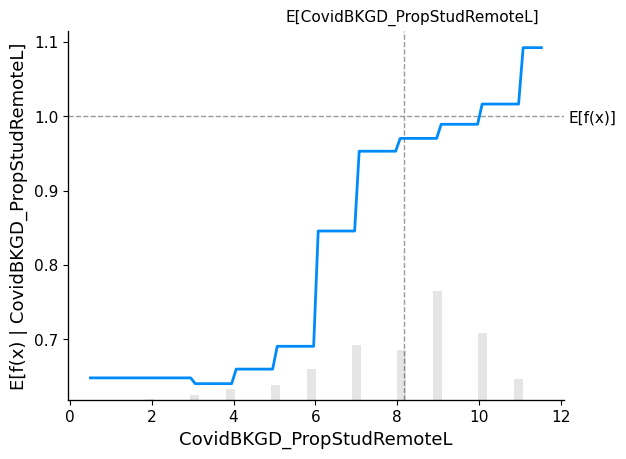}\label{figure3j}}
\subfloat[\scriptsize{SAR3-barrier: internet}]{\includegraphics[width=0.24\textwidth]{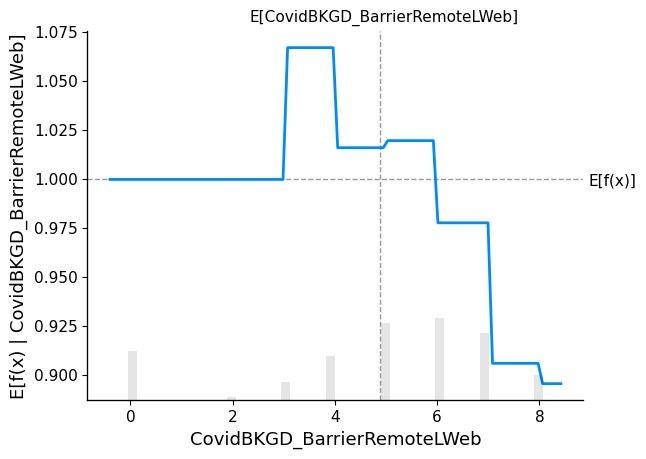}\label{figure3k}}
\subfloat[\scriptsize{SAR3-barrier: system$/$material}]{\includegraphics[width=0.24\textwidth]{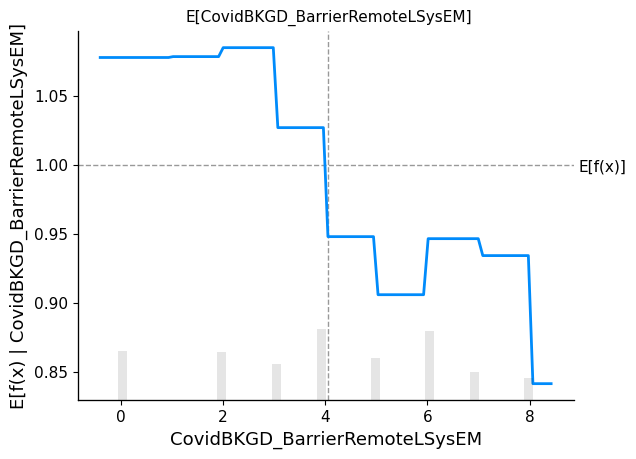}\label{figure3l}}
\\
\subfloat[\scriptsize{SAR4-closure days}]{\includegraphics[width=0.24\textwidth]{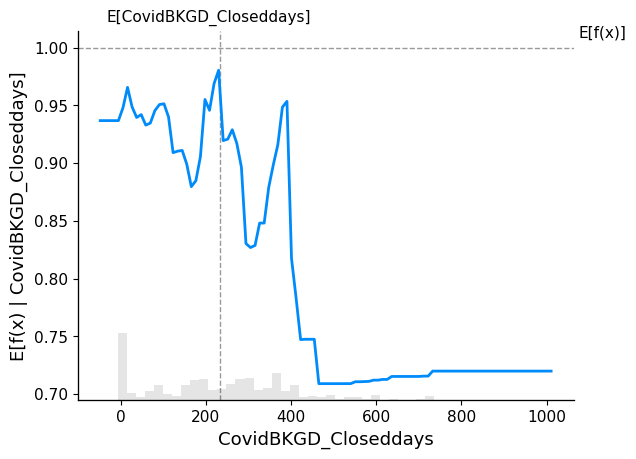}\label{figure3m}}
\subfloat[\scriptsize{SAR4-study remotely}]{\includegraphics[width=0.24\textwidth]{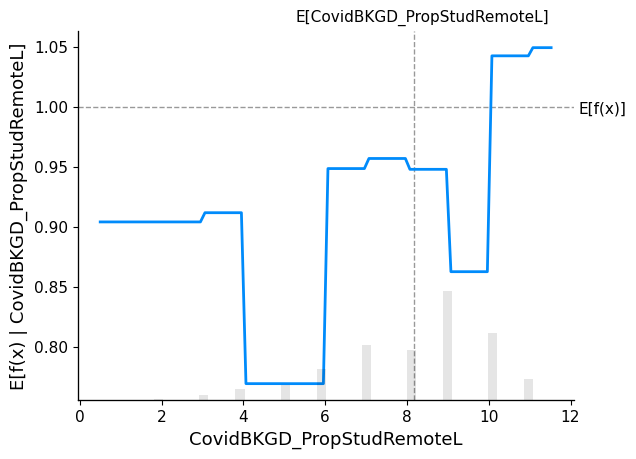}\label{figure3n}}
\subfloat[\scriptsize{SAR4-barrier: internet}]{\includegraphics[width=0.24\textwidth]{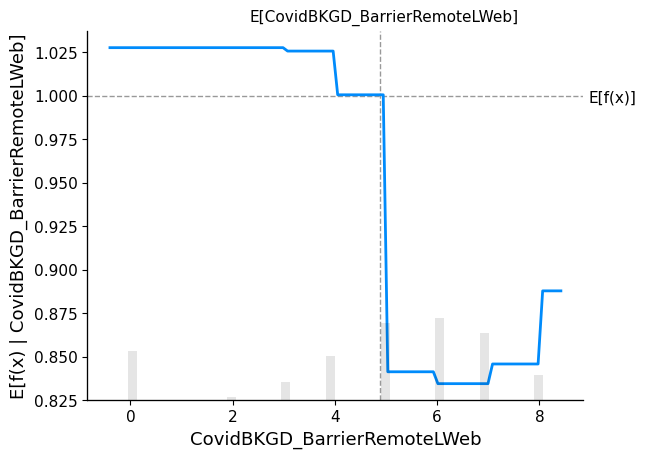}\label{figure3o}}
\subfloat[\scriptsize{SAR4-barrier: system$/$material}]{\includegraphics[width=0.24\textwidth]{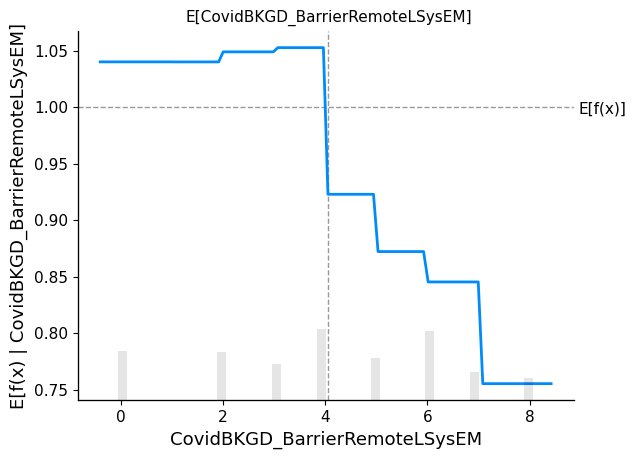}\label{figure3p}}
\caption{Partial dependence plots for COVID-19 background variables}
\label{figure3}
\end{figure}

This section offers a more detailed analysis of variables related to the delivery of education during the pandemic and their impact on SAR probabilities. This is done by inspecting estimated relative probabilities for COVID background variables which are displayed in the partial dependence plots of Figure \ref{figure3}. Note that here marginal effects of these variables are plotted, which are transformed as odd-ratios (OR) for a better interpretability.

\medskip
\begin{center}
  [Figure \ref{figure3} here]
\end{center}
\medskip

Firstly, partial dependence plots for number of days of school closure for all SAR outcomes (Figures \ref{figure3a}, \ref{figure3e}, \ref{figure3i} and \ref{figure3m}) show an uniformly decreasing tendency on the SAR's likelihood for longer schools shutdowns. An increment of 170 days of schools shutdowns, from the LAC region's mean days of school closure ($\approx$ 230 days) to 400 days, leads to reductions on estimated odd-ratio ($\widehat{\text{OR}}$) between 10\% to 25\%, therefore showing the deterioration of the chances of students being academic resilient for students attending Latin American schools with longer closures during the pandemic. For instance, SAR2's $\widehat{\text{OR}}$ is equal to 0.98 at the mean of 230 days, but decreasing to 0.85 at 400 days; and for SAR4 the corresponding values for $\widehat{\text{OR}}$ are 0.98 and 0.72. Secondly, having larger cohorts of students engaged in remote learning have contextual effects by boosting the chances of a student being academic resilient. This is shown by an increasing tendency of $\widehat{\text{OR}}$ (SAR1 to SAR3; Figures \ref{figure3b}, \ref{figure3f} and \ref{figure3j}), with $\widehat{\text{OR}} >1$  when the covariate CovidBKGD\_PropStudRemoteL takes the value 10 (which is equal to a 81\% to 90\% student's cohort attending distance learning in a given school). Thirdly, the larger the connectivity, systemic and material barriers for students' engagement with remote learning (plots in the last two columns of Figure \ref{figure3}), the lower the chances of a student to be into the SAR group, which is expected because of the inverse association between barriers and achievement. Overall, this analysis suggests how the pandemic powerfully shaped SAR probabilities in the LAC region.

\subsubsection{Personality trait variables}
\label{section513}

\begin{figure}[ht!]
\centering
\subfloat[\scriptsize{SAR1-curiosity}]{\includegraphics[width=0.33\textwidth]{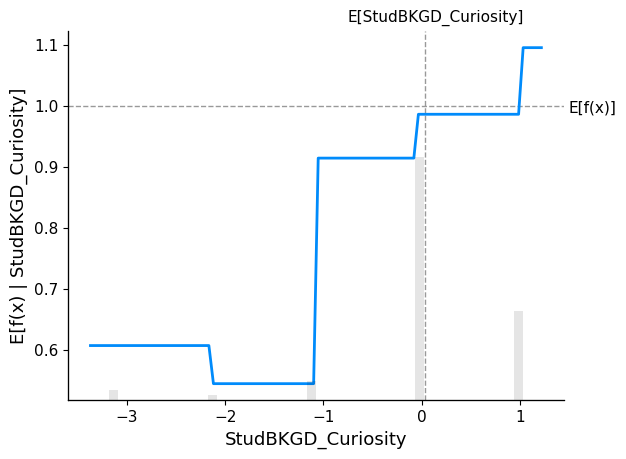}\label{figure4a}}
\subfloat[\scriptsize{SAR1-perseverance}]{\includegraphics[width=0.33\textwidth]{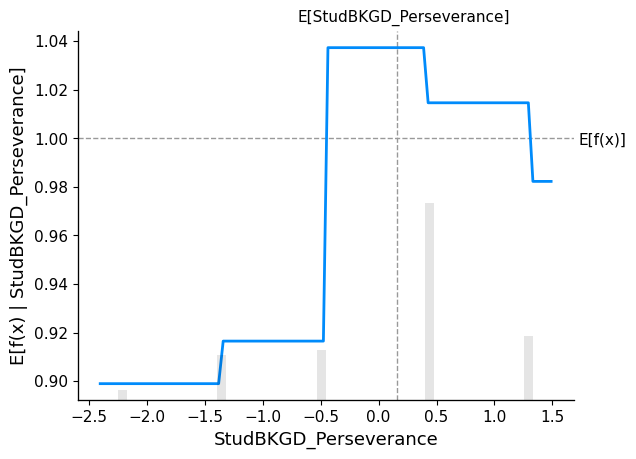}\label{figure4b}}
\subfloat[\scriptsize{SAR2-curiosity}]{\includegraphics[width=0.33\textwidth]{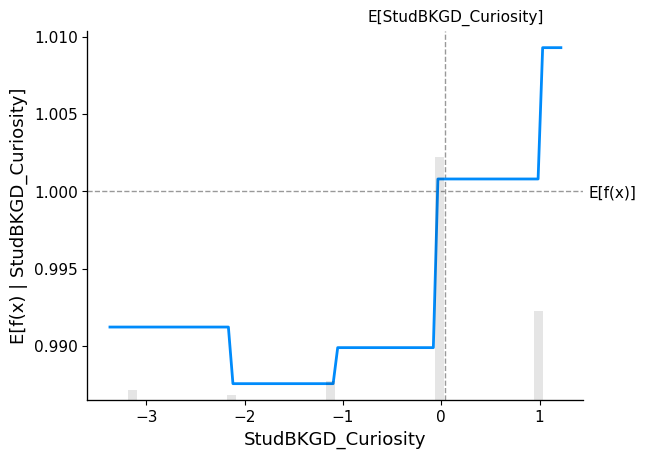}\label{figure4c}}
\\
\subfloat[\scriptsize{SAR2-perseverance}]{\includegraphics[width=0.33\textwidth]{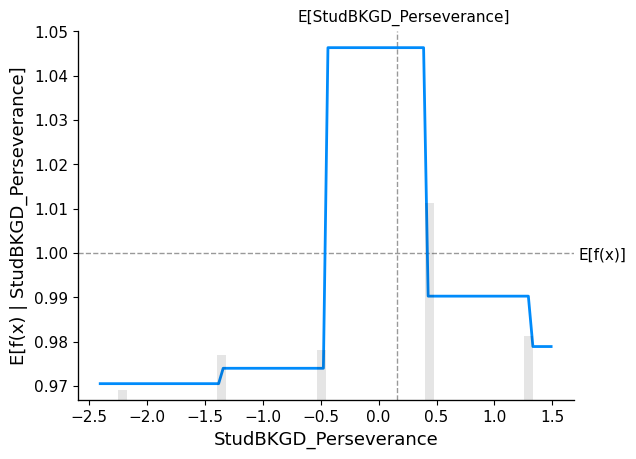}\label{figure4d}}
\subfloat[\scriptsize{SAR3-curiosity}]{\includegraphics[width=0.33\textwidth]{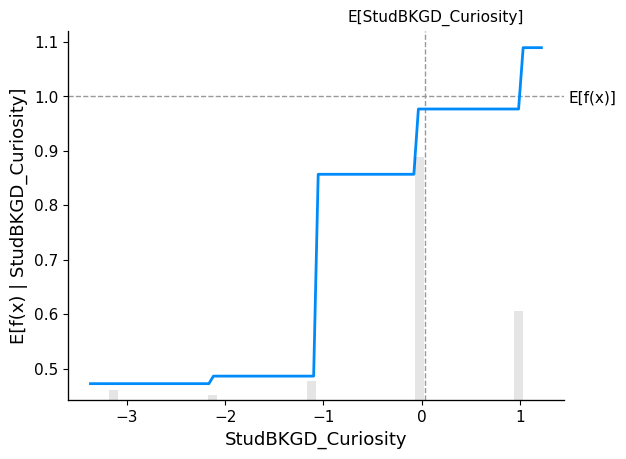}\label{figure4e}}
\subfloat[\scriptsize{SAR3-perseverance}]{\includegraphics[width=0.33\textwidth]{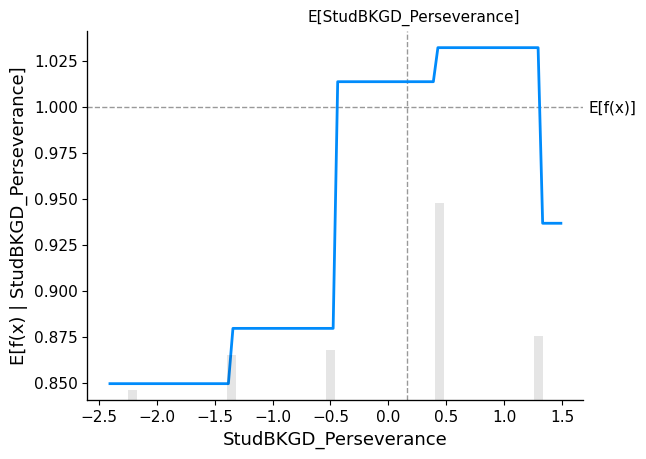}\label{figure4f}}
\\
\subfloat[\scriptsize{SAR4-curiosity}]{\includegraphics[width=0.33\textwidth]{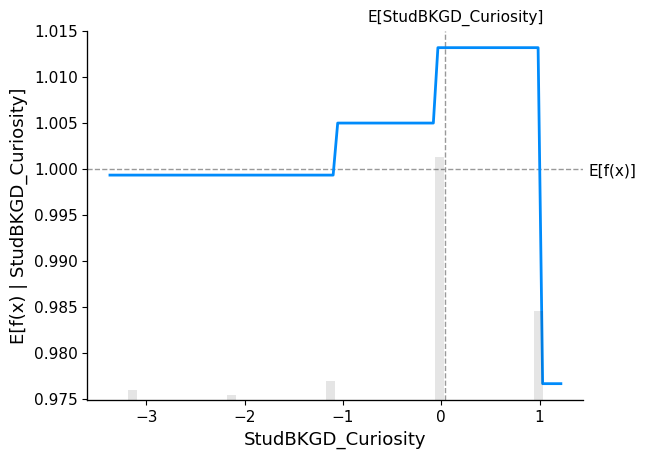}\label{figure4g}}
\subfloat[\scriptsize{SAR4-perseverance}]{\includegraphics[width=0.33\textwidth]{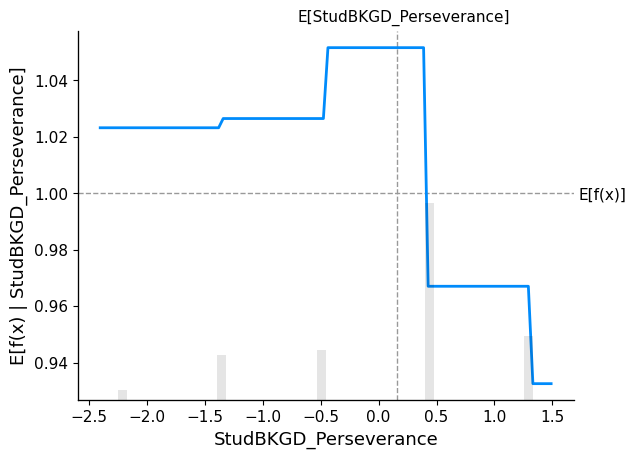}\label{figure4h}}
\caption{Partial dependence plots for soft skills (curiosity and perseverance)}
\label{figure4}
\end{figure}

Personality traits have been found to be protective factors for resilience against adversity. \cite{tudor17}'s review identifies positive affect, self-esteem, extraversion, social support, and optimism. Here, we examine two traits: students' curiosity and perseverance, which are measured as indices, with higher values showcasing a students superior extent of these personality characteristics. Estimated odd ratios across these two variables scales are shown Figure \ref{figure4}, confirming that personality traits matter for SAR in the LAC region.

\medskip
\begin{center}
  [Figure \ref{figure4} here]
\end{center}
\medskip

With regards to curiosity, for students with positive values for its index (=1), $\widehat{\text{OR}}$ are over one (=1.10). In other words, students with the highest extent of curiosity have a 10\% larger probability of being academic resilient compared to those students with an index equal to zero; and this gap widens for those students with low curiosity (negative values for the index) as their $\widehat{\text{OR}}$ are around 0.5-0.6 (for SAR1 and SAR3; Figures \ref{figure4a} and \ref{figure4e}). For SAR2 and SAR4 the impact of curiosity on predicting SAR is smaller (around 1\%), though it is still increasing across the scale (Figures \ref{figure4c} and Figures \ref{figure4g}). Likewise, the higher the perseverance a student possesses, the higher his$/$her chances of being academic resilient; in particular, when the perseverance index takes the value of +0.5, $\widehat{\text{OR}} \approx 1.04$ (Figures \ref{figure4b}, \ref{figure4d} and \ref{figure4h}).


\subsection{Sub-sample estimations}
\label{section52}

\begin{figure}[ht!]
\centering
\subfloat[\footnotesize{Private-public (SAR1)}]{\includegraphics[width=0.49\textwidth]{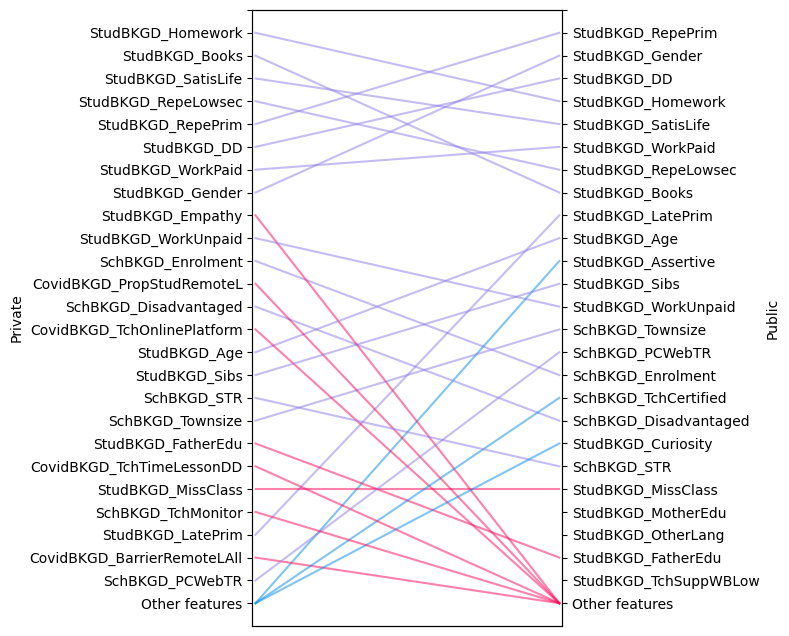}\label{figure5a}}
\subfloat[\footnotesize{Private-public (SAR2)}]{\includegraphics[width=0.49\textwidth]{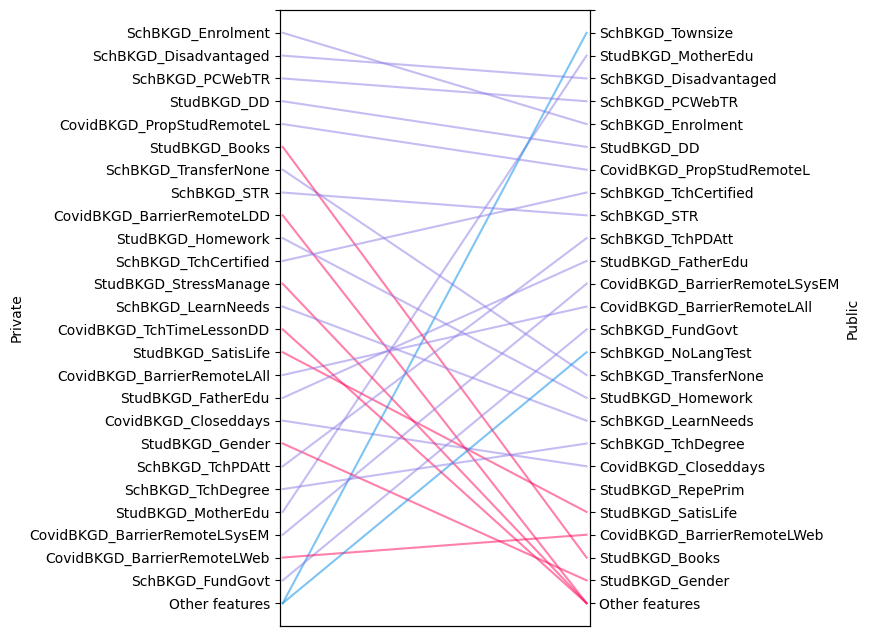}\label{figure5b}}
\\
\subfloat[\footnotesize{Urban-rural (SAR1)}]{\includegraphics[width=0.49\textwidth]{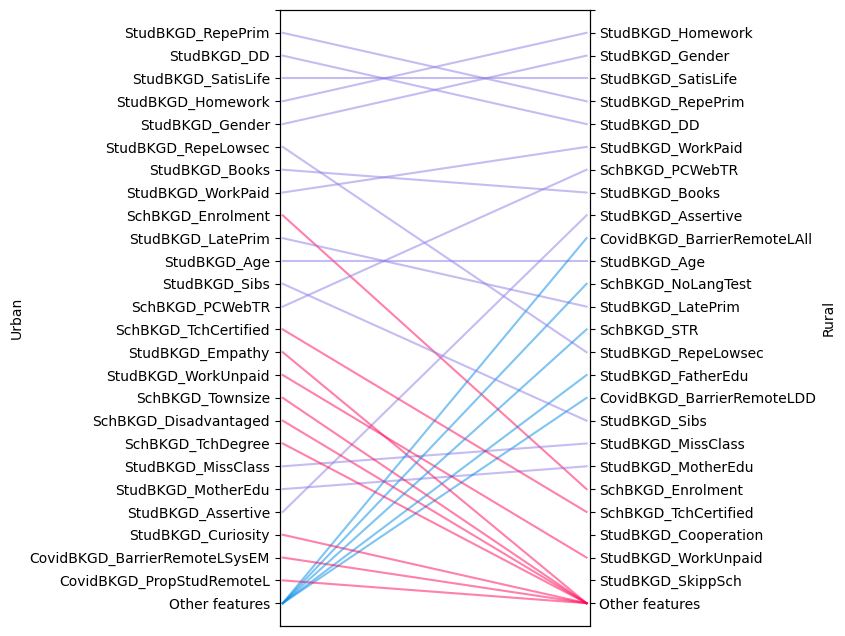}\label{figure5c}}
\subfloat[\footnotesize{Urban-rural (SAR2)}]{\includegraphics[width=0.49\textwidth]{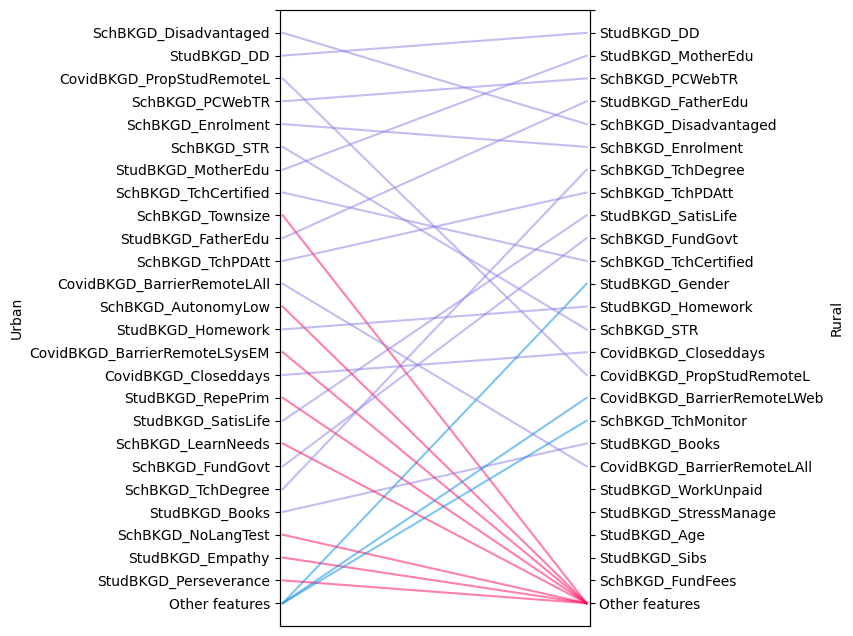}\label{figure5d}}
\caption{Sub-samples determinants comparison. Outcomes: SAR1 and SAR2}
\label{figure5}
\medskip
\begin{minipage}{0.99\textwidth}
\footnotesize{
Notes: (1) Country plots show the first 25 covariates in order of associations given by absolute values of average SHAP values for each covariate.
}
\end{minipage}
\end{figure}

Given the highly segregated nature of Latin American education systems (\citealp{acosta23,delprato15,murillo17,murillo23}), here we go deeper than the whole sample analysis by inspecting specific factors of SAR by school type  and by school location. Figure \ref{figure5} shows the relative global feature importance of the 25 overlapping features of the private, public, urban and rural schools models for the two main dependent variables: SAR1 and SAR2.\footnote{Ranked determinants by the same school sub-groups for the remaining two outcomes, SAR3 and SAR4, are included in Figure \ref{figureB1} of the Appendix.}

\medskip
\begin{center}
  [Figure \ref{figure5} here]
\end{center}
\medskip

To begin with, private-public SAR determinants comparison are shown in Figures \ref{figure5a} (SAR1) and \ref{figure5b} (SAR2), where private schools' estimates are shown on the left and public schools' estimates are displayed on the right. The Spearman's correlation coefficient of the private and public school models' covariates importance is 0.6485 (p-value = 0.0001) based on the SAR1 dependent variable, showing the significant positive correlation between the ranking of the overlapping features in private and public sub-samples; though, for SAR2, this correlation is much smaller and lacking statistical significance ($\rho$ = 0.32, p-value = 0.1185). A more in-depth comparison of determinants for these two sub-samples for SAR1 (Figure \ref{figure5a}) shows some degree of similarity, with minor rank changes among top-ranked covariates (with the exception of primary repetition and gender which move up in importance, being the top two determinants in public schools), but with some lower-ranked student's and school's characteristics significantly raising their relevance in public schools (e.g., StudBKGD\_Assertive, SchBKGD\_PCWebTR, SchBKGD\_TchCertified). For SAR2 (Figure \ref{figure5b}), with bigger variation on determinants (lower $\rho$), various covariates are substantially more prominent when defining the chances of a students being academic resilient, such as: parental education (especially mother's education), teachers PD attendance, barriers for remote learning during COVID and the prevalence of minority language students.

Estimated determinant's ranking of the GBT models by school location are displayed in Figures \ref{figure5c} and \ref{figure5d}. It is worth mentioning that for the SAR1 model ($\rho$ = 0.7915, p-value = 0.000, for features ranking in the two sub-samples), repetition and stock of digital devices for learning are slightly more important in urban than in rural schools whereas, on the oppositive direction, gender and student's engagement with homework are more relevant in rural schools settings (Figure \ref{figure5c}). Similarly, key school inputs (namely: student$/$teacher ratio of connected PC and the STR) are highly ranked for SAR chances in rural schools, as well as the extent of COVID-19 barriers for remote learning, father education (as mother education is similarly ranked in the two sub-samples) and student's assertiveness. When modelling SAR's likelihood controlling for households' and schools' average socio-economic status for SAR2 (Figure \ref{figure5d}), with a correlation coefficient of urban and rural school models' covariates importance of 0.6115 (p-value = 0.0015), urban-rural determinants gaps relocate towards family inputs related to achievement (and, in turn, for academic resilience) in the case of rural students compared to urban students. Out of the top five ranked characteristics, for instance, only one belongs into the family domain in urban schools but three in rural schools (mother and father education and number of digital devices). Besides that, the chances of rural school's students to be academically resilient seem to be relatively more powerfully shaped by teaching quality (SchBKGD\_TchDegree and SchBKGD\_TchPDAtt), instead of raw measures of teaching stock (i.e., STR) in the case of students attending urban schools.


\subsection{Experiment–local explanations. Students at extremes of SHAP index distribution}
\label{section53}

\begin{figure}[hp!]
\centering
\subfloat[\footnotesize{SAR1 (profile: high)}]{\includegraphics[width=0.49\textwidth]{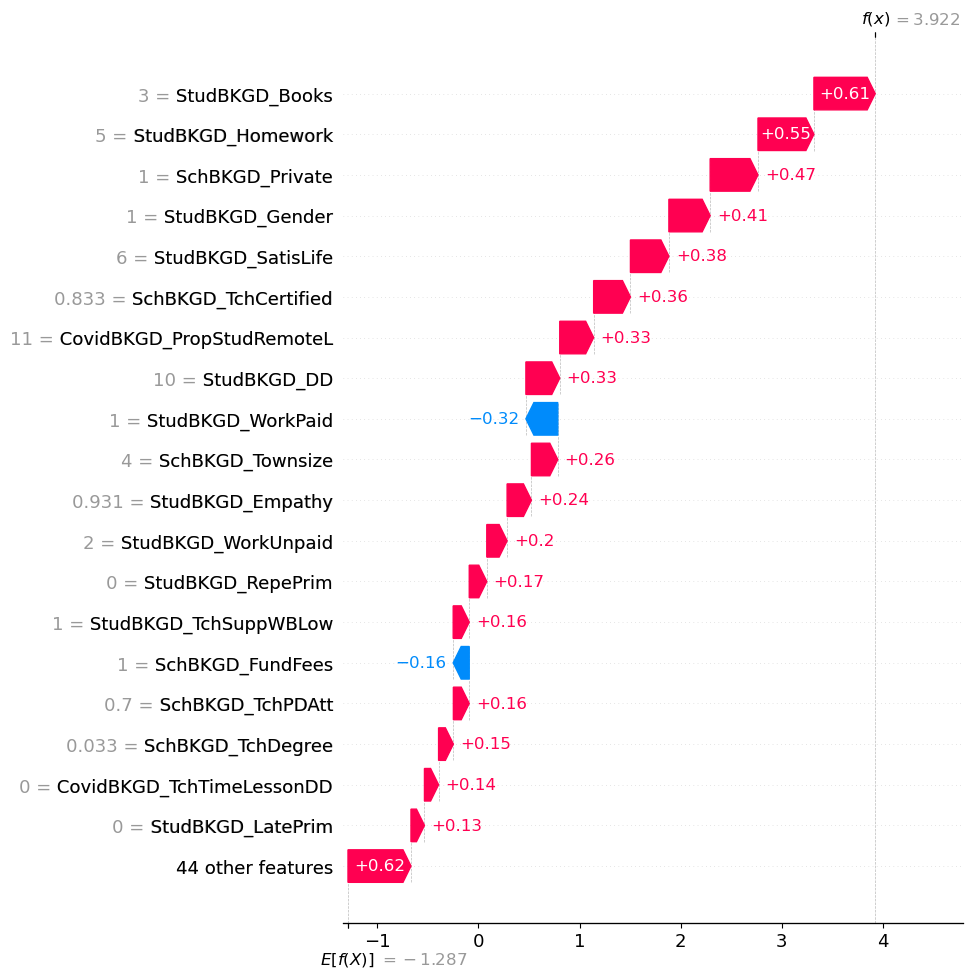}\label{figure6a}}
\subfloat[\footnotesize{SAR1 (profile: low)}]{\includegraphics[width=0.49\textwidth]{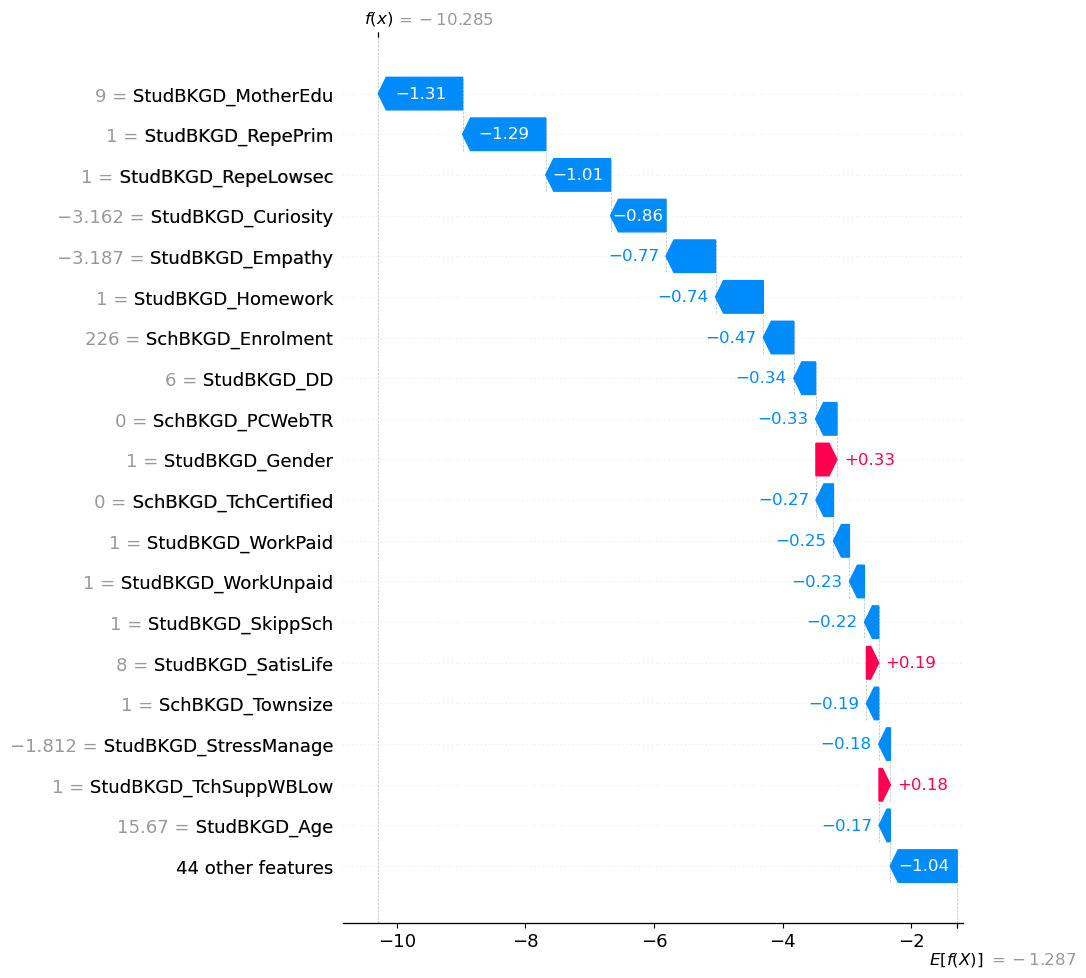}\label{figure6b}}
\\
\subfloat[\footnotesize{SAR2 (profile: high)}]{\includegraphics[width=0.49\textwidth]{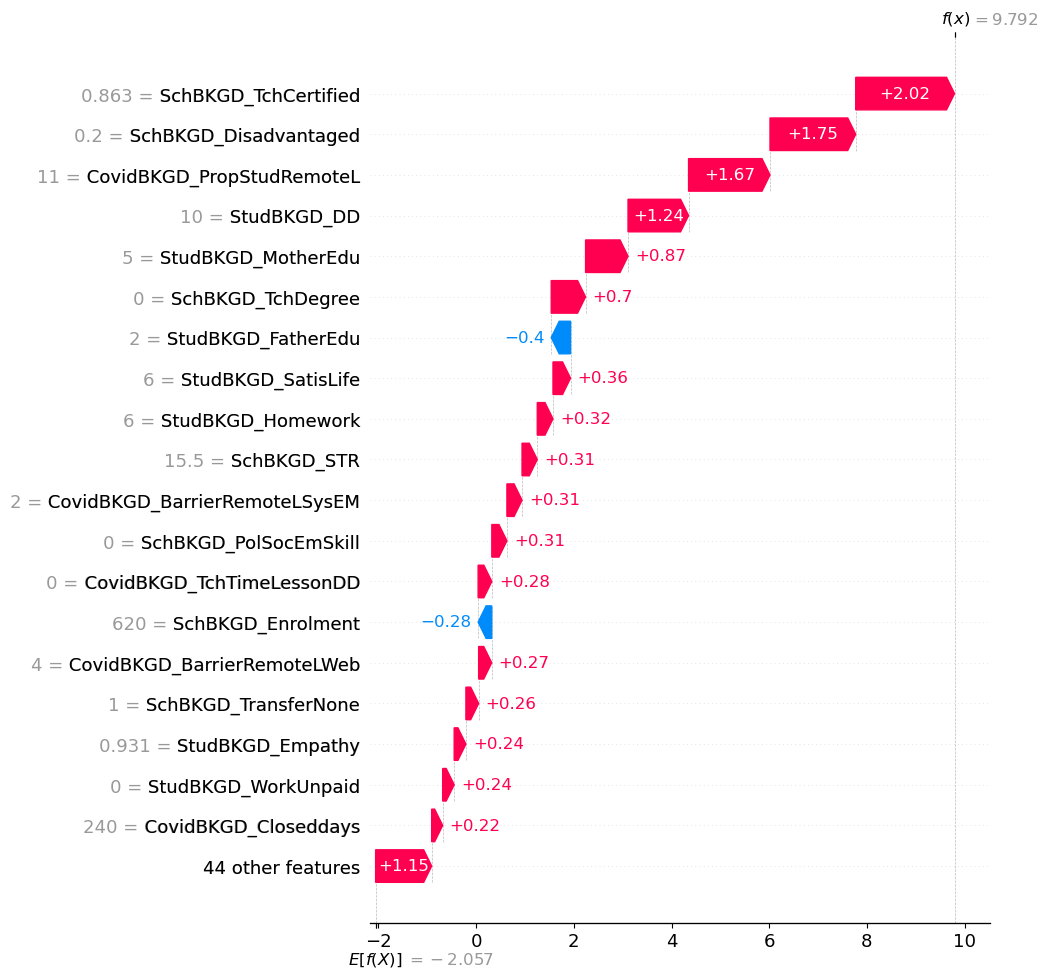}\label{figure6c}}
\subfloat[\footnotesize{SAR2 (profile: low)}]{\includegraphics[width=0.49\textwidth]{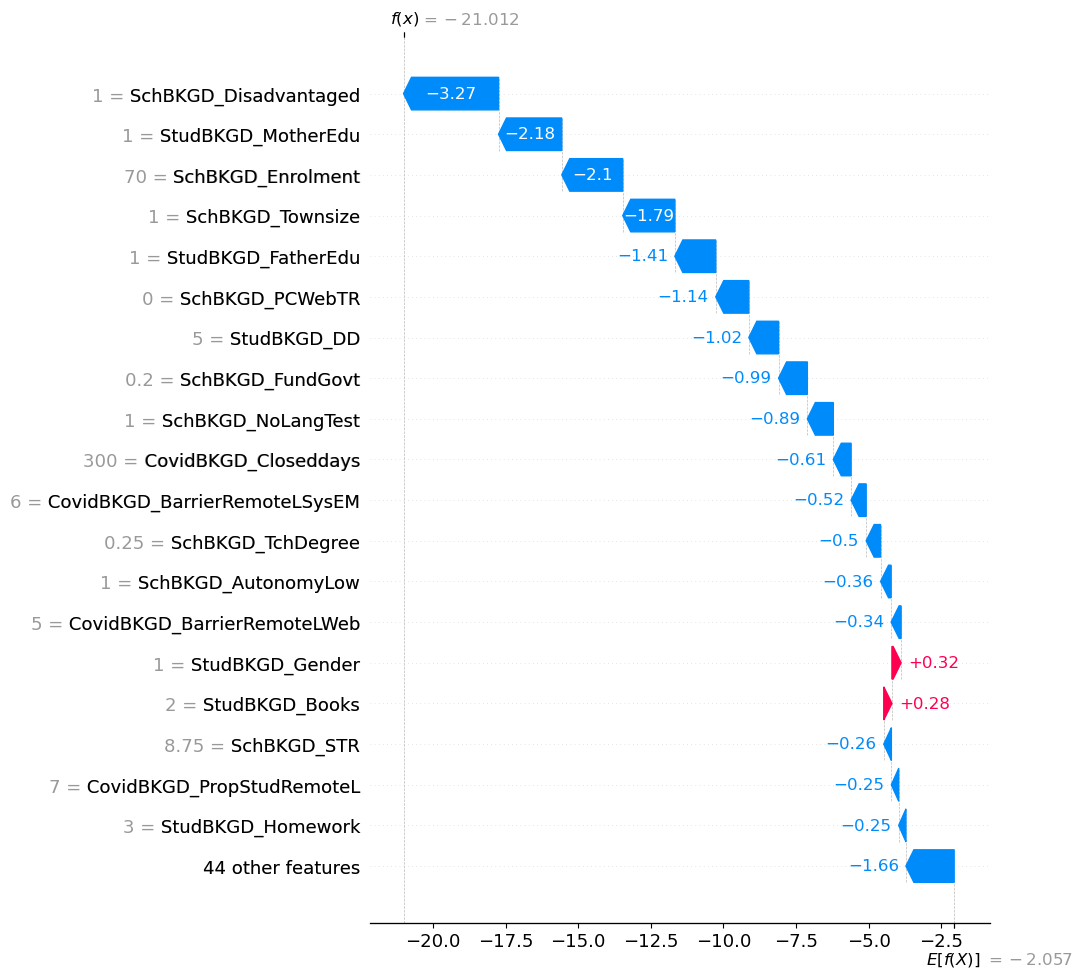}\label{figure6d}}
\caption{Covariates individual SHAP values for students with the lowest and highest average SHAP values contributions (indicators: SAR1 and SAR2)}
\label{figure6}
\medskip
\begin{minipage}{0.99\textwidth}
\footnotesize{
Notes: (1) See Table A1 for variables' acronyms, definitions and associated labels to values (in grey, left to covariate name).
}
\end{minipage}
\end{figure}

In this section, we focus on local explanations of SHAP values which are given by specific observations. For this, we proceed in three steps: (i) we calculate the mean sample contribution of SHAP values for each student and sort them out; (ii) we select students with the minimum and maximum contributions so that to profile the impact of covariates at extremes; and (iii) once student pairs are identified, we plot top determinants for each student, with plots showing the specific value each covariate takes. SAR1 and SAR2 results\footnote{In Figure \ref{figureB2}, the same local analysis is displayed for the SAR3 and SAR4 indicators.} for this analysis is included in Figure \ref{figure6}.

\medskip
\begin{center}
  [Figure \ref{figure6} here]
\end{center}
\medskip

The local explanation exercise for the SAR1 outcome where we compare the profile of two students, one with highest chance ($\Phi^{\text{max}}_{i}$ = +5.21) and the other with lowest chance ($\Phi^{\text{min}}_{i}$ = -8.99) to be academic resilient are shown in top two plots of Figure \ref{figure6}. A student with the highest probability of being academic resilient attends a private school, has around 26-100 (=3) books, is engaged fully in homework (=5, spending 3 to 4 hours per week), is male, his satisfaction with life is above average (=6), has 10 digital devices for learning, he works just one day per week, he is a not a repeater and show positive values in terms of empathy (see: Figure \ref{figure6a}). Also, this student attends a (private) school with high rates of certified teachers (83.3\%) and teachers attending PD (=70\%), and nearly all (=11, over 90\%) students attending remote learning educational activities. On the contrary, a student who is most likely to fall into the non-resilient group (Figure \ref{figure6b}), is a repeater (at primary and at lower secondary), has adverse personality traits (all indices for curiosity, empathy, and stress management are negative), does little homework (=1, $\le$ 30 minutes), and has only 6 digital devices.

We carry out the same exercise for the model with SAR2 as dependent variable, which is shown in Figure \ref{figure6c} (the chosen student with the largest probability of falling into the SAR group; $\Phi^{\text{max}}_{i}$ = +11.84) and in Figure \ref{figure6d} (a student with the smallest chances of falling into the SAR group; $\Phi^{\text{min}}_{i}$ = -18.95). Consider the first ten ranked covariates in terms of impacts. A higher prevalence of certified teachers (86\%), only 20\% disadvantaged students in school, nearly the total cohort studying remotely during the pandemic, having a larger number of digital devices and fairly educated mother (=5, upper secondary), above average satisfaction with life and regularly doing homework, and more teachers available per student (STR = 15.5), are all features linked to achieve academic resilience based on SAR2. Though, in the opposite way, a student with non-existent chances of reaching the SAR group attends a hugely disadvantaged school, has low-educated parents (StudBKGD\_MotherEdu and StudBKGD\_FatherEdu =1, uncompleted primary) and a lower (=5) number of digital devices, comes from a small school with enrolment less than 100 where all students have a minority language background and little government funding (20\%), and without PC connected to internet. Note, too, the additional gap on the number of days a school remained closed during the pandemic: 240 days for a SAR student and 300 days for a NSAR student.


\section{Discussion}
\label{section6}

With the backdrop of having the longest period of pandemic-related school closures ($\approx$ 270 days) and poor connectivity for remote learning (\citealp{cepal22,munguia23}), Latin America's inequality-driven learning crisis has been intensified by the COVID-19 pandemic through its disproportionate larger impact among students from poorest households (\citealp{lustig23}). Due to the pandemic, families' income shocks have led to an increasing poverty in the LAC region, with estimates indicating a raise of poverty rates between 23\%-39\%, which compounded established drivers of learning poverty in the region. Students in the bottom 10\% of the income distribution lost nearly four times more years of education than students in the top 10\% of the income distribution (\citealp{bracco25}). Still, within this gloomy post-pandemic education scenario, some disadvantaged students were able to reach reasonable competency levels in different subjects. An analysis of this group of academic resilient students, whose performance is high against the odds given their poorer background (\citealp{rudd21,xenofontos23}), is a potential policy catalyst for learning recovery as knowing which are the pull up factors for students to be more likely to fall into the academic resilient group (versus those poorer students who are not academic resilient) can be used as a baseline to aid those poorer students at the bottom of the learning distribution.

Underpinned by this hypothesis, and  relying on the latest PISA wave for 2022 for nine LAC countries, we delve into the determinants of students academic resilience based on four alternative definitions which all use as working sample the 40\% poorer students of the LAC region. We do so by making use of recent advancements on interpretable ML modelling (\citealp{lundberg17}), i.e., the SHAP method, to determine academic resilience's main drivers. We obtain a global version of estimates, both at the regional level and by sub-samples, as well as a local version of estimates. Ultimately, the paper seeks to assess equality of opportunity in education and its determinants for the LAC region as a policy tool to boost upward social mobility for the most deprived students.


\subsection{Main findings}
\label{section61}

On the one hand, from the global explanatory analysis (using the whole sample), leading findings can be summarised as follows. First, covariates with the largest impacts (regardless of the direction of associations) for the SAR1 indicator (which is based on the level 2 achievement threshold) are from to the student’s group. These are students' features such as stock of digital devices and books at home, gender, homework, life satisfaction, primary and lower secondary repetition, and engagement with paid work, with personal traits (assertiveness, curiosity and empathy) having smaller predictive power as the private school dummy. For the SAR2 indicator (a estimated probability on the top 40\%, net of SES effects at the student and school levels 2, plus controlling for school and country effects in a 3-level multilevel framework), more powerful predictors are from the school domain instead (e.g.,  rate of disadvantaged students at school, school size, the ratio of PC connected to the internet and STR). Second, the alternative indicators (i.e., SAR3 and SAR4) which place a further restriction (either focusing on most unequal schools or excluding the most efficient school based on random effects), the array of most powerful determinants shift towards the complementary domains.

Third, as regards to COVID-19 background and personality variables, estimated odd-ratios $\widehat{\text{OR}}$  indicate a strong influence on the probability of students' academic resilience for pandemic covariates and some association for soft skills such as the degree of a student's curiosity and perseverance. For example, there is an inverse associations of SAR's likelihood and length of schools closure (e.g., an increment of the number of days a school remained close from 170 to 400 leads to a reduction on the OR of academic resilience between 10\%-25\%); and, equally, an increment on the barriers for engagement in remote learning is negatively related with SAR probabilities. Furthermore, sub-sample results (private versus public schools, and urban versus rural schools) point towards some discrepancies across determinants. For instance, based on SAR1, student's assertiveness and rate of certified teachers and PC connected to internet are more prominent for students attending public schools --as well as gender and repetition-- and mother education, PD rate of teachers and remote learning barriers in the case of the SAR2 model. In the more disadvantaged rural schools,  gender and student's engagement with homework are more relevant than in urban schools, as well as vital school inputs such as STR and COVID-19 barriers for remote learning (SAR1 model).

On the other hand, local estimations for students placed at the extremes of SHAP index distribution reveal how closely related are students and school inputs with SAR. The personal characteristics in the profile of a student with the highest chance of being academic resilient for the SAR1 model is a student who attends a private school who has not repeated, has a fair number of books and digital devices at home, hardly working for paid, and who is engaged with homework, attends a school that is well-resourced in terms of teaching quality (high rates of certified teachers of 83.3\% and teachers attending PD equals to 70\%). Though, reversely, the profile of a student with the minimal chance of being academic resilient is a repeater with negative personality traits (in terms of empathy, and stress management), and who does very little homework and has low-educated parents who did not finish primary (for the SAR2 model).


\subsection{Limitations and future research}
\label{section62}

The analysis of the paper has some limitations. Estimates denote associations but not causal claims because of the cross-section nature of the PISA data employed and, as a result, estimates reported should be considered with caution. Even though we carried out some sub-sample analysis on the drivers of academic resilience, a first future line of investigation could be a more thorough analysis of the role of intersectionality  (combinations of student$/$family covariates with school covariates) on the likelihood of a student being academic resilient. Another venue of future research is to look how different conceptualisations of academic resilience and their determinants are aligned with specific educational countries policies targeting the most disadvantaged students.


\section{Conclusions}
\label{section7}

On the back of the learning crisis of the Latin American and the Caribbean (LAC) region, this article offered a detailed analysis of factors underpinning students who are academically resilient; that is, students whose performance is high against the odds given their poorer background.  From a policy perspective, the paper's analysis can be framed as learning recovery mechanism for the LAC region because, knowing which are the pull up factors for students to be more likely to fall into the academic resilient group (versus those poorer students who are not academic resilient), can be employed as a baseline to aid those poorer students at the bottom of the learning distribution. Specifically, we used the latest PISA wave (year: 2022) for nine LAC countries to identify leading factors behind students academic resilience using four definitions using recent advancements on explainable machine learning methods. The paper contributed to the literature in different ways. First, we relied on four different versions of the student academic resilient (SAR) indicator (the dependent variable); and, too, we introduced two novel SAR indicators. Second, as far as we are aware, this is the first study investigating the determinants of students academic resilience for the LAC region. Third, the empirical analysis also focuses on drivers by education sub-systems. Finally,  the paper relied on novel explainable artificial intelligence (XAI) methods (Shapley values), adding to scarce body of research on SAR based on learning surveys for other regions (e.g., \citealp{agasisti18,cheung24}).


\newpage
\printendnotes


\clearpage
\setlength\bibsep{0pt}
\bibliographystyle{elsarticle-harv}
\bibliography{references}

\begin{thebibliography}{69}
\expandafter\ifx\csname natexlab\endcsname\relax\def\natexlab#1{#1}\fi
\providecommand{\url}[1]{\texttt{#1}}
\providecommand{\href}[2]{#2}
\providecommand{\path}[1]{#1}
\providecommand{\DOIprefix}{doi:}
\providecommand{\ArXivprefix}{arXiv:}
\providecommand{\URLprefix}{URL: }
\providecommand{\Pubmedprefix}{pmid:}
\providecommand{\doi}[1]{\href{http://dx.doi.org/#1}{\path{#1}}}
\providecommand{\Pubmed}[1]{\href{pmid:#1}{\path{#1}}}
\providecommand{\bibinfo}[2]{#2}
\ifx\xfnm\relax \def\xfnm[#1]{\unskip,\space#1}\fi
\bibitem[{Acosta(2023)}]{acosta23}
\bibinfo{author}{Acosta, F.}, \bibinfo{year}{2023}.
\newblock \bibinfo{title}{Between expansion and segmentation: revisiting old
  and new disparities in secondary education in latin america}.
\newblock \bibinfo{journal}{International Journal of Inclusive Education} ,
  \bibinfo{pages}{1--18}.
\bibitem[{Agasisti et~al.(2018)Agasisti, Avvisati, Borgonovi and
  Longobardi}]{agasisti18}
\bibinfo{author}{Agasisti, T.}, \bibinfo{author}{Avvisati, F.},
  \bibinfo{author}{Borgonovi, F.}, \bibinfo{author}{Longobardi, S.},
  \bibinfo{year}{2018}.
\newblock \bibinfo{title}{Academic resilience: What schools and countries do to
  help disadvantaged students succeed in pisa} .
\bibitem[{Agasisti et~al.(2021)Agasisti, Avvisati, Borgonovi and
  Longobardi}]{agasisti21}
\bibinfo{author}{Agasisti, T.}, \bibinfo{author}{Avvisati, F.},
  \bibinfo{author}{Borgonovi, F.}, \bibinfo{author}{Longobardi, S.},
  \bibinfo{year}{2021}.
\newblock \bibinfo{title}{What school factors are associated with the success
  of socio-economically disadvantaged students? an empirical investigation
  using pisa data}.
\newblock \bibinfo{journal}{Social Indicators Research} \bibinfo{volume}{157},
  \bibinfo{pages}{749--781}.
\bibitem[{Agasisti et~al.(2017)Agasisti, Longobardi and Regoli}]{agasisti17}
\bibinfo{author}{Agasisti, T.}, \bibinfo{author}{Longobardi, S.},
  \bibinfo{author}{Regoli, A.}, \bibinfo{year}{2017}.
\newblock \bibinfo{title}{A cross-country panel approach to exploring the
  determinants of educational equity through pisa data}.
\newblock \bibinfo{journal}{Quality \& Quantity} \bibinfo{volume}{51},
  \bibinfo{pages}{1243--1260}.
\bibitem[{Allan et~al.(2014)Allan, McKenna and Dominey}]{allan14}
\bibinfo{author}{Allan, J.F.}, \bibinfo{author}{McKenna, J.},
  \bibinfo{author}{Dominey, S.}, \bibinfo{year}{2014}.
\newblock \bibinfo{title}{Degrees of resilience: profiling psychological
  resilience and prospective academic achievement in university inductees}.
\newblock \bibinfo{journal}{British Journal of Guidance \& Counselling}
  \bibinfo{volume}{42}, \bibinfo{pages}{9--25}.
\bibitem[{Arias~Ortiz et~al.(2023)Arias~Ortiz, Bos, Giambruno and
  Zoido}]{arias23}
\bibinfo{author}{Arias~Ortiz, E.}, \bibinfo{author}{Bos, M.},
  \bibinfo{author}{Giambruno, C.}, \bibinfo{author}{Zoido, P.},
  \bibinfo{year}{2023}.
\newblock \bibinfo{title}{Latin America and the Caribbean in PISA 2022: How
  Many Students are Low Performers?}
\newblock \bibinfo{type}{Technical Report}. Inter-American Development Bank.
\bibitem[{Arias~Ortiz et~al.(2024)Arias~Ortiz, Due{\~n}as, Giambruno and
  L{\'o}pez}]{arias24}
\bibinfo{author}{Arias~Ortiz, E.}, \bibinfo{author}{Due{\~n}as, X.},
  \bibinfo{author}{Giambruno, C.}, \bibinfo{author}{L{\'o}pez, {\'A}.},
  \bibinfo{year}{2024}.
\newblock \bibinfo{title}{The State of Education in Latin America and the
  Caribbean: Learning Assessments}.
\newblock \bibinfo{type}{Technical Report}. Inter-American Development Bank.
\bibitem[{Attanasio et~al.(2025)Attanasio, de~La~O, Ferreira, Ib{\'a}{\~n}ez
  and Messina}]{attanasio25}
\bibinfo{author}{Attanasio, O.}, \bibinfo{author}{de~La~O, A.L.},
  \bibinfo{author}{Ferreira, F.H.}, \bibinfo{author}{Ib{\'a}{\~n}ez, A.M.},
  \bibinfo{author}{Messina, J.}, \bibinfo{year}{2025}.
\newblock \bibinfo{title}{Inequality in latin america and the caribbean: a
  wide-ranging review}.
\bibitem[{Azevedo et~al.(2021)Azevedo, Hasan, Goldemberg, Geven and
  Iqbal}]{azevedo21}
\bibinfo{author}{Azevedo, J.P.}, \bibinfo{author}{Hasan, A.},
  \bibinfo{author}{Goldemberg, D.}, \bibinfo{author}{Geven, K.},
  \bibinfo{author}{Iqbal, S.A.}, \bibinfo{year}{2021}.
\newblock \bibinfo{title}{Simulating the potential impacts of covid-19 school
  closures on schooling and learning outcomes: A set of global estimates}.
\newblock \bibinfo{journal}{The World Bank Research Observer}
  \bibinfo{volume}{36}, \bibinfo{pages}{1--40}.
\bibitem[{Betth{\"a}user et~al.(2023)Betth{\"a}user, Bach-Mortensen and
  Engzell}]{betthauser23}
\bibinfo{author}{Betth{\"a}user, B.A.}, \bibinfo{author}{Bach-Mortensen, A.M.},
  \bibinfo{author}{Engzell, P.}, \bibinfo{year}{2023}.
\newblock \bibinfo{title}{A systematic review and meta-analysis of the evidence
  on learning during the covid-19 pandemic}.
\newblock \bibinfo{journal}{Nature human behaviour} \bibinfo{volume}{7},
  \bibinfo{pages}{375--385}.
\bibitem[{Bracco et~al.(2025)Bracco, Ciaschi, Gasparini, Marchionni and
  Neidh{\"o}fer}]{bracco25}
\bibinfo{author}{Bracco, J.}, \bibinfo{author}{Ciaschi, M.},
  \bibinfo{author}{Gasparini, L.}, \bibinfo{author}{Marchionni, M.},
  \bibinfo{author}{Neidh{\"o}fer, G.}, \bibinfo{year}{2025}.
\newblock \bibinfo{title}{The impact of covid-19 on education in latin america:
  Long-run implications for poverty and inequality}.
\newblock \bibinfo{journal}{Review of Income and Wealth} \bibinfo{volume}{71},
  \bibinfo{pages}{e12687}.
\bibitem[{Brunori et~al.(2024)Brunori, Ferreira and Neidh{\"o}fer}]{brunori24}
\bibinfo{author}{Brunori, P.}, \bibinfo{author}{Ferreira, F.H.},
  \bibinfo{author}{Neidh{\"o}fer, G.}, \bibinfo{year}{2024}.
\newblock \bibinfo{title}{Inequality of opportunity and intergenerational
  persistence in Latin America}.
\newblock \bibinfo{type}{Technical Report} \bibinfo{number}{17202}. IZA
  Discussion Papers.
\bibitem[{CEPAL(2022)}]{cepal22}
\bibinfo{author}{CEPAL}, \bibinfo{year}{2022}.
\newblock \bibinfo{title}{Panorama social de américa latina y el caribe 2022:
  la transformación de la educación como base para el desarrollo sostenible}.
\bibitem[{Chen et~al.(2023)Chen, Covert, Lundberg and Lee}]{chen23}
\bibinfo{author}{Chen, H.}, \bibinfo{author}{Covert, I.C.},
  \bibinfo{author}{Lundberg, S.M.}, \bibinfo{author}{Lee, S.I.},
  \bibinfo{year}{2023}.
\newblock \bibinfo{title}{Algorithms to estimate shapley value feature
  attributions}.
\newblock \bibinfo{journal}{Nature Machine Intelligence} \bibinfo{volume}{5},
  \bibinfo{pages}{590--601}.
\bibitem[{Chen and Guestrin(2016)}]{chen16}
\bibinfo{author}{Chen, T.}, \bibinfo{author}{Guestrin, C.},
  \bibinfo{year}{2016}.
\newblock \bibinfo{title}{Xgboost: A scalable tree boosting system}, in:
  \bibinfo{booktitle}{Proceedings of the 22nd acm sigkdd international
  conference on knowledge discovery and data mining}, pp.
  \bibinfo{pages}{785--794}.
\bibitem[{Cheung(2017)}]{cheung17}
\bibinfo{author}{Cheung, K.c.}, \bibinfo{year}{2017}.
\newblock \bibinfo{title}{The effects of resilience in learning variables on
  mathematical literacy performance: A study of learning characteristics of the
  academic resilient and advantaged low achievers in shanghai, singapore, hong
  kong, taiwan and korea}.
\newblock \bibinfo{journal}{Educational Psychology} \bibinfo{volume}{37},
  \bibinfo{pages}{965--982}.
\bibitem[{Cheung et~al.(2024)Cheung, Sit, Zheng, Lam, Mak and Ieong}]{cheung24}
\bibinfo{author}{Cheung, K.c.}, \bibinfo{author}{Sit, P.s.},
  \bibinfo{author}{Zheng, J.q.}, \bibinfo{author}{Lam, C.c.},
  \bibinfo{author}{Mak, S.k.}, \bibinfo{author}{Ieong, M.k.},
  \bibinfo{year}{2024}.
\newblock \bibinfo{title}{A machine-learning model of academic resilience in
  the times of the covid-19 pandemic: Evidence drawn from 79
  countries/economies in the pisa 2022 mathematics study}.
\newblock \bibinfo{journal}{British Journal of Educational Psychology}
  \bibinfo{volume}{94}, \bibinfo{pages}{1224--1244}.
\bibitem[{Choi and Sung(2024)}]{choi24}
\bibinfo{author}{Choi, Y.}, \bibinfo{author}{Sung, J.}, \bibinfo{year}{2024}.
\newblock \bibinfo{title}{Do key predictors of academic resilience differ
  across cultures? evidence from korea and the us}.
\newblock \bibinfo{journal}{Youth \& Society} \bibinfo{volume}{56},
  \bibinfo{pages}{1237--1262}.
\bibitem[{Clavel et~al.(2022)Clavel, Garc{\'\i}a~Crespo and Sanz
  San~Miguel}]{clavel22}
\bibinfo{author}{Clavel, J.G.}, \bibinfo{author}{Garc{\'\i}a~Crespo, F.J.},
  \bibinfo{author}{Sanz San~Miguel, L.}, \bibinfo{year}{2022}.
\newblock \bibinfo{title}{Rising above their circumstances: what makes some
  disadvantaged east and south-east asian students perform far better in
  science than their background predicts?}
\newblock \bibinfo{journal}{Asia Pacific Journal of Education}
  \bibinfo{volume}{42}, \bibinfo{pages}{714--729}.
\bibitem[{Delprato et~al.(2015)Delprato, K{\"o}seleci and
  Antequera}]{delprato15}
\bibinfo{author}{Delprato, M.}, \bibinfo{author}{K{\"o}seleci, N.},
  \bibinfo{author}{Antequera, G.}, \bibinfo{year}{2015}.
\newblock \bibinfo{title}{Educaci{\'o}n para todos en am{\'e}rica latina:
  Evoluci{\'o}n del impacto de la desigualdad escolar en los resultados
  educativos}.
\newblock \bibinfo{journal}{Revista latinoamericana de educaci{\'o}n comparada}
  \bibinfo{volume}{6}, \bibinfo{pages}{45--75}.
\bibitem[{Erberber et~al.(2015)Erberber, Stephens, Mamedova, Ferguson and
  Kroeger}]{erberber15}
\bibinfo{author}{Erberber, E.}, \bibinfo{author}{Stephens, M.},
  \bibinfo{author}{Mamedova, S.}, \bibinfo{author}{Ferguson, S.},
  \bibinfo{author}{Kroeger, T.}, \bibinfo{year}{2015}.
\newblock \bibinfo{title}{Socioeconomically disadvantaged students who are
  academically successful: Examining academic resilience cross-nationally.
  policy brief no. 5.}
\newblock \bibinfo{journal}{International Association for the Evaluation of
  Educational Achievement} .
\bibitem[{Ersozlu et~al.(2024)Ersozlu, Taheri and Koch}]{ersozlu24}
\bibinfo{author}{Ersozlu, Z.}, \bibinfo{author}{Taheri, S.},
  \bibinfo{author}{Koch, I.}, \bibinfo{year}{2024}.
\newblock \bibinfo{title}{A review of machine learning methods used for
  educational data}.
\newblock \bibinfo{journal}{Education and Information Technologies} ,
  \bibinfo{pages}{1--21}.
\bibitem[{Escalante~Mateos et~al.(2020)Escalante~Mateos, Fern{\'a}ndez-Zabala,
  Go{\~n}i~Palacios and Izar-de-la-Fuente D{\'\i}az-de Cerio}]{escalante20}
\bibinfo{author}{Escalante~Mateos, N.}, \bibinfo{author}{Fern{\'a}ndez-Zabala,
  A.}, \bibinfo{author}{Go{\~n}i~Palacios, E.},
  \bibinfo{author}{Izar-de-la-Fuente D{\'\i}az-de Cerio, I.},
  \bibinfo{year}{2020}.
\newblock \bibinfo{title}{School climate and perceived academic performance:
  Direct or resilience-mediated relationship?}
\newblock \bibinfo{journal}{Sustainability} \bibinfo{volume}{13},
  \bibinfo{pages}{68}.
\bibitem[{Fern{\'a}ndez et~al.(2024)Fern{\'a}ndez, Pag{\'e}s, Szekely and
  Acevedo}]{fernandez24}
\bibinfo{author}{Fern{\'a}ndez, R.}, \bibinfo{author}{Pag{\'e}s, C.},
  \bibinfo{author}{Szekely, M.}, \bibinfo{author}{Acevedo, I.},
  \bibinfo{year}{2024}.
\newblock \bibinfo{title}{Education inequalities in Latin America and the
  Caribbean}.
\newblock \bibinfo{type}{Technical Report}. National Bureau of Economic
  Research.
\bibitem[{Frisby et~al.(2020)Frisby, Hosek and Beck}]{frisby20}
\bibinfo{author}{Frisby, B.N.}, \bibinfo{author}{Hosek, A.M.},
  \bibinfo{author}{Beck, A.C.}, \bibinfo{year}{2020}.
\newblock \bibinfo{title}{The role of classroom relationships as sources of
  academic resilience and hope}.
\newblock \bibinfo{journal}{Communication Quarterly} \bibinfo{volume}{68},
  \bibinfo{pages}{289--305}.
\bibitem[{Gabrielli et~al.(2022)Gabrielli, Longobardi and
  Strozza}]{gabrielli22}
\bibinfo{author}{Gabrielli, G.}, \bibinfo{author}{Longobardi, S.},
  \bibinfo{author}{Strozza, S.}, \bibinfo{year}{2022}.
\newblock \bibinfo{title}{The academic resilience of native and
  immigrant-origin students in selected european countries}.
\newblock \bibinfo{journal}{Journal of Ethnic and Migration Studies}
  \bibinfo{volume}{48}, \bibinfo{pages}{2347--2368}.
\bibitem[{Garc{\'\i}a-Crespo et~al.(2021)Garc{\'\i}a-Crespo,
  Fern{\'a}ndez-Alonso and Mu{\~n}iz}]{garcia21}
\bibinfo{author}{Garc{\'\i}a-Crespo, F.J.},
  \bibinfo{author}{Fern{\'a}ndez-Alonso, R.}, \bibinfo{author}{Mu{\~n}iz, J.},
  \bibinfo{year}{2021}.
\newblock \bibinfo{title}{Academic resilience in european countries: The role
  of teachers, families, and student profiles}.
\newblock \bibinfo{journal}{Plos one} \bibinfo{volume}{16},
  \bibinfo{pages}{e0253409}.
\bibitem[{Garc{\'\i}a~Crespo et~al.(2019)Garc{\'\i}a~Crespo, Gali{\'a}n,
  Fern{\'a}ndez~Alonso, Mu{\~n}iz et~al.}]{garcia19}
\bibinfo{author}{Garc{\'\i}a~Crespo, F.J.}, \bibinfo{author}{Gali{\'a}n, B.},
  \bibinfo{author}{Fern{\'a}ndez~Alonso, R.}, \bibinfo{author}{Mu{\~n}iz, J.},
  et~al., \bibinfo{year}{2019}.
\newblock \bibinfo{title}{Resiliencia educativa en comprensi{\'o}n lectora:
  factores determinantes en pirls-europa}.
\newblock \bibinfo{journal}{Revista De Educaci{\'o}n, 384} .
\bibitem[{Garc{\'\i}a~Crespo et~al.(2022)Garc{\'\i}a~Crespo,
  Su{\'a}rez~{\'A}lvarez, Fern{\'a}ndez~Alonso, Mu{\~n}iz et~al.}]{garcia22}
\bibinfo{author}{Garc{\'\i}a~Crespo, F.J.},
  \bibinfo{author}{Su{\'a}rez~{\'A}lvarez, J.},
  \bibinfo{author}{Fern{\'a}ndez~Alonso, R.}, \bibinfo{author}{Mu{\~n}iz, J.},
  et~al., \bibinfo{year}{2022}.
\newblock \bibinfo{title}{Academic resilience in mathematics and science:
  Europe timss-2019 data}.
\newblock \bibinfo{journal}{Psicothema} .
\bibitem[{Hastie et~al.(2009)Hastie, Tibshirani and Friedman}]{hastie09}
\bibinfo{author}{Hastie, T.}, \bibinfo{author}{Tibshirani, R.},
  \bibinfo{author}{Friedman, J.}, \bibinfo{year}{2009}.
\newblock \bibinfo{title}{The Elements of Statistical Learning: Data Mining,
  Inference, and Prediction}.
\newblock \bibinfo{publisher}{Springer}.
\bibitem[{Hilbert et~al.(2021)Hilbert, Coors, Kraus, Bischl, Lindl, Frei, Wild,
  Krauss, Goretzko and Stachl}]{hilbert21}
\bibinfo{author}{Hilbert, S.}, \bibinfo{author}{Coors, S.},
  \bibinfo{author}{Kraus, E.}, \bibinfo{author}{Bischl, B.},
  \bibinfo{author}{Lindl, A.}, \bibinfo{author}{Frei, M.},
  \bibinfo{author}{Wild, J.}, \bibinfo{author}{Krauss, S.},
  \bibinfo{author}{Goretzko, D.}, \bibinfo{author}{Stachl, C.},
  \bibinfo{year}{2021}.
\newblock \bibinfo{title}{Machine learning for the educational sciences}.
\newblock \bibinfo{journal}{Review of Education} \bibinfo{volume}{9},
  \bibinfo{pages}{e3310}.
\bibitem[{Ke et~al.(2017)Ke, Meng, Finley, Wang, Chen, Ma, Ye and Liu}]{ke17}
\bibinfo{author}{Ke, G.}, \bibinfo{author}{Meng, Q.}, \bibinfo{author}{Finley,
  T.}, \bibinfo{author}{Wang, T.}, \bibinfo{author}{Chen, W.},
  \bibinfo{author}{Ma, W.}, \bibinfo{author}{Ye, Q.}, \bibinfo{author}{Liu,
  T.Y.}, \bibinfo{year}{2017}.
\newblock \bibinfo{title}{Lightgbm: A highly efficient gradient boosting
  decision tree}.
\newblock \bibinfo{journal}{Advances in neural information processing systems}
  \bibinfo{volume}{30}.
\bibitem[{Krauss et~al.(2017)Krauss, Do and Huck}]{krauss17}
\bibinfo{author}{Krauss, C.}, \bibinfo{author}{Do, X.A.},
  \bibinfo{author}{Huck, N.}, \bibinfo{year}{2017}.
\newblock \bibinfo{title}{Deep neural networks, gradient-boosted trees, random
  forests: Statistical arbitrage on the s\&p 500}.
\newblock \bibinfo{journal}{European Journal of Operational Research}
  \bibinfo{volume}{259}, \bibinfo{pages}{689--702}.
\bibitem[{Kumari and Srivastava(2017)}]{kumari17}
\bibinfo{author}{Kumari, R.}, \bibinfo{author}{Srivastava, S.K.},
  \bibinfo{year}{2017}.
\newblock \bibinfo{title}{Machine learning: A review on binary classification}.
\newblock \bibinfo{journal}{International Journal of Computer Applications}
  \bibinfo{volume}{160}.
\bibitem[{Lundberg et~al.(2020)Lundberg, Erion, Chen, DeGrave, Prutkin, Nair,
  Katz, Himmelfarb, Bansal and Lee}]{lundberg20}
\bibinfo{author}{Lundberg, S.M.}, \bibinfo{author}{Erion, G.},
  \bibinfo{author}{Chen, H.}, \bibinfo{author}{DeGrave, A.},
  \bibinfo{author}{Prutkin, J.M.}, \bibinfo{author}{Nair, B.},
  \bibinfo{author}{Katz, R.}, \bibinfo{author}{Himmelfarb, J.},
  \bibinfo{author}{Bansal, N.}, \bibinfo{author}{Lee, S.I.},
  \bibinfo{year}{2020}.
\newblock \bibinfo{title}{From local explanations to global understanding with
  explainable ai for trees}.
\newblock \bibinfo{journal}{Nature machine intelligence} \bibinfo{volume}{2},
  \bibinfo{pages}{56--67}.
\bibitem[{Lundberg and Lee(2017)}]{lundberg17}
\bibinfo{author}{Lundberg, S.M.}, \bibinfo{author}{Lee, S.I.},
  \bibinfo{year}{2017}.
\newblock \bibinfo{title}{A unified approach to interpreting model
  predictions}, in: \bibinfo{booktitle}{Proceedings of the 31st International
  Conference on Neural Information Processing Systems},
  \bibinfo{publisher}{Curran Associates Inc.}, \bibinfo{address}{Red Hook, NY,
  USA}. p. \bibinfo{pages}{4768–4777}.
\bibitem[{Lustig et~al.(2023)Lustig, Pabon, Neidh{\"o}fer and
  Tommasi}]{lustig23}
\bibinfo{author}{Lustig, N.}, \bibinfo{author}{Pabon, V.M.},
  \bibinfo{author}{Neidh{\"o}fer, G.}, \bibinfo{author}{Tommasi, M.},
  \bibinfo{year}{2023}.
\newblock \bibinfo{title}{Short and long-run distributional impacts of covid-19
  in latin america}.
\newblock \bibinfo{journal}{Econom{\'\i}a} \bibinfo{volume}{22},
  \bibinfo{pages}{96--116}.
\bibitem[{Luthar et~al.(2000)Luthar, Cicchetti and Becker}]{luthar00}
\bibinfo{author}{Luthar, S.S.}, \bibinfo{author}{Cicchetti, D.},
  \bibinfo{author}{Becker, B.}, \bibinfo{year}{2000}.
\newblock \bibinfo{title}{The construct of resilience: A critical evaluation
  and guidelines for future work}.
\newblock \bibinfo{journal}{Child development} \bibinfo{volume}{71},
  \bibinfo{pages}{543--562}.
\bibitem[{Martin et~al.(2022)Martin, Burns, Collie, Cutmore, MacLeod and
  Donlevy}]{martin22}
\bibinfo{author}{Martin, A.J.}, \bibinfo{author}{Burns, E.C.},
  \bibinfo{author}{Collie, R.J.}, \bibinfo{author}{Cutmore, M.},
  \bibinfo{author}{MacLeod, S.}, \bibinfo{author}{Donlevy, V.},
  \bibinfo{year}{2022}.
\newblock \bibinfo{title}{The role of engagement in immigrant students’
  academic resilience}.
\newblock \bibinfo{journal}{Learning and Instruction} \bibinfo{volume}{82},
  \bibinfo{pages}{101650}.
\bibitem[{Martin and Marsh(2006)}]{martin06}
\bibinfo{author}{Martin, A.J.}, \bibinfo{author}{Marsh, H.W.},
  \bibinfo{year}{2006}.
\newblock \bibinfo{title}{Academic resilience and its psychological and
  educational correlates: A construct validity approach}.
\newblock \bibinfo{journal}{Psychology in the Schools} \bibinfo{volume}{43},
  \bibinfo{pages}{267--281}.
\bibitem[{Munguia(2023)}]{munguia23}
\bibinfo{author}{Munguia, N.}, \bibinfo{year}{2023}.
\newblock \bibinfo{title}{Covid-19 and its influence on sustainable development
  goal 4: Latin america and caribbean region}, in: \bibinfo{booktitle}{SDGs in
  the Americas and Caribbean Region}. \bibinfo{publisher}{Springer}, pp.
  \bibinfo{pages}{1--17}.
\bibitem[{Murillo and Garrido(2017)}]{murillo17}
\bibinfo{author}{Murillo, F.J.}, \bibinfo{author}{Garrido, C.M.},
  \bibinfo{year}{2017}.
\newblock \bibinfo{title}{Estimaci{\'o}n de la magnitud de la segregaci{\'o}n
  escolar en am{\'e}rica latina}.
\newblock \bibinfo{journal}{Magis: Revista Internacional de Investigaci{\'o}n
  en Educaci{\'o}n} \bibinfo{volume}{9}, \bibinfo{pages}{11--30}.
\bibitem[{Murillo et~al.(2023)Murillo, Mart{\'\i}nez-Garrido and
  Gra{\~n}a}]{murillo23}
\bibinfo{author}{Murillo, J.}, \bibinfo{author}{Mart{\'\i}nez-Garrido, C.},
  \bibinfo{author}{Gra{\~n}a, R.}, \bibinfo{year}{2023}.
\newblock \bibinfo{title}{Segregaci{\'o}n escolar por nivel socioecon{\'o}mico
  en educaci{\'o}n primaria en am{\'e}rica latina y el caribe}.
\newblock \bibinfo{journal}{REICE-Revista Iberoamericana sobre Calidad,
  Eficacia y Cambio en Educaci{\'o}n} .
\bibitem[{Naidu et~al.(2023)Naidu, Zuva and Sibanda}]{naidu23}
\bibinfo{author}{Naidu, G.}, \bibinfo{author}{Zuva, T.},
  \bibinfo{author}{Sibanda, E.M.}, \bibinfo{year}{2023}.
\newblock \bibinfo{title}{A review of evaluation metrics in machine learning
  algorithms}, in: \bibinfo{booktitle}{Computer Science On-line Conference},
  \bibinfo{organization}{Springer}. pp. \bibinfo{pages}{15--25}.
\bibitem[{Natekin and Knoll(2013)}]{natekin13}
\bibinfo{author}{Natekin, A.}, \bibinfo{author}{Knoll, A.},
  \bibinfo{year}{2013}.
\newblock \bibinfo{title}{Gradient boosting machines, a tutorial}.
\newblock \bibinfo{journal}{Frontiers in neurorobotics} \bibinfo{volume}{7},
  \bibinfo{pages}{21}.
\bibitem[{Neidh{\"o}fer et~al.(2021)Neidh{\"o}fer, Lustig and
  Tommasi}]{neidhofer21}
\bibinfo{author}{Neidh{\"o}fer, G.}, \bibinfo{author}{Lustig, N.},
  \bibinfo{author}{Tommasi, M.}, \bibinfo{year}{2021}.
\newblock \bibinfo{title}{Intergenerational transmission of lockdown
  consequences: prognosis of the longer-run persistence of covid-19 in latin
  america}.
\newblock \bibinfo{journal}{The Journal of Economic Inequality}
  \bibinfo{volume}{19}, \bibinfo{pages}{571--598}.
\bibitem[{Ntamwiza and Bwire(2025)}]{ntamwiza25}
\bibinfo{author}{Ntamwiza, J.M.}, \bibinfo{author}{Bwire, H.},
  \bibinfo{year}{2025}.
\newblock \bibinfo{title}{Predicting biking preferences in kigali city: A
  comparative study of traditional statistical models and ensemble machine
  learning models}.
\newblock \bibinfo{journal}{Transport Economics and Management}
  \bibinfo{volume}{3}, \bibinfo{pages}{78--91}.
\bibitem[{OECD(2011)}]{oecd11}
\bibinfo{author}{OECD}, \bibinfo{year}{2011}.
\newblock \bibinfo{title}{Against the odds: Disadvantaged students who succeed
  in school}.
\newblock \bibinfo{publisher}{ERIC Clearinghouse}.
\bibitem[{OECD(2016)}]{oecd16}
\bibinfo{author}{OECD}, \bibinfo{year}{2016}.
\newblock \bibinfo{title}{PISA 2015 results: Policies and practices for
  successful schools}.
\newblock \bibinfo{publisher}{OECD}.
\bibitem[{OECD(2023)}]{oecd23a}
\bibinfo{author}{OECD}, \bibinfo{year}{2023}.
\newblock \bibinfo{title}{PISA 2022 Results (Volume I): The State of Learning
  and Equity in Education}.
\newblock \bibinfo{type}{Technical Report}.
\newblock \DOIprefix\doi{https://doi.org/10.1787/53f23881-en}.
\bibitem[{Qiu et~al.(2022)Qiu, Chen, Dincer, Lundberg, Kaeberlein and
  Lee}]{qiu22}
\bibinfo{author}{Qiu, W.}, \bibinfo{author}{Chen, H.}, \bibinfo{author}{Dincer,
  A.B.}, \bibinfo{author}{Lundberg, S.}, \bibinfo{author}{Kaeberlein, M.},
  \bibinfo{author}{Lee, S.I.}, \bibinfo{year}{2022}.
\newblock \bibinfo{title}{Interpretable machine learning prediction of
  all-cause mortality}.
\newblock \bibinfo{journal}{Communications medicine} \bibinfo{volume}{2},
  \bibinfo{pages}{125}.
\bibitem[{Rainio et~al.(2024)Rainio, Teuho and Kl{\'e}n}]{rainio24}
\bibinfo{author}{Rainio, O.}, \bibinfo{author}{Teuho, J.},
  \bibinfo{author}{Kl{\'e}n, R.}, \bibinfo{year}{2024}.
\newblock \bibinfo{title}{Evaluation metrics and statistical tests for machine
  learning}.
\newblock \bibinfo{journal}{Scientific Reports} \bibinfo{volume}{14},
  \bibinfo{pages}{6086}.
\bibitem[{Raschka and Mirjalili(2019)}]{raschka19}
\bibinfo{author}{Raschka, S.}, \bibinfo{author}{Mirjalili, V.},
  \bibinfo{year}{2019}.
\newblock \bibinfo{title}{Python machine learning: Machine learning and deep
  learning with Python, scikit-learn, and TensorFlow 2}.
\newblock \bibinfo{publisher}{Packt publishing ltd}.
\bibitem[{Rudd et~al.(2021)Rudd, Meissel and Meyer}]{rudd21}
\bibinfo{author}{Rudd, G.}, \bibinfo{author}{Meissel, K.},
  \bibinfo{author}{Meyer, F.}, \bibinfo{year}{2021}.
\newblock \bibinfo{title}{Measuring academic resilience in quantitative
  research: A systematic review of the literature}.
\newblock \bibinfo{journal}{Educational research review} \bibinfo{volume}{34},
  \bibinfo{pages}{100402}.
\bibitem[{Rutter(2006)}]{rutter06}
\bibinfo{author}{Rutter, M.}, \bibinfo{year}{2006}.
\newblock \bibinfo{title}{Implications of resilience concepts for scientific
  understanding}.
\newblock \bibinfo{journal}{Annals of the new York Academy of Sciences}
  \bibinfo{volume}{1094}, \bibinfo{pages}{1--12}.
\bibitem[{Rutter(2012)}]{rutter12}
\bibinfo{author}{Rutter, M.}, \bibinfo{year}{2012}.
\newblock \bibinfo{title}{Resilience as a dynamic concept}.
\newblock \bibinfo{journal}{Development and psychopathology}
  \bibinfo{volume}{24}, \bibinfo{pages}{335--344}.
\bibitem[{Shapley(1953)}]{shapley53}
\bibinfo{author}{Shapley, L.S.}, \bibinfo{year}{1953}.
\newblock \bibinfo{title}{A value for n-person games}.
\newblock \bibinfo{journal}{Contribution to the Theory of Games}
  \bibinfo{volume}{2}.
\bibitem[{{\v{S}}trumbelj and Kononenko(2014)}]{vstrumbelj14}
\bibinfo{author}{{\v{S}}trumbelj, E.}, \bibinfo{author}{Kononenko, I.},
  \bibinfo{year}{2014}.
\newblock \bibinfo{title}{Explaining prediction models and individual
  predictions with feature contributions}.
\newblock \bibinfo{journal}{Knowledge and information systems}
  \bibinfo{volume}{41}, \bibinfo{pages}{647--665}.
\bibitem[{Tudor and Spray(2017)}]{tudor17}
\bibinfo{author}{Tudor, K.E.}, \bibinfo{author}{Spray, C.M.},
  \bibinfo{year}{2017}.
\newblock \bibinfo{title}{Approaches to measuring academic resilience: A
  systematic review}.
\newblock \bibinfo{journal}{International Journal of Research Studies in
  Education} \bibinfo{volume}{7}.
\bibitem[{Vicente et~al.(2021)Vicente, Pastor and Soler}]{vicente21}
\bibinfo{author}{Vicente, I.}, \bibinfo{author}{Pastor, J.M.},
  \bibinfo{author}{Soler, {\'A}.}, \bibinfo{year}{2021}.
\newblock \bibinfo{title}{Improving educational resilience in the oecd
  countries: Two convergent paths}.
\newblock \bibinfo{journal}{Journal of Policy Modeling} \bibinfo{volume}{43},
  \bibinfo{pages}{1149--1166}.
\bibitem[{Wang et~al.(2022)Wang, King and Leung}]{wang22}
\bibinfo{author}{Wang, F.}, \bibinfo{author}{King, R.B.},
  \bibinfo{author}{Leung, S.O.}, \bibinfo{year}{2022}.
\newblock \bibinfo{title}{Beating the odds: Identifying the top predictors of
  resilience among hong kong students}.
\newblock \bibinfo{journal}{Child Indicators Research} \bibinfo{volume}{15},
  \bibinfo{pages}{1921--1944}.
\bibitem[{Wang and Gordon(2012)}]{wang12}
\bibinfo{author}{Wang, M.C.}, \bibinfo{author}{Gordon, E.W.},
  \bibinfo{year}{2012}.
\newblock \bibinfo{title}{Educational resilience in inner-city America:
  Challenges and prospects}.
\newblock \bibinfo{publisher}{Routledge}.
\bibitem[{Wills and Hofmeyr(2019)}]{wills19}
\bibinfo{author}{Wills, G.}, \bibinfo{author}{Hofmeyr, H.},
  \bibinfo{year}{2019}.
\newblock \bibinfo{title}{Academic resilience in challenging contexts: Evidence
  from township and rural primary schools in south africa}.
\newblock \bibinfo{journal}{International Journal of Educational Research}
  \bibinfo{volume}{98}, \bibinfo{pages}{192--205}.
\bibitem[{Windle et~al.(2011)Windle, Bennett and Noyes}]{windle11}
\bibinfo{author}{Windle, G.}, \bibinfo{author}{Bennett, K.M.},
  \bibinfo{author}{Noyes, J.}, \bibinfo{year}{2011}.
\newblock \bibinfo{title}{A methodological review of resilience measurement
  scales}.
\newblock \bibinfo{journal}{Health and quality of life outcomes}
  \bibinfo{volume}{9}, \bibinfo{pages}{1--18}.
\bibitem[{Xenofontos and Mouroutsou(2023)}]{xenofontos23}
\bibinfo{author}{Xenofontos, C.}, \bibinfo{author}{Mouroutsou, S.},
  \bibinfo{year}{2023}.
\newblock \bibinfo{title}{Resilience in mathematics education research: a
  systematic review of empirical studies}.
\newblock \bibinfo{journal}{Scandinavian Journal of Educational Research}
  \bibinfo{volume}{67}, \bibinfo{pages}{1041--1055}.
\bibitem[{Yavuz and Kutlu(2016)}]{yavuz16}
\bibinfo{author}{Yavuz, H.{\c{C}}.}, \bibinfo{author}{Kutlu, {\"O}.},
  \bibinfo{year}{2016}.
\newblock \bibinfo{title}{Investigation of the factors affecting the academic
  resilience of economically disadvantaged high school students}.
\newblock \bibinfo{journal}{Education and Science} \bibinfo{volume}{41}.
\bibitem[{Ye et~al.(2021)Ye, Strietholt and Bl{\"o}meke}]{ye21}
\bibinfo{author}{Ye, W.}, \bibinfo{author}{Strietholt, R.},
  \bibinfo{author}{Bl{\"o}meke, S.}, \bibinfo{year}{2021}.
\newblock \bibinfo{title}{Academic resilience: Underlying norms and validity of
  definitions}.
\newblock \bibinfo{journal}{Educational Assessment, Evaluation and
  Accountability} \bibinfo{volume}{33}, \bibinfo{pages}{169--202}.
\bibitem[{Zhang and Cutumisu(2024)}]{zhang24}
\bibinfo{author}{Zhang, Y.}, \bibinfo{author}{Cutumisu, M.},
  \bibinfo{year}{2024}.
\newblock \bibinfo{title}{Predicting the mathematics literacy of resilient
  students from high-performing economies: A machine learning approach}.
\newblock \bibinfo{journal}{Studies in Educational Evaluation}
  \bibinfo{volume}{83}, \bibinfo{pages}{101412}.
\bibitem[{Zheng et~al.(2024)Zheng, Cheung and Sit}]{zheng24}
\bibinfo{author}{Zheng, J.q.}, \bibinfo{author}{Cheung, K.c.},
  \bibinfo{author}{Sit, P.s.}, \bibinfo{year}{2024}.
\newblock \bibinfo{title}{A systematic review of academic resilience in east
  asia: Evidence from the large-scale assessment research}.
\newblock \bibinfo{journal}{Psychology in the Schools} \bibinfo{volume}{61},
  \bibinfo{pages}{1238--1254}.

\end{thebibliography}


\newpage

\newpage
\clearpage

\begin{table}[htbp]
\begin{threeparttable}
\caption{Student academic resilience indicators}
\begin{tabular}{llllll}
\hline
 & Whole Sample & School – public & School – private & School – rural & School – urban \\
 \hline
 & (1) & (2) & (3) & (4) & (5) \\
SAR1 & 0.212 & 0.194 & 0.400 & 0.135 & 0.259 \\
SAR2 & 0.117 & 0.095 & 0.348 & 0.044 & 0.162 \\
SAR3 & 0.112 & 0.103 & 0.211 & 0.072 & 0.137 \\
SAR4 & 0.052 & 0.037 & 0.213 & 0.023 & 0.070 \\
 &   &   &   &   &  \\
N & 17,058  & 15,599  & 1,459  &  6,505 & 10,553 \\
\hline
\end{tabular}
\begin{tablenotes}[para,flushleft]
\medskip
\footnotesize{
Notes: (1) Students academic resilience rates (SAR) are calculated for the two bottom families SES quintiles and are defined as follows: (i) SAR1: Level 2 and above. (ii) SAR2: Prediction top 40\% (multilevel model with family and school SES). (iii) SAR3: Level 2 and above plus corr SES top half distribution. (iv) SAR4: Prediction top 40\% excluding highest 20\% of highest schools random intercepts.
}
\end{tablenotes}
\label{table1}
\end{threeparttable}
\end{table}

\begin{sidewaystable}
\begin{threeparttable}
\caption{Covariates differences by student  academic resilience status – students variables}
\begin{tabular}{llllllllll}
\hline
 & \multicolumn{4}{c}{SAR1}  & &  \multicolumn{4}{c}{SAR2}   \\
  \cmidrule{2-5}\cmidrule{7-10}
 & SAR – no & SAR – yes & Difference &  &  & SAR – no & SAR – yes & Difference &  \\
 & (1) & (2) & (3) &  &  & (4) & (5) & (6) &  \\
Age & 15.805 & 15.869 & 0.06 & *** &  & 15.813 & 15.857 & 0.04 & *** \\
Gender-male & 0.465 & 0.569 & 0.10 & *** &  & 0.484 & 0.508 & 0.02 & ** \\
Satisfied with life & 7.164 & 6.699 & -0.47 & *** &  & 7.115 & 6.697 & -0.42 & *** \\
Assertiveness (index) & -0.112 & 0.006 & 0.12 & *** &  & -0.095 & -0.027 & 0.07 & *** \\
Cooperation (index) & 0.179 & 0.141 & -0.04 & *** &  & 0.175 & 0.140 & -0.03 & * \\
Curiosity (index) & 0.002 & 0.158 & 0.16 & *** &  & 0.024 & 0.115 & 0.09 & *** \\
Emotional control (index) & 0.002 & 0.158 & 0.16 & *** &  & 0.024 & 0.115 & 0.09 & *** \\
Empathy (index) & -0.061 & 0.149 & 0.21 & *** &  & -0.045 & 0.203 & 0.25 & *** \\
Perseverance (index) & 0.173 & 0.171 & 0.00 &  &  & 0.178 & 0.126 & -0.05 & ** \\
Stress management (index) & 0.060 & 0.097 & 0.04 & ** &  & 0.060 & 0.123 & 0.06 & *** \\
Homework-time spent & 2.976 & 3.729 & 0.75 & *** &  & 3.070 & 3.629 & 0.56 & *** \\
Number of digital devices at home & 6.441 & 7.797 & 1.36 & *** &  & 6.552 & 8.058 & 1.51 & *** \\
Number of books at home & 1.038 & 1.432 & 0.39 & *** &  & 1.078 & 1.452 & 0.37 & *** \\
Immigrant & 0.054 & 0.033 & -0.02 & *** &  & 0.051 & 0.042 & -0.01 & * \\
Speak other language & 0.058 & 0.013 & -0.04 & *** &  & 0.053 & 0.017 & -0.04 & *** \\
Late entrant in primary & 0.252 & 0.137 & -0.11 & *** &  & 0.236 & 0.161 & -0.07 & *** \\
Repeated in primary & 0.228 & 0.036 & -0.19 & *** &  & 0.205 & 0.055 & -0.15 & *** \\
Repeated in lower secondary & 0.172 & 0.042 & -0.13 & *** &  & 0.156 & 0.057 & -0.10 & *** \\
Missed school (more than 3 months in a row) & 0.101 & 0.027 & -0.07 & *** &  & 0.092 & 0.032 & -0.06 & *** \\
Skipped a day of school in last 2 weeks  & 0.290 & 0.219 & -0.07 & *** &  & 0.280 & 0.238 & -0.04 & *** \\
Teacher support for well being (index, low) & 0.278 & 0.350 & 0.07 & *** &  & 0.284 & 0.361 & 0.08 & *** \\
Sense of belonging to school (index, low) & 0.168 & 0.170 & 0.00 &  &  & 0.170 & 0.158 & -0.01 &  \\
School climate – bad (index, low) & 0.486 & 0.520 & 0.03 & *** &  & 0.489 & 0.521 & 0.03 & *** \\
Safety in school and surroundings area (index, low) & 0.478 & 0.427 & -0.05 & *** &  & 0.469 & 0.450 & -0.02 &  \\
Work unpaid days & 3.105 & 3.020 & -0.09 & ** &  & 3.104 & 2.962 & -0.14 & *** \\
Work paid days & 1.464 & 0.822 & -0.64 & *** &  & 1.384 & 0.905 & -0.48 & *** \\
Mother educational level & 3.543 & 3.689 & 0.15 & *** &  & 3.524 & 3.951 & 0.43 & *** \\
Father educational level & 3.600 & 3.733 & 0.13 & *** &  & 3.592 & 3.905 & 0.31 & *** \\
Number of siblings & 2.348 & 2.022 & -0.33 & *** &  & 2.319 & 1.976 & -0.34 & *** \\
 &  &  &  &  &  &  &  &  &  \\
N & 13,445 & 3,613 &  &  &  & 15,064 & 1,994 &  &  \\
\hline
\end{tabular}
\label{table2}
\begin{tablenotes}[para,flushleft]
\medskip
\footnotesize{
(1) Significance levels: * p $\leq$ 0.10, ** p $\leq$ 0.05, ***p $\leq$ 0.01.
}
\end{tablenotes}
\end{threeparttable}
\end{sidewaystable}

\begin{sidewaystable}
\begin{threeparttable}
\caption{Covariates differences by student academic resilience status – school and COVID-19 variables}
\begin{tabular}{llllllllll}
\hline
 & \multicolumn{4}{c}{SAR1}  & &  \multicolumn{4}{c}{SAR2}   \\
  \cmidrule{2-5}\cmidrule{7-10}
 & SAR – no & SAR – yes & Difference &  &  & SAR – no & SAR – yes & Difference &  \\
 & (1) & (2) & (3) &  &  & (4) & (5) & (6) \\
\textit{Panel A – School variables } &  &  &  &  &  &  &  &  \\
Urban & 0.582 & 0.756 & 0.17 & *** &  & 0.587 & 0.856 & 0.27*** \\
Private & 0.065 & 0.162 & 0.10 & *** &  & 0.063 & 0.255 & 0.19*** \\
Community size & 2.699 & 3.151 & 0.45 & *** &  & 2.717 & 3.381 & 0.66*** \\
Competition  & 0.722 & 0.808 & 0.09 & *** &  & 0.730 & 0.820 & 0.09*** \\
Government funding (\%) & 0.816 & 0.779 & -0.04 & *** &  & 0.812 & 0.778 & -0.03*** \\
Fees funding (\%) & 0.116 & 0.152 & 0.04 & *** &  & 0.116 & 0.178 & 0.06*** \\
Students w/ a second language (\%) & 0.086 & 0.045 & -0.04 &  &  & 0.084 & 0.030 & -0.05*** \\
Students w/ learning needs (\%) & 0.045 & 0.044 & 0.00 & *** &  & 0.045 & 0.040 & 0.00*** \\
Disadvantaged students (\%) & 0.494 & 0.425 & -0.07 &  &  & 0.498 & 0.344 & -0.15*** \\
Enrolment & 763 & 1017 & 254 & *** &  & 775 & 1129 & 354*** \\
STR & 19.656 & 20.963 & 1.31 & *** &  & 19.865 & 20.439 & 0.57 \\
Teachers attended PD programme in last 3 months (\%) & 0.402 & 0.414 & 0.01 & * &  & 0.403 & 0.412 & 0.01 \\
Certified teachers (\%) & 0.603 & 0.615 & 0.01 &  &  & 0.605 & 0.609 & 0.00 \\
Teachers w/ a degree (\%) & 0.607 & 0.555 & -0.05 & *** &  & 0.604 & 0.529 & -0.07*** \\
PC connected web-teacher ratio & 0.256 & 0.291 & 0.03 & *** &  & 0.255 & 0.333 & 0.08*** \\
No admission criteria used & 0.168 & 0.194 & 0.03 & *** &  & 0.167 & 0.226 & 0.06*** \\
No transfer criteria used & 0.510 & 0.528 & 0.02 & * &  & 0.508 & 0.556 & 0.05*** \\
Social and emotional skills policy  & 0.721 & 0.728 & 0.01 &  &  & 0.722 & 0.726 & 0.00 \\
Methods to monitor the practice of teachers & 2.508 & 2.518 & 0.01 &  &  & 2.498 & 2.599 & 0.10*** \\
Policy on ability grouping  & 0.379 & 0.378 & 0.00 &  &  & 0.381 & 0.360 & -0.02* \\
ICT infrastructure per student (index, low) & 0.497 & 0.611 & 0.11 & *** &  & 0.507 & 0.626 & 0.12*** \\
Autonomy (index, low) & 0.591 & 0.582 & -0.01 &  &  & 0.597 & 0.532 & -0.07*** \\
Improvements and quality policies (index, low) & 0.469 & 0.472 & 0.00 &  &  & 0.469 & 0.469 & 0.00 \\
Staff and studs support for inclusivity (index, low) & 0.489 & 0.493 & 0.00 &  &  & 0.489 & 0.491 & 0.00 \\
Parents involvement fostered by school (index, low) & 0.478 & 0.545 & 0.07 & *** &  & 0.484 & 0.556 & 0.07*** \\
 &  &  &  &  &  &  &  &  \\
\textit{Panel B – School COVID-19 variables } &  &  &  &  &  &  &  &  \\
Proportion of studs attending distance learning & 7.942 & 8.483 & 0.54 & *** &  & 7.922 & 9.113 & 1.19*** \\
Teachers have skills for instruction & 0.802 & 0.823 & 0.02 & *** &  & 0.801 & 0.851 & 0.05*** \\
Teachers have time to integrate digital devices in lessons  & 0.565 & 0.515 & -0.05 & *** &  & 0.554 & 0.557 & 0.00 \\
Teachers have learning support platform & 0.519 & 0.632 & 0.11 & *** &  & 0.520 & 0.716 & 0.20*** \\
Remote instruction hindered by digital devices access & 4.880 & 4.497 & -0.38 & *** &  & 4.889 & 4.117 & -0.77*** \\
Remote instruction hindered by internet access & 5.002 & 4.590 & -0.41 & *** &  & 5.016 & 4.145 & -0.87*** \\
Remote instruction hindered by system and material & 4.184 & 3.659 & -0.52 & *** &  & 4.190 & 3.183 & -1.01*** \\
Remote instruction hindered by all factors (index) & 0.317 & 0.051 & -0.27 & *** &  & 0.321 & -0.206 & -0.53*** \\
Number of days school closed & 234.6 & 225.6 & -9.0 & *** &  & 235.7 & 210.3 & -25.4*** \\
 &  &  &  &  &  &  &  &  \\
N & 13,445 & 3,613 &  &  &  & 15,064 & 1,994 &  \\
\hline
\end{tabular}
\label{table3}
\begin{tablenotes}[para,flushleft]
\medskip
\footnotesize{
(1) Significance levels: * p $\leq$ 0.10, ** p $\leq$ 0.05, ***p $\leq$ 0.01.
}
\end{tablenotes}
\end{threeparttable}
\end{sidewaystable}

\begin{table}[htbp]
\begin{threeparttable}
\caption{Comparison of ML models – parameters search. Whole sample}
\begin{tabular}{llll}
\hline
Grid search  & Chosen parameters & AUROC score & AUPRC score \\
\hline
 & (1) & (2) & (3) \\
1. Logit &  &\\
Penalty (L1, L2) &  &  \\
C = (0.1, 1, 10) &  &  \\
SAR1 & L1, 1 & 0.804 & 0.514 \\
SAR2 & L1, 1 & 0.816 & 0.365 \\
SAR3 & L1, 0.1 & 0.754 & 0.253 \\
SAR4 & L2, 0.1 & 0.774 & 0.142 \\
 &  &  \\
2. Neural networks &  &  \\
Hidden layer sizes &  &  \\
\{200\}, \{100, 100\}, \{200, 100, 50\} &  &  \\
Activation (relu, tanh, logistic) &  &  \\
SAR1 & \{200\}, logistic & 0.867 & 0.644 \\
SAR2 & \{200, 100, 50\}, logistic & 0.925 & 0.689 \\
SAR3 & \{200, 100, 50\}, tanh & 0.827 & 0.368 \\
SAR4 & \{200\}, logistic & 0.920 & 0.530 \\
 &  &  \\
3. Gradient boosted trees  &  &  \\
number of estimators &  &  \\
(100, 500, 1000, 5000) &  &  \\
subsample ratio (0.5, 0.7, 0.9) &  &  \\
trees of max depth (3, 5, 7, 9) &  &  \\
learning rate (0.001, 0.01, 0.1) &  &  \\
SAR1 & 5000, 0.5, 3, 0.01 & \textbf{0.911} & \textbf{0.742} \\
SAR2 & 1000, 0.9, 5, 0.1 & \textbf{0.989} & \textbf{0.940} \\
SAR3 & 500, 0.7, 7, 0.1 & \textbf{0.910} & \textbf{0.650} \\
SAR4 & 5000, 0.9, 7, 0.1 & \textbf{0.979}  & \textbf{0.872} \\
\hline
\end{tabular}
\label{table4}
\end{threeparttable}
\end{table}


\appendix


\renewcommand{\thesubsection}{\Alph{subsection}}
\setcounter{figure}{0}
\renewcommand{\thefigure}{A\arabic{figure}}
\setcounter{table}{0}
\renewcommand{\thetable}{A\arabic{table}}

\begin{table}[htbp]
\scriptsize{
\caption{Variables description}
\begin{tabular}{ll}
\hline
Variables & Definition \\
\hline
\textit{Panel A – student variables} &  \\
StudBKGD\_Age & Student – age \\
StudBKGD\_Assertive & Student – assertiveness, index \\
StudBKGD\_BelongSchLow & Student – sense of belonging to school (0-High, 1-Low) \\
StudBKGD\_Books & Number of books at home \\
StudBKGD\_Cooperation & Student – cooperation, index \\
StudBKGD\_Curiosity & Student – curiosity, index \\
StudBKGD\_DD & Number of digital devices at home \\
StudBKGD\_EmotionControl & Student – emotional control, index \\
StudBKGD\_Empathy & Student – empathy, index \\
StudBKGD\_FatherEdu & Father educational level \\
StudBKGD\_Gender & Student – gender (0-Female, 1-Male) \\
StudBKGD\_Homework & Student – homework in all subjects-time spent \\
StudBKGD\_Immigrant & Student – immigrant (0-No, 1-Yes) \\
StudBKGD\_LatePrim & Student – late entrant in primary (0-No, 1-Yes) \\
StudBKGD\_MissClass & Student – missed school $\ge$ 3 months in a row (0-No, 1-Yes) \\
StudBKGD\_MotherEdu & Mother educational level \\
StudBKGD\_OtherLang & Student – speak other language (0-No, 1-Yes) \\
StudBKGD\_Perseverance & Student – perseverance, index \\
StudBKGD\_RepeLowsec & Student – repeated in lower secondary (0-No, 1-Yes) \\
StudBKGD\_RepePrim & Student – repeated in primary (0-No, 1-Yes) \\
StudBKGD\_SatisLife & Student – satisfied with life \\
StudBKGD\_SchClimaBadLow & Student – bad school climate (0-High, 1-Low) \\
StudBKGD\_SchSafeLackLow & Student – lack of school safety and in the area (0-High, 1-Low) \\
StudBKGD\_Sibs & Number of siblings \\
StudBKGD\_SkippSch & Student – skipped 1 day of school in last 2 weeks  (0-No, 1-Yes) \\
StudBKGD\_StressManage & Student – stress management, index \\
StudBKGD\_TchSuppWBLow & Student – teacher support well being (0-High, 1-Low) \\
StudBKGD\_WorkPaid & Student – work paid days \\
StudBKGD\_WorkUnpaid & Student – work unpaid days \\
\textit{Panel B – school variables} &  \\
SchBKGD\_AbilityGroup & School – policy on ability grouping (0-No, 1-Yes) \\
SchBKGD\_AdmissionNone & School – no admission criteria used (0-No, 1-Yes) \\
SchBKGD\_AutonomyLow & School – autonomy (0-High, 1-Low) \\
SchBKGD\_Competition & School – competition (0-No, 1-Yes) \\
SchBKGD\_Disadvantaged & School – disadvantaged studs (\%) \\
SchBKGD\_Enrolment & School – enrolment \\
SchBKGD\_FundFees & School – fees funding (\%) \\
SchBKGD\_FundGovt & School – government funding (\%) \\
SchBKGD\_ICTInfraLow & School – ICT infrastructure per stud (0-High, 1-Low) \\
SchBKGD\_ICTInfra & School – ICT infrastructure per stud \\
SchBKGD\_LearnNeeds & School – studs w/ a second language (\%) \\
SchBKGD\_NoLangTest & School – studs w/ learning needs (\%) \\
SchBKGD\_ParentInvolLow & School – parents involvement fostered by school (0-High, 1-Low) \\
SchBKGD\_PCWebTR & School – PC connected web-teacher ratio \\
SchBKGD\_PolSocEmSkill & School – social and emotional skills policy (0-No, 1-Yes) \\
SchBKGD\_Private & School – private (0-No, 1-Yes) \\
SchBKGD\_QualPolLow & School – improvements and quality policies (0-High, 1-Low) \\
SchBKGD\_STR & School – STR \\
SchBKGD\_SuppInclusityLow & School – staff and studs support for inclusivity (0-High, 1-Low) \\
SchBKGD\_TchCertified & School – certified teachers (\%) \\
SchBKGD\_TchDegree & School – teachers w/ a degree (\%) \\
SchBKGD\_TchMonitor & School – methods to monitor the practice of teachers \\
SchBKGD\_TchPDAtt & School – teachers attended PD programme in last 3 months (\%) \\
SchBKGD\_TestStud & School – testing studs learning \\
SchBKGD\_Townsize & School – community size \\
SchBKGD\_TransferNone & School – no transfer criteria used (0-No, 1-Yes) \\
SchBKGD\_Urban & School – urban (0-No, 1-Yes) \\
CovidBKGD\_BarrierRemoteLDD & Covid – remote instruction hindered by digital devices access \\
CovidBKGD\_BarrierRemoteLSysEM & Covid – remote instruction hindered by system and educational material \\
CovidBKGD\_BarrierRemoteLWeb & Covid – remote instruction hindered by internet access \\
CovidBKGD\_Closeddays & Covid – number of days school closed \\
CovidBKGD\_PropStudRemoteL & Covid – proportion of studs attending distance learning \\
CovidBKGD\_TchOnlinePlatform & Covid – teachers have learning support platform (0-SD or D, 1-SA or A) \\
CovidBKGD\_TchSkillsRemoteL & Covid – teachers have skills for instruction (0-SD or D, 1-SA or A) \\
CovidBKGD\_TchTimeLessonDD & Covid – teachers have time to integrate DD in lessons (0-SD or D, 1-SA or A) \\
CovidBKGD\_BarrierRemoteLAll & Covid – remote instruction hindered by all factors \\
\hline
\end{tabular}
\label{tableA1}
}
\end{table}


\newpage
\renewcommand{\thesubsection}{\Alph{subsection}}
\setcounter{figure}{0}
\renewcommand{\thefigure}{B\arabic{figure}}

\clearpage
\begin{figure}[ht!]
\centering
\subcaptionbox{Private-public (SAR3)}{\includegraphics[width=.49\textwidth]{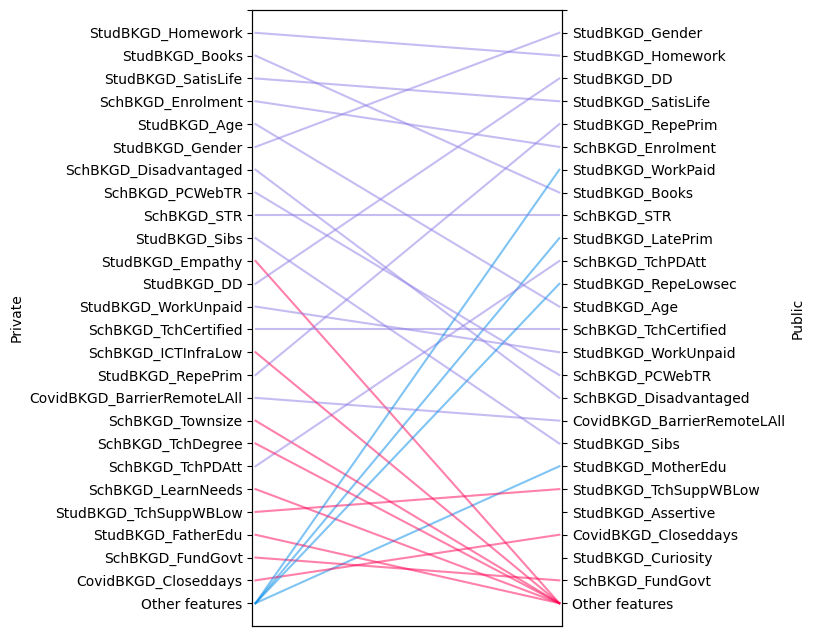}}
\subcaptionbox{Private-public (SAR4)}{\includegraphics[width=.49\textwidth]{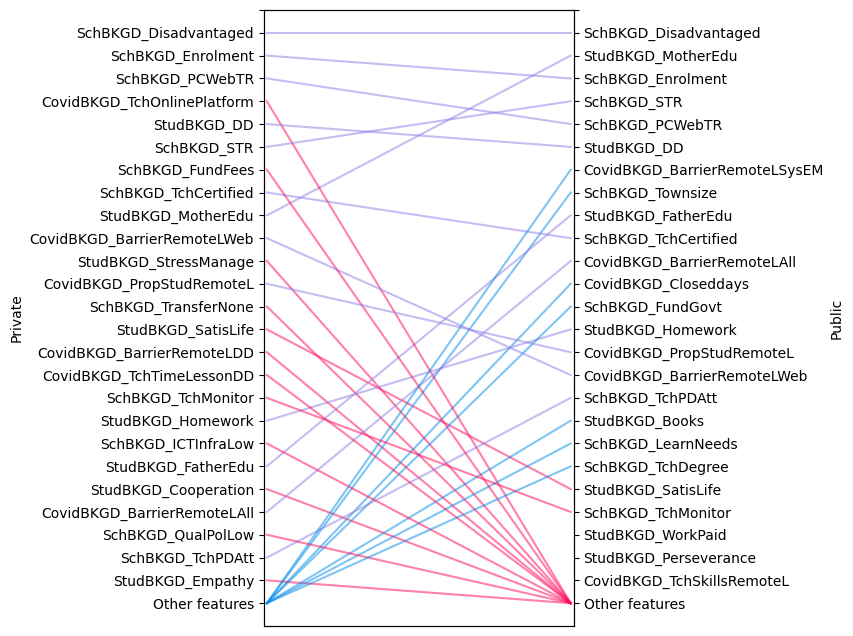}}

\subcaptionbox{Urban-rural (SAR3)}{\includegraphics[width=.49\textwidth]{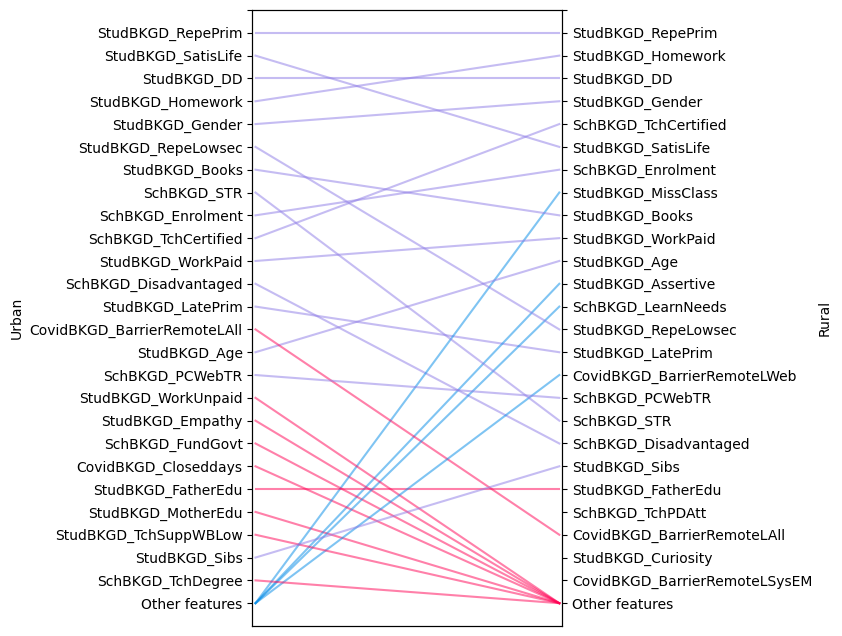}}
\subcaptionbox{Urban-rural (SAR4)}{\includegraphics[width=.49\textwidth]{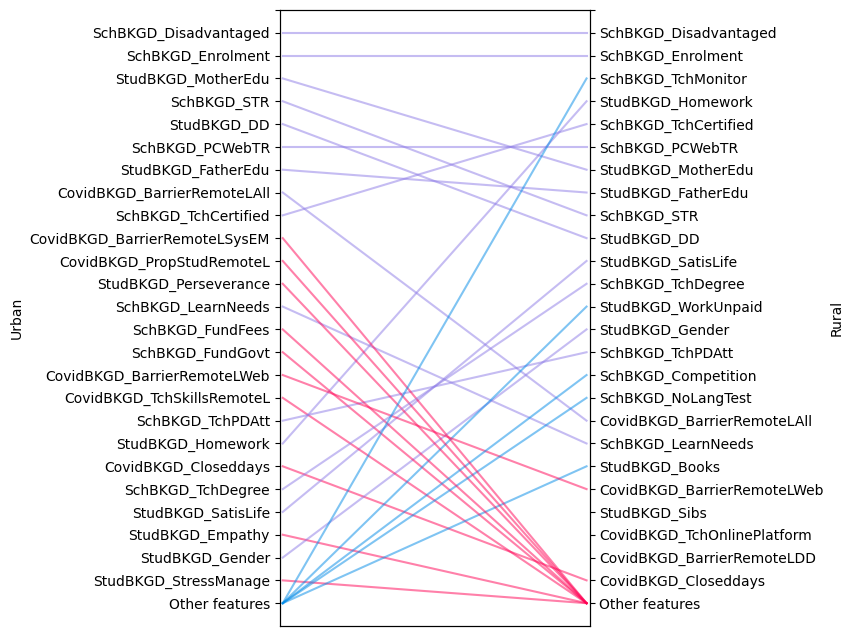}}
\caption{Sub-samples determinants comparison. Outcomes: SAR3 and SAR4}
\label{figureB1}
      \medskip
\begin{minipage}{0.99\textwidth}
\footnotesize{
Notes: (1) Country plots show the first 25 covariates in order of associations given by absolute values of average SHAP values for each covariate.
}
\end{minipage}
\end{figure}

\begin{figure}[hp!]
\centering
\subcaptionbox{SAR3 (profile: high)}{\includegraphics[width=.49\textwidth]{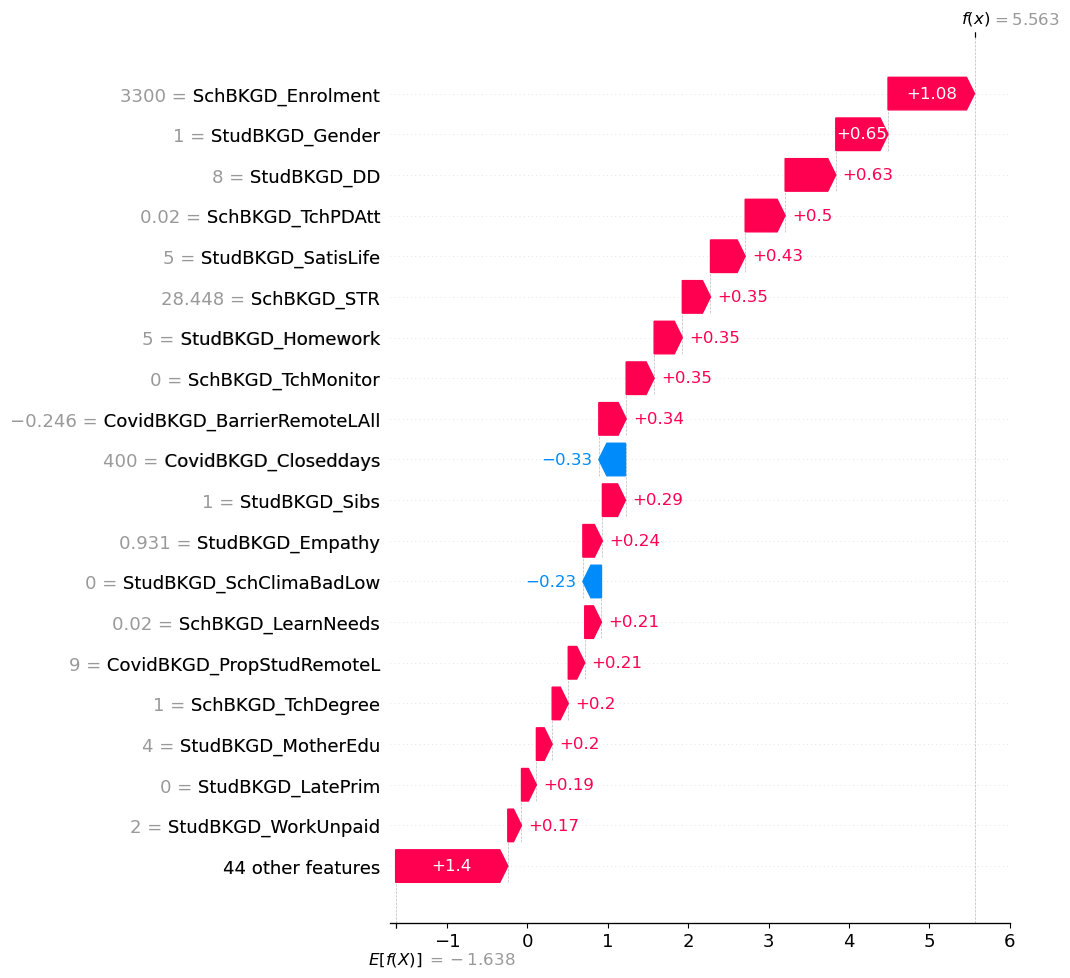}}
\subcaptionbox{SAR3 (profile: low)}{\includegraphics[width=.49\textwidth]{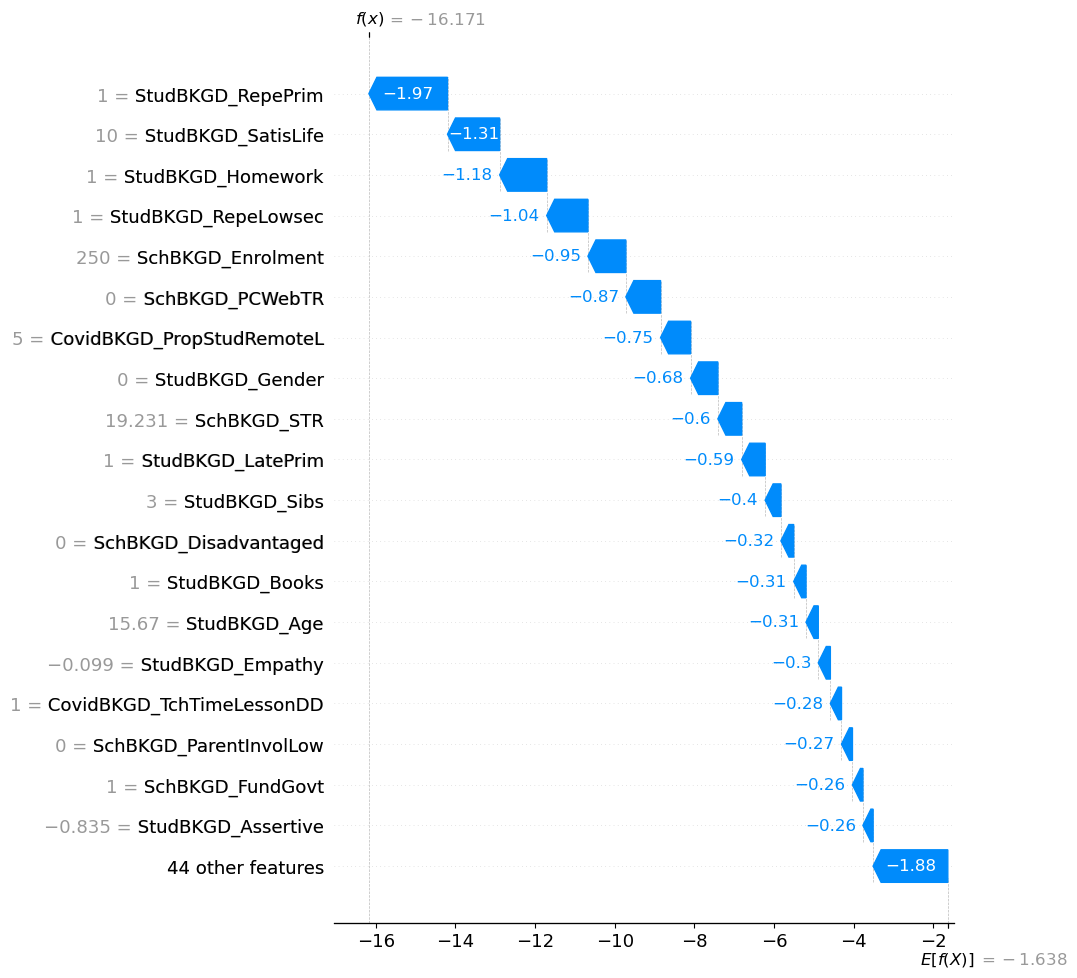}}

\subcaptionbox{SAR4 (profile: high)}{\includegraphics[width=.49\textwidth]{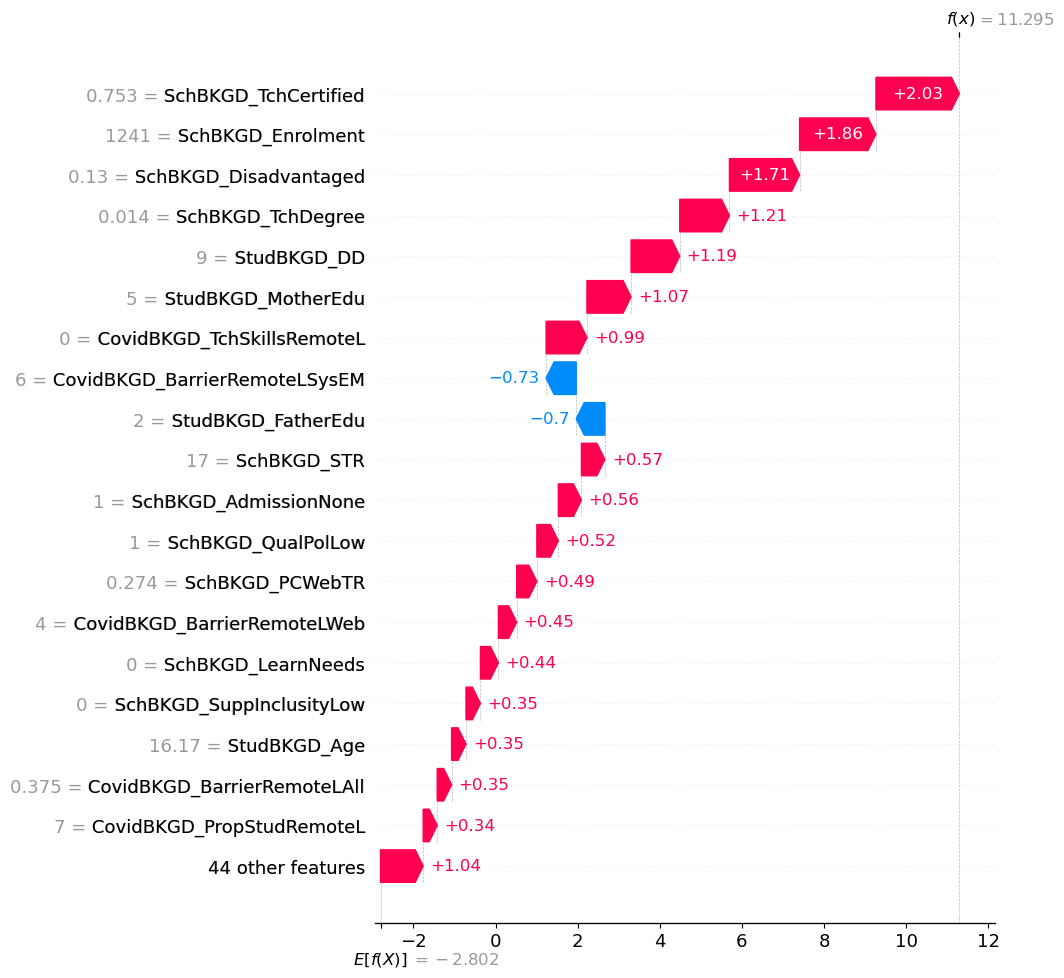}}
\subcaptionbox{SAR4 (profile: low)}{\includegraphics[width=.49\textwidth]{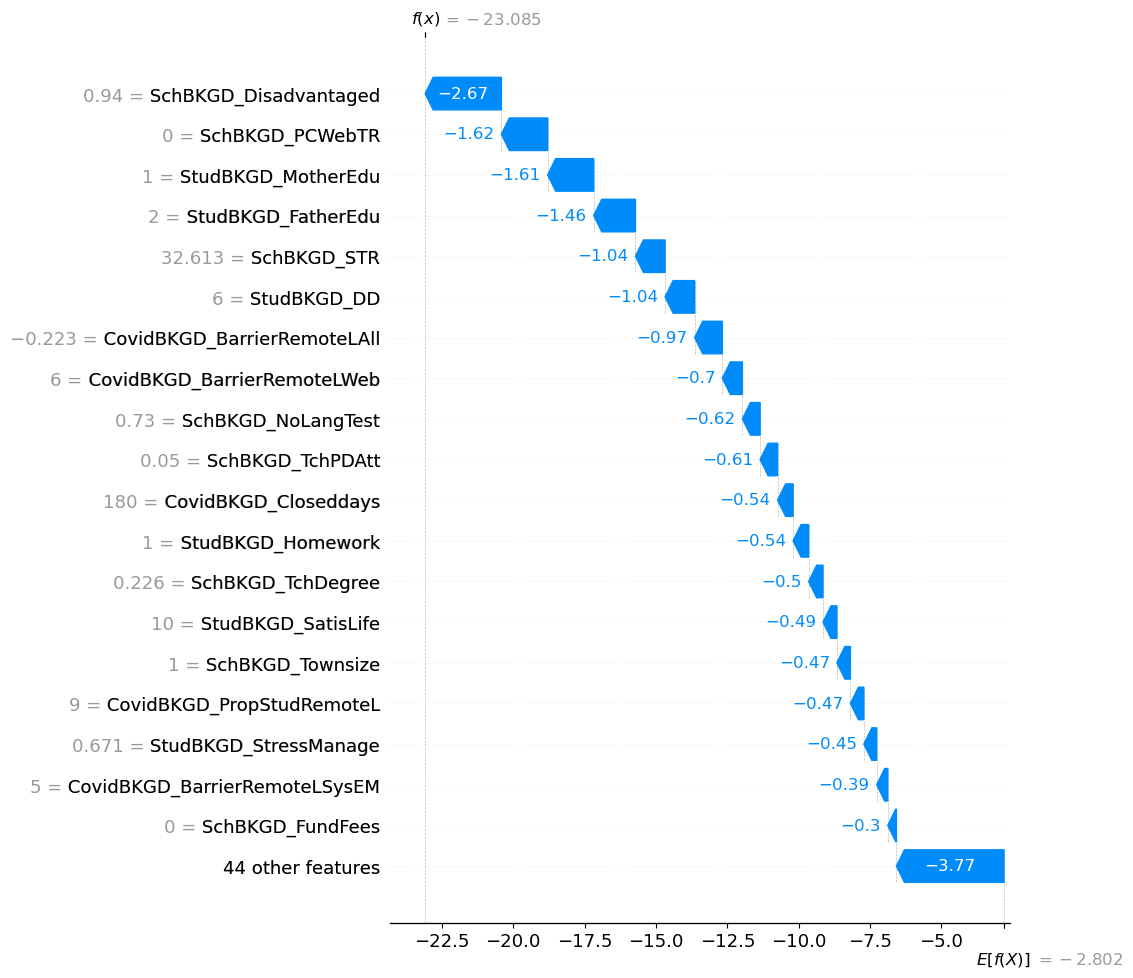}}
\caption{Covariates individual SHAP values for students with the lowest and highest average SHAP values contributions (indicators: SAR3 and SAR4)}
\label{figureB2}
\medskip
\begin{minipage}{0.99\textwidth}
\footnotesize{
Notes: (1) See Table \ref{tableA1} for variables' acronyms, definitions and associated labels to values (in grey, left to covariate name).
}
\end{minipage}
\end{figure}


\newpage
\renewcommand{\thesubsection}{\Alph{subsection}}
\setcounter{figure}{0}
\renewcommand{\thefigure}{C\arabic{figure}}
\setcounter{table}{0}
\renewcommand{\thetable}{C\arabic{table}}


\begin{sidewaystable}
\caption{Comparison of ML models – parameters search. Sub-samples}
\begin{tabular}{llllllll}
\hline
  & Chosen parameters & AUROC score & AUPRC score &  & Chosen parameters & AUROC score & AUPRC score \\
\hline
 & (1) & (2) & (3) &    & (4) & (5) & (6)   \\
  \cmidrule{2-4}\cmidrule{6-8}
 & \multicolumn{3}{c}{Private schools}  & &  \multicolumn{3}{c}{Public schools}   \\
  \cmidrule{2-4}\cmidrule{6-8}

1. Logit &  &  &  &  &  &  \\
SAR1 & L1, 10 & 0.7178 & 0.5787 &  & L1, 0.1 & 0.7760 & 0.4439 \\
SAR2 & L1, 10 & 0.7492 & 0.5633 &  & L1, 1 & 0.7862 & 0.2616 \\
SAR3 & L1, 0.1 & 0.5102 & 0.2515 &  & L1, 0.1 & 0.7181 & 0.2067 \\
SAR4 & L1, 0.1 & 0.7408 & 0.3540 &  & L2, 10 & 0.7187 & 0.0661 \\
 &  &  &  &  &  &  \\
2. Neural networks &  &  &  &  &  &  \\
SAR1 & \{100, 100\}, logistic & 0.7885 & 0.6650 &  & \{200\}, logistic & 0.8497 & 0.5772 \\
SAR2 & \{200\}, tanh & 0.9049 & 0.8535 &  & \{100, 100\}, logistic & 0.9185 & 0.6462 \\
SAR3 & \{200\}, tanh & 0.7547 & 0.5310 &  & \{200\}, logistic & 0.8433 & 0.4117 \\
SAR4 & \{100, 100\}, logistic & 0.8594 & 0.5813 &  & \{200\}, logistic & 0.9272 & 0.5016 \\
 &  &  &  &  &  &  \\
3. Gradient boosted trees  &  &  &  &  &  &  \\
SAR1 & 1000, 0.9, 3, 0.01 & \textbf{0.8580} & \textbf{0.8026} &  & 5000, 0.7, 3, 0.01 & \textbf{0.8983} & \textbf{0.6852} \\
SAR2 & 5000, 0.5, 9, 0.1 & \textbf{0.9489} & \textbf{0.8993} &  & 5000, 0.7, 3, 0.1 & \textbf{0.9918} & \textbf{0.9423} \\
SAR3 & 500, 0.9, 7, 0.1 & \textbf{0.8313} & \textbf{0.6325} &  & 5000, 0.7, 9, 0.1 & \textbf{0.9006} & \textbf{0.5411} \\
SAR4 &  5000, 0.9, 9, 0.01 & \textbf{0.9315} & \textbf{0.8226} & & 5000, 0.9, 3, 0.1 & \textbf{0.9839} & \textbf{0.8562} \\

  \cmidrule{2-4}\cmidrule{6-8}  & \multicolumn{3}{c}{Urban schools}  & &  \multicolumn{3}{c}{Rural schools}   \\
  \cmidrule{2-4}\cmidrule{6-8}

1. Logit &  &  &  &  &  &  \\
SAR1 & L1, 0.1 & 0.7903 & 0.5418 &  & L1, 0.1 & 0.7836 & 0.3215 \\
SAR2 & L1, 1 & 0.7882 & 0.3757 &  & L1, 0.1 & 0.7472 & 0.1358 \\
SAR3 & L1, 0.1 & 0.7232 & 0.2277 &  & L1, 0.1 & 0.7430 & 0.1765 \\
SAR4 & L1, 0.1 & 0.7533 & 0.1547 &  & L2, 10 & 0.7533 & 0.0932 \\
 &  &  &  &  &  &  \\
2. Neural networks &  &  &  &  &  &  \\
SAR1 & \{200\}, logistic & 0.8592 & 0.6622 &  & \{200, 100, 50\}, relu & 0.7934 & 0.4084 \\
SAR2 & \{200\}, logistic & 0.9418 & 0.7778 &  & \{200\}, logistic & 0.9214 & 0.4697 \\
SAR3 & \{200\}, logistic & 0.8458 & 0.4591 &  & \{200\}, tanh & 0.8305 & 0.2966 \\
SAR4 & \{200\}, logistic & 0.9079 & 0.5492 &  & \{200\}, tanh & 0.9098 & 0.6269 \\
 &  &  &  &  &  &  \\
3. Gradient boosted trees  &  &  &  &  &  &  \\
SAR1 & 1000, 0.9,7, 0.01 & \textbf{0.8955} & \textbf{0.7333} &  & 5000, 0.5, 3, 0.01 & \textbf{0.8907} & \textbf{0.6046} \\
SAR2 & 1000, 0.9, 5, 0.1 & \textbf{0.9905} & \textbf{0.9537} &  & 5000, 0.5, 7, 0.1 & \textbf{0.9741} & \textbf{0.8787} \\
SAR3 & 500, 0.7, 7, 0.1 & \textbf{0.9103} & \textbf{0.6035} &  & 5000, 0.9, 3, 0.01 & \textbf{0.9131} & \textbf{0.5567} \\
SAR4 & 5000, 0.9, 3, 0.1 & \textbf{0.9780} & \textbf{0.8783} &  & 5000, 0.9, 3, 0.1 & \textbf{0.9552} & \textbf{0.8159} \\
\hline
\end{tabular}
\label{tableC1}
\end{sidewaystable}

\begin{figure}
\centering
\includegraphics[width=0.99\textwidth]{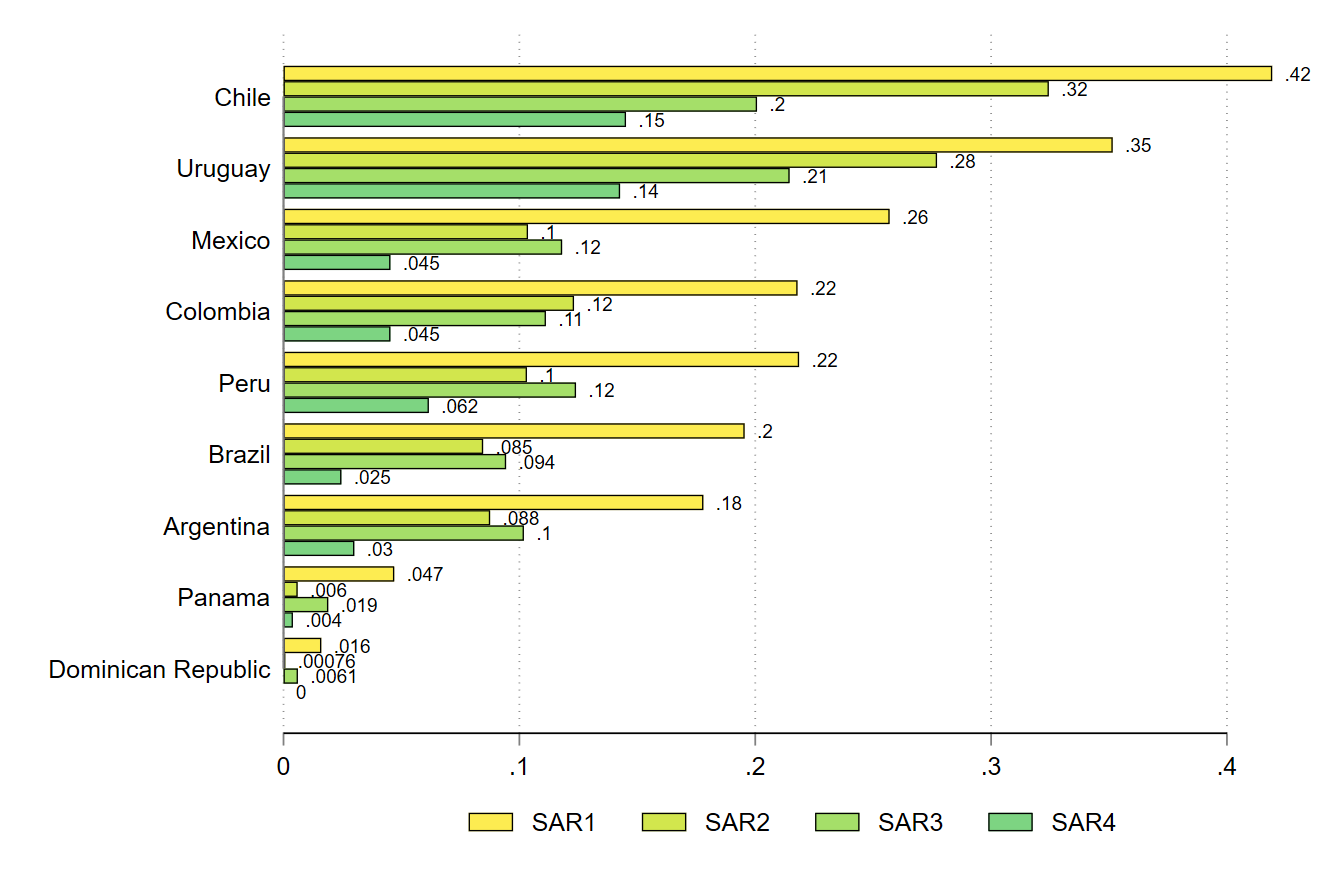}
\caption{Students academic resilience rates per country}
\label{figureC1}
\medskip
\begin{minipage}{0.99\textwidth}
\footnotesize{
Notes: (1) Students academic resilience rates (SAR) are calculated for the two bottom families SES quintiles and are defined as follows: (i) SAR1: Level 2 and above. (ii) SAR2: Prediction top 40\% (multilevel model with family and school SES). (iii) SAR3: Level 2 and above plus corr SES top half distribution. (iv) SAR4: Prediction top 40\% excluding highest 20\% of highest schools random intercepts.
}
\end{minipage}
\end{figure}

\begin{figure}[ht!]
\centering
\subcaptionbox{Chile-Dominican Rep.}{\includegraphics[width=.49\textwidth]{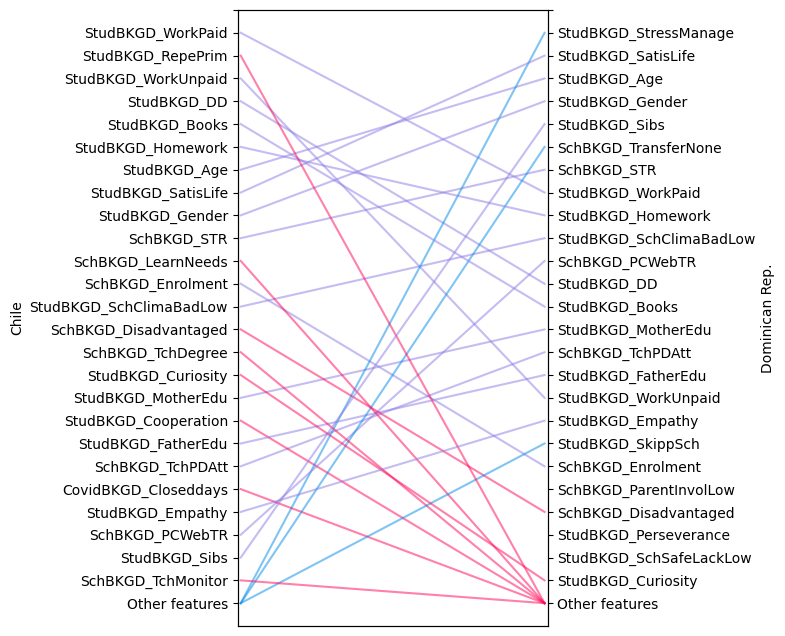}}
\subcaptionbox{Uruguay-Panama}{\includegraphics[width=.49\textwidth]{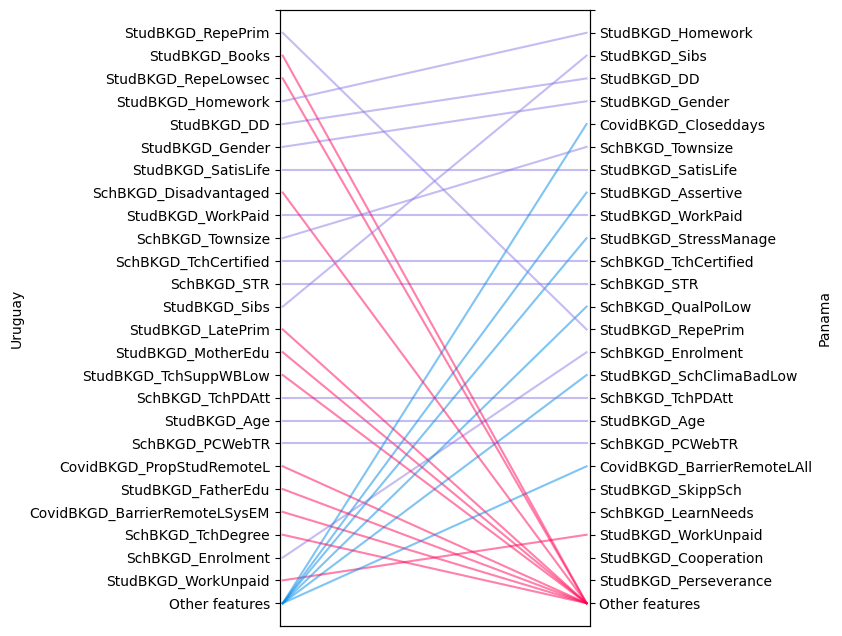}}

\subcaptionbox{Mexico-Argentina}{\includegraphics[width=.49\textwidth]{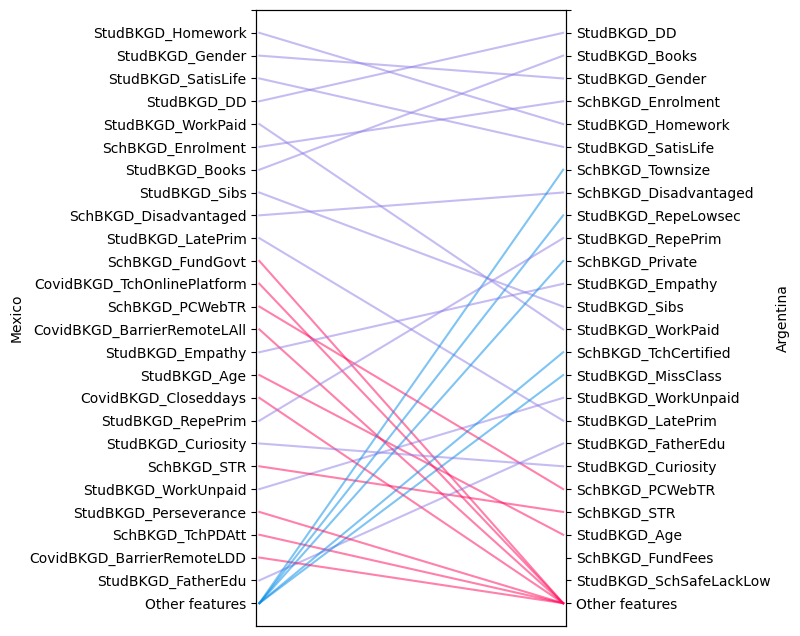}}
\subcaptionbox{Peru-Brazil}{\includegraphics[width=.49\textwidth]{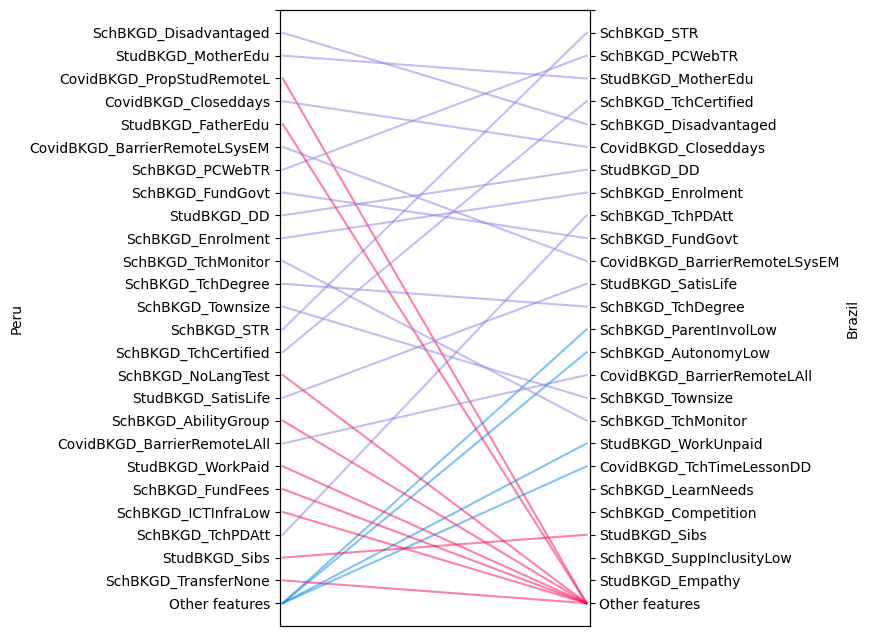}}

\subcaptionbox{Colombia-Dominican Rep.}{\includegraphics[width=.49\textwidth]{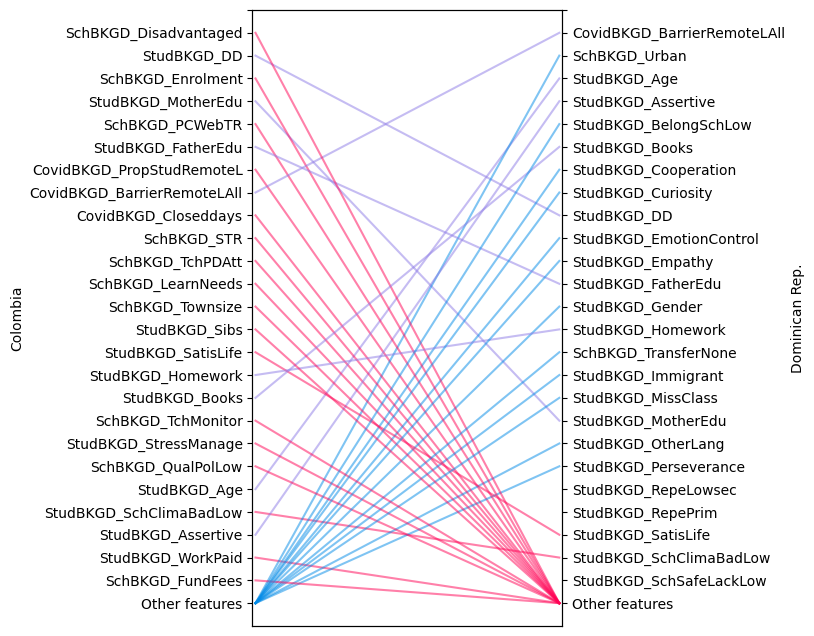}}
\caption{Countries determinants comparison. Outcome: SAR1}
\label{figureC2}
\medskip
\begin{minipage}{0.99\textwidth}
\footnotesize{
Notes: (1) Country plots show the first 25 covariates in order of associations given by absolute values of average SHAP values for each covariate.
}
\end{minipage}
\end{figure}

\begin{figure}[ht!]
\centering
\subcaptionbox{Chile-Dominican Rep.}{\includegraphics[width=.49\textwidth]{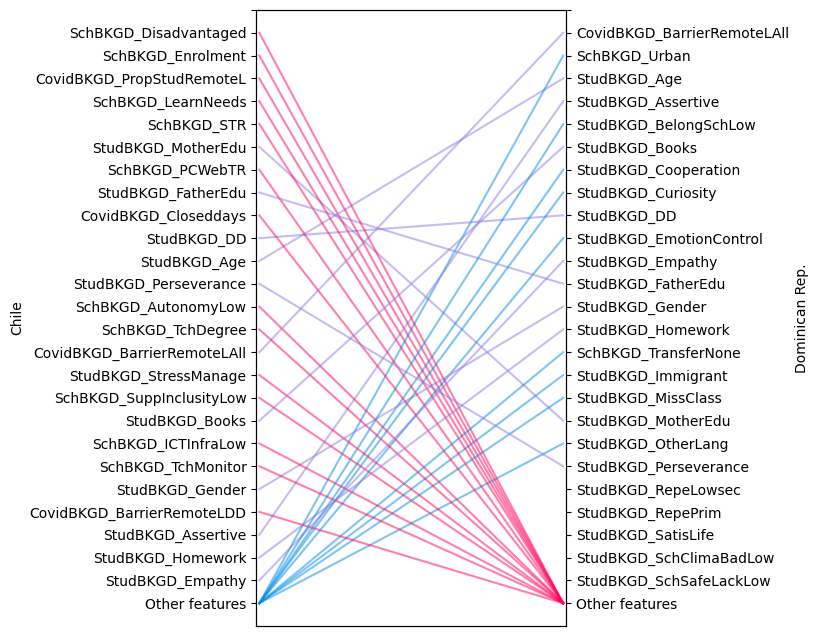}}
\subcaptionbox{Uruguay-Panama}{\includegraphics[width=.49\textwidth]{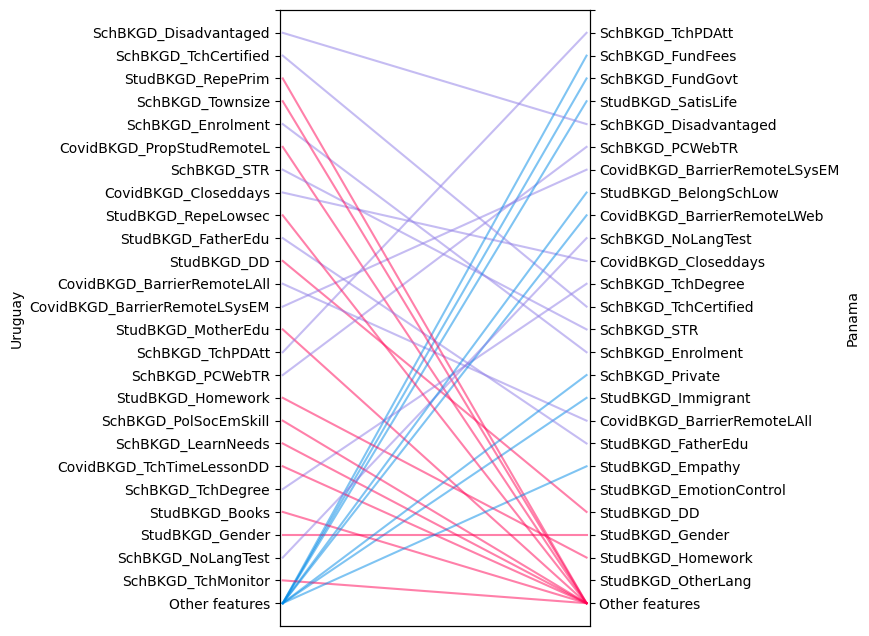}}

\subcaptionbox{Colombia-Brazil}{\includegraphics[width=.49\textwidth]{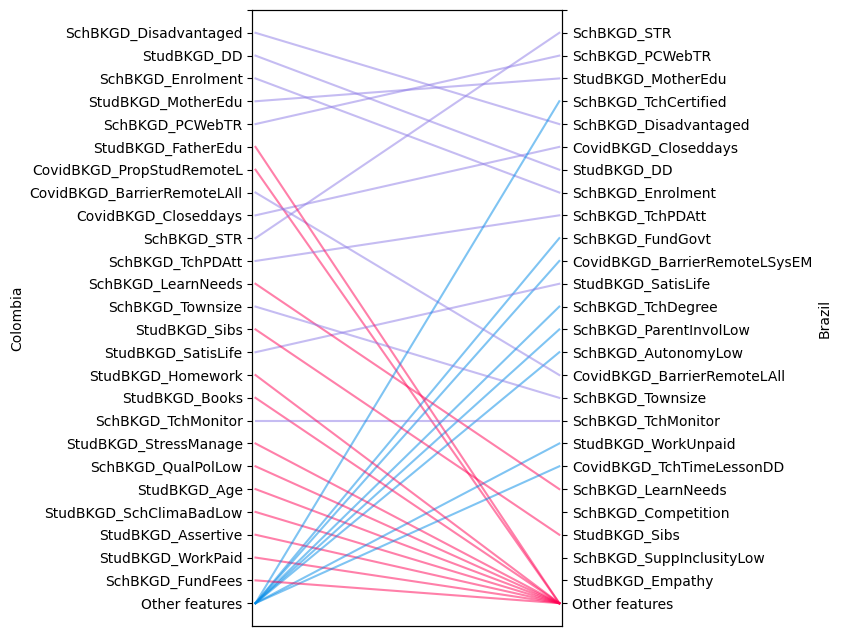}}
\subcaptionbox{Mexico-Argentina}{\includegraphics[width=.49\textwidth]{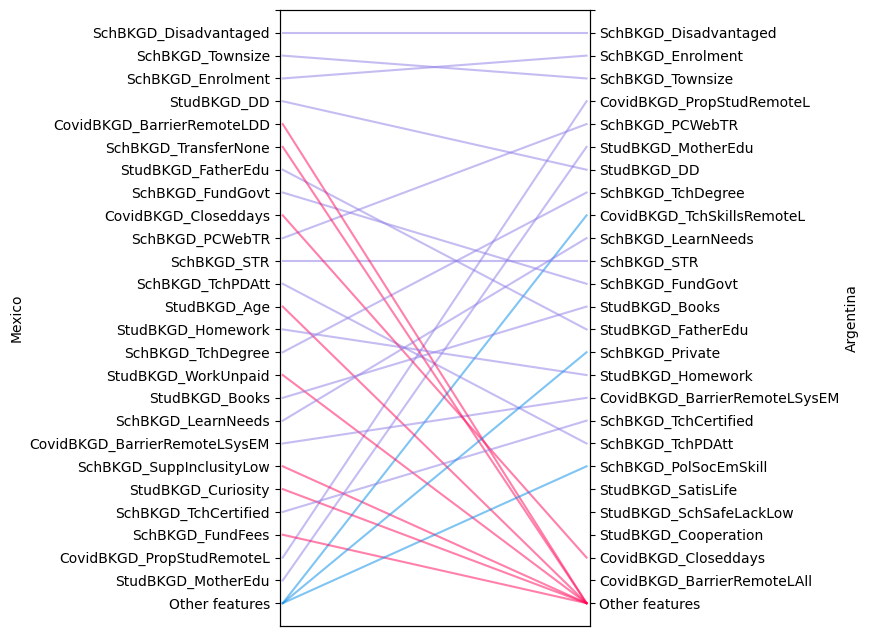}}

\subcaptionbox{Peru-Dominican Rep.}{\includegraphics[width=.49\textwidth]{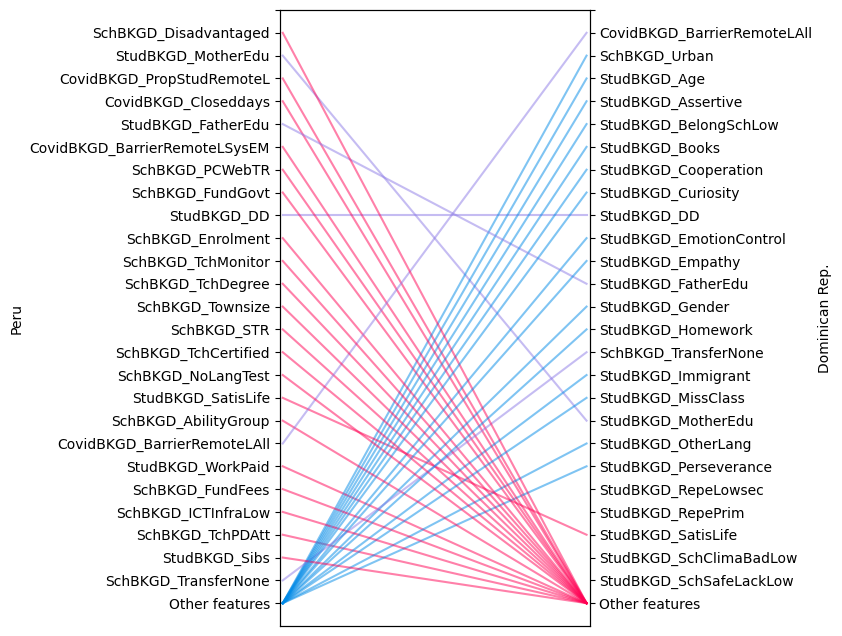}}
\caption{Countries determinants comparison. Outcome: SAR2}
\label{figureC3}
\medskip
\begin{minipage}{0.99\textwidth}
\footnotesize{
Notes: (1) Country plots show the first 25 covariates in order of associations given by absolute values of average SHAP values for each covariate.
}
\end{minipage}
\end{figure}

\end{document}